\documentclass[aps,prb,twocolumn]{revtex4}

\usepackage{graphicx}
\usepackage{color}
\usepackage{amsmath}
\usepackage{amssymb}
\usepackage{bm}

\graphicspath{{./figureExports/}}

\usepackage[english]{babel}

\usepackage{lipsum}

\newcommand{\diff}{\text{d}}

\begin{document}

  \title{Symmetry breaking and (pseudo)spin polarization in Veselago lenses for massless Dirac fermions}

  \author{K. J. A. Reijnders, M. I. Katsnelson}
  \affiliation{Radboud University, Institute for Molecules and Materials,
  Heyendaalseweg 135, 6525 AJ Nijmegen, The Netherlands}
  
  \email{K.Reijnders@science.ru.nl}

  \date{\today}

  \begin{abstract}
    We study Veselago lensing of massless Dirac fermions by \emph{n-p} junctions for electron sources with a certain polarization. This polarization corresponds to pseudospin for graphene and to real spin for topological insulators.
    Both for a point source and for injection into a sample through a narrow lead, we find that polarization leads to spatial symmetry breaking. For the Green's function, this results in a vertical displacement, or even complete vanishing of the main focus, depending on the exact polarization. For injection through a lead, it leads to a difference between the amounts of current emitted with positive and negative transversal momenta.
    We study both systems in detail using the semiclassical approximation. By comparing the results to the exact solutions, we establish that semiclassical methods provide a very effective way to study these systems. For the Green's function, we derive an easy-to-use analytical formula for the vertical displacement of the main focus. For current injection through a lead, we use semiclassical methods to identify two different scattering regimes.
  \end{abstract}

  \maketitle

  \section{Introduction}
  
  Focussing is an effect well-known in optics, where light rays are refracted by a lens to create spots of high intensity. A particular kind of lens was proposed by Veselago,~\cite{Veselago68} who investigated lenses made of materials with a negative refractive index. Such lenses have already been realized in metamaterials,~\cite{Smith00,Houck03,Grbic04} chiral metamaterials~\cite{Tretyakov03,Pendry04,Zhang09,Xiong10} and photonic crystals,~\cite{Parimi04,Cubukcu03} and can be used to produce an image with subwavelength resolution.~\cite{Pendry00,Grbic04}
  Soon after the discovery of graphene, Cheianov \emph{et al.}~\cite{Cheianov07} realized that \mbox{\emph{n-p} junctions} in this material would be ideally suited to create an electronic analog of a Veselago lens.

  Graphene is a two-dimensional gapless semiconductor, whose low-energy charge carriers are governed by the Dirac equation.~\cite{Wallace47,McClure57,Slonczewski58,Semenoff84,CastroNeto09,Katsnelson13}
  This results in peculiar behavior of its electrons, most notably Klein tunneling~\cite{Klein29,Katsnelson06,Cheianov06,Shytov08,Tudorovskiy12,Reijnders13}: an electron normally incident on a potential barrier is transmitted with unit probability. The transmission probability decreases as the angle of incidence increases, meaning that an electron beam is collimated.~\cite{Cheianov06,Shytov08,Tudorovskiy12,Reijnders13}
  A few years after its discovery, Klein tunneling was shown experimentally,~\cite{Young09,Stander09} and recent experiments show that it is still in center of attention today.~\cite{Chen16,Lee16b,Gutierrez16} Notably, the angular dependence of the transmission coefficient has recently been measured.~\cite{Chen16}
  
  A graphene \emph{n-p} junction exhibits (Veselago) lensing, because for electrons the group velocity is parallel to the phase velocity, whereas for holes the group velocity is opposite to the phase velocity.~\cite{Cheianov07} Klein tunneling is crucial in this process, because it makes the \emph{n-p} interface highly transparent to electrons.
  Recently, two experimental groups have demonstrated Veselago lensing in graphene samples. In the first experiment,~\cite{Lee15} the authors measured ballistic transport accross a graphene device, and found an increase in the background-subtracted current in the bipolar regime. In the second experiment,~\cite{Chen16} transverse magnetic focussing was employed to show Veselago lensing, allowing the authors to simultaneously measure the angle-dependent transmission coefficient.
  
  Theoretical papers on the subject have considered the Green's function for a quasi-one-dimensional \emph{n-p} junction,~\cite{Cheianov07} or have looked at circular \emph{n-p} junctions.~\cite{Cserti07,Peterfalvi10,Wu14} In the latter case, semiclassical considerations were also presented,~\cite{Peterfalvi10,Wu14} though the semiclassical approximation to the wavefunction near the main focus was not computed. Another study~\cite{Choi14} considered a Veselago lens in a graphene nanoribbon, and showed that the geometrical phase that is acquired when scattering off a zigzag edge influences the interference pattern. Finally, a numerical study~\cite{Milovanovic15} was conducted where the authors considered \emph{n-p} junctions in graphene samples of realistic size, with current entering from a narrow lead. The authors compared their findings to a semiclassical billiard model,~\cite{Beenakker89,Milovanovic13} and generally found good agreement.
  
  In this paper, we perform a theoretical study of Veselago lenses formed by quasi-one-dimensional \emph{n-p} junctions. We mainly consider the Green's function, although at the end of the paper we also briefly consider the situation where current flows into a sample through a lead that is attached on one of its sides. 
  Our emphasis is not so much on the classical focussing, but rather on the matrix character of the Dirac Hamiltonian and on how it influences the interference pattern. In particular, we consider the case where the point source or incoming wave has a certain sublattice or pseudospin polarization, meaning that the current is not equally distributed among the two graphene sublattices. The fact that we are dealing with spinors makes this problem different from optical problems, where one is usually concerned with the Helmholtz equation.
  
  We study these interference effects using the semiclassical approximation, which is valid when we have a small parameter in our problem. Earlier studies~\cite{Tudorovskiy12,Reijnders13} have shown that this requires either large length scales or high energies. The first step of our analysis is to carefully review the classical problem, for which we need a few elements of the general theory of caustics and wave fronts, a theory known as catastrophe theory.~\cite{Berry80,Poston78,Arnold82,Arnold75} We then apply the stationary phase approximation~\cite{Fedoryuk77,Guillemin77,Maslov81} to our solution. However, the simplest form of this approximation fails near the main focus, where our primary interest lies.
  In order to quantitatively study interference effects near the main focus, we therefore employ the Pearcey approximation.~\cite{Pearcey46,Connor81b,Dobrokhotov14} We also briefly consider the uniform approximation,~\cite{Ursell72,Connor81b} which, in a different form, was successfully applied to graphene for a rather large semiclassical parameter.~\cite{Reijnders13} Because we compare the various approximations with the exact solution, our study can also be considered as a benchmark for the application of various semiclassical methods to graphene.
  
  One of our interests is to see if pseudospin polarization could lead to symmetry breaking between the $K$ and $K'$-valleys in graphene. 
  If this is the case, then it may provide another way of creating valley polarization in graphene.~\cite{Rycerz07} 
  Since charge carriers in both valleys obey the same classical Hamiltonian,~\cite{Katsnelson13} it is clear that the valley polarization we are looking for can only result from quantum interference.
  Therefore, it is unlikely that a polarization of 100\% could be realized in our system. Such a polarization could for instance be detected using the valley Hall effect~\cite{Xiao07,Gorbachev14} or second harmonic generation.~\cite{Wehling15}  
  
  We believe that there may be ways to realize such a sublattice polarization in graphene experimentally. Firstly, one could inject electrons on a single site using a scanning tunneling microscope (STM) with an atomically sharp tip. Secondly, one could consider a device where electrons tunnel into a graphene layer through hexagonal boron nitride (h--BN). Because the strengths of the carbon--nitrogen and carbon--boron interactions differ,~\cite{Sachs11,Bokdam14,VanWijk14} this could lead to an asymmetry between graphene's sublattices. In this context, we note that it has recently been shown experimentally that a device with a few layers of h--BN between two layers of (bilayer) graphene can be used to manipulate the valley and pseudospin state of Dirac electrons.~\cite{Wallbank16}
  We believe that a graphene sample with current flowing in through a graphene lead at one of its sides would be easier to realize. Here, one could create an initial (i.e. in the lead) sublattice polarization by using a substrate that acts differently on both sublattices, giving rise to a mass term in the Dirac equation.~\cite{Katsnelson13}
  
  Regarding the experimental realization of high-energy states in graphene, which would be needed to study the deep semiclassical limit, we note that it is possible to create hole-doped states with energies around 0.5 -- 0.6 eV. This can be achieved by molecular doping~\cite{Bae10,Wehling08,Nair13} with HNO$_3$ or NO$_2$, but can also be reached on a SiO$_2$/Si substrate after proton irradiation.~\cite{Lee16}.   
  Electron doping can for instance be achieved using aniline~\cite{Liu11}, with which one can reach energies of about 0.25~eV. Doping graphene with alkali metals, such as lithium, very high electron doping above 1~eV can be achieved,~\cite{Khademi16} which can also induce superconductivity.~\cite{Profeta12,Ludbrook15}
  
  In our theoretical considerations, we consider a sharp \emph{n-p} junction. Although semiclassical tunneling has been extensively studied for smooth \emph{n-p} junctions,~\cite{Shytov08,Tudorovskiy12,Reijnders13} studying the Green's function for such a junction is far from straightforward. Although considering a sharp barrier is not very realistic from an experimental point of view,~\cite{Chen16} we do not believe that this will significantly influence the main results. An indication for this is given by the aforementioned numerical study,~\cite{Milovanovic15} where the authors found that in going from a sharp barrier to a smooth barrier the main features of the results were preserved. One notable effect should be the broadening of the main focus.~\cite{Phong16,Milovanovic15}
  
  Finally, let us briefly discuss the relation between Veselago lensing in (chiral) metamaterials and in graphene.   
  Whereas in metamaterials negative refraction typically occurs in a narrow frequency band around a resonance, in graphene it occurs for the full range of energies for which the Dirac equation is applicable, which means for energies until about 1~eV.~\cite{Katsnelson13}
  Since, within the Dirac approximation, the classical Hamiltonians of the two valleys in graphene are equal, the classical trajectories in both valleys coincide. Therefore, if (pseudo)spin polarization leads to valley polarization, this has to happen because of quantum interference. The situation is quite different in chiral metamaterials, where the refractive index is different for left-handed and right-handed circularly polarized light.~\cite{Tretyakov03,Pendry04,Zhang09,Xiong10} Hence, the rays, which are the analogs of the classical trajectories, are different for both types of handedness. Furthermore, since one refractive index is typically negative, whilst the other one is positive, the classical rays are focussed for only one handedness and a well-defined polarization can be created.
  
  Although graphene will be our main example in this paper, we stress that the behavior of its charge carriers is not unique. Another class of materials whose electrons follow the massless Dirac equation is formed by the two-dimensional surfaces of three-dimensional topological insulators.~\cite{Hasan10,Qi11,Moore10,Hasan11,Bansil16}
  We are then dealing with real spin instead of pseudospin and using a spin-polarized STM one can inject a single spin.
  Therefore, we will generally refer to charge carriers governed by the Dirac equation as massless Dirac fermions and clearly indicate it when we specialize to the case of graphene.  
  
  The paper is organized in the following way: in section~\ref{sec:veselago}, we discuss the basic equations that describe the Green's function of an electronic Veselago lens for massless Dirac fermions. Subsequently, we discuss classical focussing and caustics in section~\ref{sec:caustics}, and quantum interference and symmetry breaking in section~\ref{sec:symm-breaking}. In section~\ref{sec:semiclassical-evaluation}, we discuss the semiclassical evaluation of the Green's function, and compare various approximations with the exact solution. A semiclassical derivation of the vertical displacement of the maximum that results from (pseudo)spin polarization is presented in section~\ref{sec:asymm-semiclassical}. The resulting formula is tested for the case of graphene. In  section~\ref{sec:curr-edge-samp}, we briefly consider the case where current enters a graphene sample through a narrow graphene lead. We successively discuss the wavefunction, symmetry breaking and the semiclassical evaluation of the wavefunction. Finally, we present our conclusions in section~\ref{sec:conclusion}.

  \section{Veselago lenses}  \label{sec:veselago}
  
  In this section, we introduce the equations that describe an electronic Veselago lens for massless Dirac fermions and review the key results from the literature. We only consider the case of the Green's function here, postponing the case where an electronic current enters the sample from one of its sides to section~\ref{sec:curr-edge-samp}. We split our considerations into three parts. In the first subsection, we define the Green's function and introduce the proper dimensionless parameters. Subsequently, we briefly review the classical focussing that was discussed in Ref.~\onlinecite{Cheianov07}. Finally, we write down the wavefunction for a Veselago lens formed by an \emph{n-p} junction with a (pseudo)spin polarized source.

  \subsection{The Green's function and dimensionless parameters}    \label{subsec:dimless-pars}
  
  The Hamiltonian for two-dimensional massless Dirac fermions is equal to~\cite{Katsnelson13}
  \begin{equation}
    \hat{H} = v_F {\bm \sigma} \cdot \hat{\mathbf{p}} + U(\mathbf{x}) 1_2 ,
    \label{eq:H-massless-Dirac}
  \end{equation}
  where $1_2$ is the two-dimensional unit matrix, the two-dimensional vector ${\bm\sigma} = (\sigma_x,\sigma_y)$ consists of the Pauli matrices and $\hat{\mathbf{p}} = -i \hbar {\bm \nabla}$ is the momentum operator. All position vectors $\mathbf{x}$ are two dimensional, i.e. $\mathbf{x}=(x,y)$. The function $U(\mathbf{x})$ represents the potential to which the charge carriers are subject and the quantity $v_F$ is the Fermi velocity. For the specific case of graphene in the nearest neighbor approximation, it is defined by $\hbar v_F = 3 t a_{CC}/2$, where $t \approx 3$~eV is the hopping parameter and $a_{CC} = 0.142$~nm is the distance between two carbon atoms.~\cite{Katsnelson13}
  
  The Green's function $G(\mathbf{x},\mathbf{x}_0)$ for the Hamiltonian~(\ref{eq:H-massless-Dirac}) is defined by
  \begin{equation}
    \left[ v_F {\bm \sigma} \cdot \hat{\mathbf{p}} + U(\mathbf{x}) 1_2 \right] G(\mathbf{x},\mathbf{x}_0) = E G(\mathbf{x},\mathbf{x}_0) + \delta(\mathbf{x}-\mathbf{x}_0) 1_2, \label{eq:Dirac-Green-dim}
  \end{equation}
  where $\mathbf{x}_0$ indicates the source from which the particles are emitted with energy $E$. For an arbitrary electron source $J(\mathbf{x})$, the equation of motion reads
  \begin{equation}
    \left[ v_F {\bm \sigma} \cdot \hat{\mathbf{p}} + U(\mathbf{x}) 1_2 \right] \Psi(\mathbf{x}) = E \Psi(\mathbf{x}) + J(\mathbf{x}), \label{eq:Dirac-dim}
  \end{equation}
  and the solution is given in terms of the Green's function as
  \begin{equation}
    \Psi(\mathbf{x}) = \int_{-\infty}^\infty \diff \mathbf{x}_0 G(\mathbf{x},\mathbf{x}_0) J(\mathbf{x}_0) . \label{eq:sol-Dirac-dim}
  \end{equation}
  In most of this paper, we will assume that we are dealing with a point source that has a certain polarization, which is pseudospin (sublattice) for the case of graphene~\cite{Katsnelson13} and true spin for the case of the two-dimensional surfaces of three-dimensional topological insulators,~\cite{Qi11} i.e.
  \begin{equation}
    J(\mathbf{x}) = \begin{pmatrix} \alpha_1 \\ \alpha_2 \end{pmatrix} \delta(\mathbf{x}-\mathbf{x}_s) . \label{eq:source}
  \end{equation}
  For convenience, we will assume that the constants $\alpha_i$ are dimensionless and that they form a vector that is normalized, i.e. $|\alpha_1|^2 +|\alpha_2|^2=1$. In practice, depending on the normalization of the source, these constants will however have a dimensionality, which can easily be incorporated into the description. It should also be noted that in the case of graphene one cannot use the continuum approximation when atomically sharp features are present. Hence, the notion of a point source implies that the diameter of the source $d_{\text{source}}$ is much larger than the interatomic distance $a$, yet much smaller than the electronic wavelength $\lambda_{el}$: $a \ll d_{\text{source}} \ll \lambda_{el}$.
  Inserting the source~(\ref{eq:source}) into Eq.~(\ref{eq:sol-Dirac-dim}), we obtain the wavefunction for our problem as
  \begin{equation}
    \Psi(\mathbf{x}) = G(\mathbf{x},\mathbf{x}_s) \begin{pmatrix} \alpha_1 \\ \alpha_2 \end{pmatrix} . \label{eq:sol-Dirac-ptJ-dim}
  \end{equation}
  
  Since we want to perform a semiclassical analysis later on, for which we need to know the true, dimensionless, semiclassical parameter, we restate the problem and its solution in dimensionless parameters. The intrinsic length scale $L$ of the problem is given by the distance from the point source to the junction, which will be introduced more precisely in the next subsection. Furthermore, let $E_0$ be the typical energy scale of the problem, which we take to be $E$ in this paper. Alternatively, one could set $E_0=U_0-E$, with $U_0$ the typical value of $U(\mathbf{x})$, without any essential difference. These definitions allow us to define the dimensionless small parameter $h=\hbar v_F/(E_0 L)$, and the dimensionless quantities $\tilde{\mathbf{x}}=\mathbf{x}/L$, $\hat{\tilde{p}}_j = -i h \partial/\partial \tilde{x}_j$, $\tilde{E}=E/E_0$ and $\tilde{U}(\tilde{\mathbf{x}})=U(\mathbf{x})/E_0$. Furthermore, we define $\tilde{G}(\tilde{\mathbf{x}},\tilde{\mathbf{x}}_0)= E_0 L^2 G(\mathbf{x},\mathbf{x}_0)$. Taking into account that $\delta(\mathbf{x}-\mathbf{x}_0)= \delta(\tilde{\mathbf{x}}-\tilde{\mathbf{x}}_0)/L^2$, we find that Eq.~(\ref{eq:Dirac-Green-dim}) becomes
  \begin{equation}
    \left[{\bm \sigma} \cdot \hat{\tilde{\mathbf{p}}} + \tilde{U}(\tilde{\mathbf{x}}) 1_2 \right] \tilde{G}(\tilde{\mathbf{x}},\tilde{\mathbf{x}}_0) = \tilde{E} \tilde{G}(\tilde{\mathbf{x}},\tilde{\mathbf{x}}_0) + \delta(\tilde{\mathbf{x}}-\tilde{\mathbf{x}}_0) 1_2, \label{eq:Dirac-Green-dimless}
  \end{equation}
  Defining $\tilde{J}(\tilde{\mathbf{x}})=L^2 J(\mathbf{x})$ and $\tilde{\Psi}(\tilde{\mathbf{x}})= E_0 L^2\Psi(\mathbf{x})$, we find that Eqs.~(\ref{eq:Dirac-dim})-(\ref{eq:sol-Dirac-ptJ-dim}) remain valid when we replace all quantities by their dimensionless counterparts. Therefore, the source $\tilde{J}(\tilde{\mathbf{x}})$ and the wavefunction $\tilde{\Psi}(\tilde{\mathbf{x}})$ are given by
  \begin{equation}
    \tilde{J}(\tilde{\mathbf{x}}) = \begin{pmatrix} \alpha_1 \\ \alpha_2 \end{pmatrix} \delta(\tilde{\mathbf{x}}-\tilde{\mathbf{x}}_s), \quad
    \tilde{\Psi}(\tilde{\mathbf{x}}) = \tilde{G}(\tilde{\mathbf{x}},\tilde{\mathbf{x}}_s) \begin{pmatrix} \alpha_1 \\ \alpha_2 \end{pmatrix} . \label{eq:source-wf-dimless}
  \end{equation}
  In the following sections, we will work almost exclusively with these redefined (dimensionless) quantities and omit the tildes. Unless explicitly stated, we will always be referring to the dimensionless quantities defined here, rather than their original counterparts.
  
  Briefly returning to the case of graphene, we remark that the Hamiltonian~(\ref{eq:H-massless-Dirac}) is only valid near one of the two conical points in the Brillouin zone, namely at the $K$-point.~\cite{Katsnelson13} Near the other conical point, the so-called $K'$-point, the Hamiltonian reads
  \begin{equation}
    \hat{H}_{K'} = v_F (\sigma_x, -\sigma_y) \cdot \hat{\mathbf{p}} + U(\mathbf{x}) 1_2 = \sigma_x \hat{H} \sigma_x .  \label{eq:H-graphene-Kpr}
  \end{equation}
  Therefore, the Green's function near the $K'$-point is related to the Green's function near the $K$-point by
  \begin{equation}
    G_{K'}(\mathbf{x},\mathbf{x}_0) = \sigma_x G(\mathbf{x},\mathbf{x}_0) \sigma_x .
    \label{eq:graphene-Green-Kpr}
  \end{equation}

  \begin{figure*}[htb]
  \includegraphics{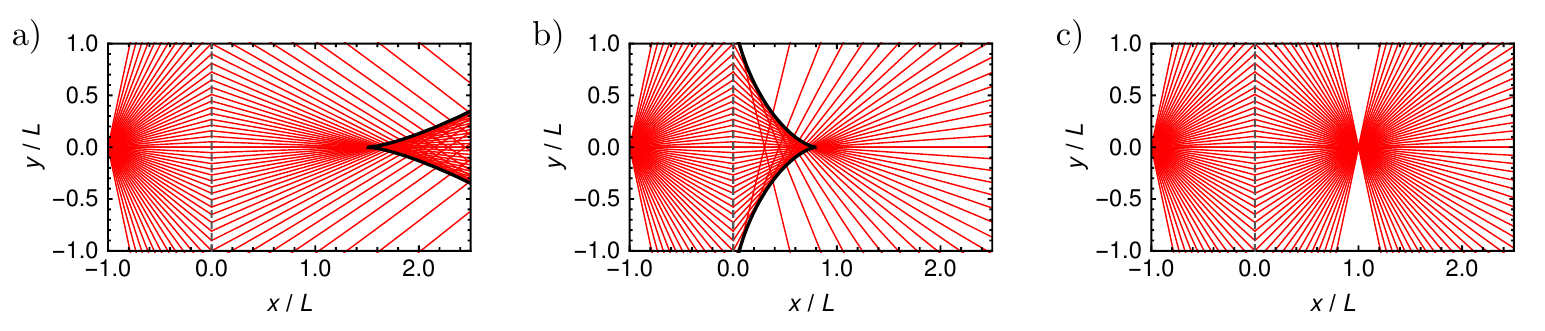}
  \caption{The classical trajectories (red lines) for massless Dirac fermions that are emitted by a point source and are incident on an \emph{n-p} junction at $x=0$ (dashed grey line). We see that the junction focusses the particles. The solid black line indicates the caustic, which is the envelope of the classical trajectories and separates the region where each point lies on a single trajectory from the region where each point lies on three trajectories. It consists of two fold lines meeting into a cusp point at $(x_{\text{cusp}},0)$. (a) For $U_0> 2E$, the cusp point $x_{\text{cusp}}>-x_s$ is the left-most point of the caustic. (b) When $U_0< 2E$, the cusp point $x_{\text{cusp}}<-x_s$ is the right-most point of the caustic. (c) For $U_0=2E$, all trajectories are focussed into a single point.}
  \label{fig:n-p-focus}
  \end{figure*}
  
  \subsection{Classical focussing}  \label{subsec:class-focussing}
  
  Before considering the focussing of the electrons, let us first consider the classical motion of massless Dirac fermions. The matrix Hamiltonian 
  \begin{equation}
    \hat{H}= {\bm \sigma} \cdot \hat{\mathbf{p}} + U(\mathbf{x}) \label{eq:H-Dirac}
  \end{equation}
  describes both electrons ($E > U(x)$) and holes ($E < U(x)$) within the same equation. One can extract the classical Hamiltonian functions that correspond to this matrix Hamiltonian by replacing the momentum operators by $c$-numbers and computing the eigenvalues.~\cite{Berlyand87,Belov06,Tudorovskiy12} We obtain two classical Hamiltonian functions,
  \begin{equation}
    H_{cl}^{\pm} = \pm | \mathbf{p} | + U(\mathbf{x}) , \label{eq:H-class}
  \end{equation}
  corresponding to electrons ($+$) and holes ($-$). This readily shows that the group velocity of electrons is parallel to their momentum, $\mathbf{v}_e = \partial H_{cl}^+/\partial \mathbf{p} = \mathbf{p}/|\mathbf{p}|$, whereas for holes the group velocity is opposite to the momentum, $\mathbf{v}_h = -\mathbf{p}/|\mathbf{p}|$.
  
  Let us now, following Ref.~\onlinecite{Cheianov07}, consider electrons emitted by a point source at position $\mathbf{x}_s=(x_s,0)$ incident on a one-dimensional \emph{n-p} junction. We assume that the potential consists of a single step at $x=0$:
  \begin{equation}
    U(\mathbf{x}) = U(x) = U_0 \Theta(x) , \label{eq:pot-step}
  \end{equation}
  where $\Theta(x)$ is the Heaviside step function and $U_0>E$. As before, in practice this means that the length scale $l_{\text{np}}$ of the potential increase satisfies $a \ll l_{\text{np}} \ll \lambda_{el}$. Then it is clear that the characteristic length scale $L$ of the system, which we used to define dimensionless parameters in the previous section, is equal to $|x_s|$.
  
  Now consider an electron incident on this potential from the left under an angle $\phi$, with momentum $\mathbf{p}_e= p_e(\cos\phi, \sin\phi)$, $p_e=E>0$. At the interface, part of this electron is reflected, whilst another part is transmitted, with momentum $\mathbf{p}_h=p_h(-\cos\theta, -\sin\theta)$. Since we consider scattering to right-moving holes, and $\mathbf{v}_h = -\mathbf{p}/|\mathbf{p}|$, we have $p_h=U_0-E>0$ for $|\theta|<\pi/2$. As the potential does not depend on $y$, the transversal momentum $p_y$ is conserved and we find the relation
  \begin{equation}
    \frac{\sin\phi}{\sin\theta} = -\frac{p_h}{p_e} = -\frac{U_0-E}{E} \equiv n, \label{eq:Snell}
  \end{equation}
  which is nothing but Snell's law for an electronic system.~\cite{Cheianov07} However, a very important characteristic of this system is that the refractive index $n$ is negative, which means that $\phi$ and $\theta$ have opposite signs. Therefore, the junction has the ability to focus electrons emitted by a source on the left-hand side, as can be seen in Fig.~\ref{fig:n-p-focus}. We will discuss this focussing in more detail in section~\ref{sec:caustics}.
  
  Finally, we note that the maximal angle $\theta$ under which electrons can be classically transmitted is $\pi/2$. For \mbox{$U_0-E<E$}, this means that electrons that are incident on the barrier under an angle larger than
  \begin{equation}
    \phi_{\text{max}} = \arcsin\left( \frac{U_0-E}{E} \right) < \frac{\pi}{2}  \label{eq:boundary-angle}
  \end{equation}
  will not be transmitted. This is related to the concept of a boundary angle in optics. For $U_0-E>E$, all electrons that are incident on the boundary can be classically transmitted, and we can set $\phi_{\text{max}}=\pi/2$.

  \subsection{The wavefunction for a polarized source}     \label{subsec:wavefunction-Veselago}
  
  Now that we have reviewed classical focussing by an \emph{n-p} junction, let us consider the wavefunction induced by the source~(\ref{eq:source-wf-dimless}).
  In appendix~\ref{app:Green}, we solve Eq.~(\ref{eq:Dirac-Green-dimless}) and obtain the Green's function~(\ref{eq:Green-np-app}). Combining this result with Eq.~(\ref{eq:source-wf-dimless}), we find that the wavefunction induced by a (pseudo)spin polarized source equals
  \begin{multline}
    \Psi(\mathbf{x}) = \frac{i}{4 \pi h^2} \int_{p_{y,\text{max}}}^{p_{y,\text{max}}} \frac{\alpha_1 e^{i\phi/2} + \alpha_2 e^{-i\phi/2}}{\cos[(\phi+\theta)/2]} \left( \begin{array}{c} e^{-i \theta/2} \\ e^{i \theta/2} \end{array} \right) \\ \times e^{i S_{np}(p_y,x,y)/h} \, \diff p_y , \label{eq:wf-Veselago}
  \end{multline}
  where
  \begin{equation}
    S_{np}(p_y,x,y) = - x_s \sqrt{E^2-p_y^2} - x \sqrt{(E-U_0)^2-p_y^2} + y p_y . \label{eq:action-np}
  \end{equation}
  is the classical action. The limits of integration in Eq.~(\ref{eq:wf-Veselago}) are determined by $p_{y,\text{max}}=E \sin\phi_{\text{max}}$, where $\phi_{\text{max}}$ was defined in the previous subsection.
  
  Of course, one can also use different source terms than~(\ref{eq:source-wf-dimless}), as was done in Ref.~\onlinecite{Cheianov07}.

  \section{Caustics} \label{sec:caustics}
  
  In the limit where the dimensionless parameter $h=\hbar v_F/(E_0 l)$, which we introduced in section~\ref{subsec:dimless-pars}, is small, the main contribution to the integral in the wavefunction~(\ref{eq:wf-Veselago}) is given by the stationary points of the action,~\cite{Fedoryuk77,Guillemin77,Maslov81} i.e. the points where $\partial S_{np}/\partial p_y$ vanishes. This means that the main contribution is given by the points that are on the classical trajectories of the system.~\cite{Arnold89} We find that they are given by
  \begin{align}
    y &= -x_s \frac{p_y}{\sqrt{E^2-p_y^2}} - x \frac{p_y}{\sqrt{(E-U_0)^2-p_y^2}} \nonumber \\
      &= -x_s \tan\phi + x \tan\theta . \label{eq:trajectories}
  \end{align}
  Naturally, these are equivalent to the trajectories that were obtained before in Ref.~\onlinecite{Cheianov07}, as reviewed in section~\ref{subsec:class-focussing}.
  
  There are also singular points, at which the second derivative $\partial^2 S_{np}/\partial p_y^2$ vanishes. These points form a curve that separates the region where each point lies on three trajectories (and hence interference takes place) from the region where each point lies on a single trajectory, as can be seen in Fig.~\ref{fig:n-p-focus}. Focussing takes places on such curves, which are known as caustics.~\cite{Berry80,Poston78,Arnold82,Arnold75} Some calculus yields that these points are defined by
  \begin{equation}
    x_{\text{cst}} = -x_s \frac{E^2}{(E-U_0)^2} \frac{((E-U_0)^2-p_y^2)^{3/2}}{(E^2-p_y^2)^{3/2}} \label{eq:x-caustic} , 
  \end{equation}
  with the corresponding $y$-value $y_{\text{cst}}$ given by Eq.~(\ref{eq:trajectories}). Alternatively, Eqs.~(\ref{eq:x-caustic}) and~(\ref{eq:trajectories}) can be cast into the form~\cite{Cheianov07}
  \begin{equation}
    y_{\text{cst}}(x_{\text{cst}}) = \pm \sqrt{\frac{\big(x_{\text{cst}}^{2/3}-x_{\text{cusp}}^{2/3}\big)^3}{n^2-1}}, \quad x_{\text{cusp}}=-|n|x_s \label{eq:yx-caustic}
  \end{equation}
  
  We can also look at the caustic from the point of view of the trajectories. If we parametrize them as $(x(t,\phi),y(t,\phi))$, then the caustic is the set of points where the Jacobian $J(t,\phi)$ vanishes. Indeed, some algebra shows that the Jacobian is proportional to the second derivative of the action:
  \begin{equation}
    J = - E \cos\phi \cos\theta \frac{\partial^2 S_{np}}{\partial p_y^2} . \label{eq:rel-jacobian-secondderiv}
  \end{equation}
  Hence the second derivative $\partial^2 S_{np}/\partial p_y^2$ vanishes if and only if the Jacobian does.
  
  Let us now consider the caustic in somewhat more detail. We first note that the transformation that sends $p_y$ to $-p_y$ reflects a trajectory in the $x$-axis, which implies that the set of trajectories is symmetric with respect to the line $y=0$. Therefore, the caustic should be symmetric with respect to the $x$-axis as well. Indeed, we see that $x_{\text{cst}}$ is invariant under reflection of $p_y$. Alternatively, we can also see directly from Eq.~(\ref{eq:yx-caustic}) that the caustic is symmetric.
  
  Second, let us consider the shape of the caustic. For general $U_0$, it consists of two so-called fold lines~\cite{Berry80,Poston78,Arnold82} meeting into a cusp point, see Fig.~\ref{fig:n-p-focus}(a). From the symmetry considerations presented above, we conclude that this cusp has to lie on the $x$-axis and therefore corresponds to $p_y=0$. Equation~(\ref{eq:x-caustic}) then implies that it is located at $x_{\text{cusp}}=-|n|x_s$. 
  
  According to catastrophe theory,~\cite{Arnold82,Poston78} a smooth change of variables can bring the action near the caustic into a certain normal form, which is a polynomial with its degree depending on the type of caustic.
  For points on the fold lines, the third derivative of $\partial^3 S_{np}/\partial p_y^3$ does not vanish, and this normal form is a third order polynomial without a quadratic term.~\cite{Arnold82,Poston78} In the Arnold classification,~\cite{Arnold82,Arnold75} this type of caustic is denoted by $A_2$. At the cusp point, denoted by $A_3$ in the Arnold classification, the third derivative vanishes as well, but the fourth derivative $\partial^4 S_{np}/\partial p_y^4$ is nonzero. It turns out that we can therefore express the action near this point as a fourth order polynomial without cubic term. In Fig.~\ref{fig:n-p-focus}, we see that we can have two types of cusp catastrophes, depending on the value of the potential $U_0$. For $U_0<2E$, we see that the cusp is the rightmost point of the caustic, whereas for $U_0>2E$, it is the leftmost point. The difference between the two types is the sign of the fourth derivative, which carries over to a plus or minus one in front of the quartic term of the normal form. For $U_0<2E$, this is a plus one, for $U_0>2E$, this is a minus one.
  
  The theory of Lagrangian singularities~\cite{Arnold82,Poston78} shows that in Hamiltonian systems in two dimensions the only generic singularities that can occur are folds and cusps. According to this theory, any other singularity will turn into one of these cases when an arbitarily small change is made to the system. However, the system that we are considering has an additional parameter that can be tuned, namely the potential strength $U_0$. As we have seen, we can change the sign of the fourth derivative of the action from positive to negative by changing the potential. In doing so, we will inevitably pass through the point where the fourth derivative vanishes, and hence through a higher order singularity. By symmetry, this higher order singularity is again located on the $x$-axis, and therefore corresponds to vanishing $p_y$. At $y=0$, the action~(\ref{eq:action-np}) is an even function of $p_y$, which means that its Taylor expansion in $p_y$ only contains terms of even order. In a generic setting, we can only expect the coefficients in front of the quadratic and the quartic terms in the expansion to vanish at this higher order singularity, since we have just two parameters, $x$ and $U_0$. This would imply a singularity corresponding to a sixth order polynomial, i.e. a two-dimensional section of the so-called butterfly catastrophe $A_5$.~\cite{Arnold82,Poston78} However, looking at the action~(\ref{eq:action-np}), we see that when $y=0$, $U_0=2E$ and $x=x_{\text{cusp}}=-x_s$ not only the second and the fourth derivative vanish, but that in fact all derivatives of $S_{np}$ with respect to $p_y$ vanish. In this very special case, the \emph{n-p} junction acts as an ideal lens and focusses all trajectories in a single point, as shown in Ref.~\onlinecite{Cheianov07} and depicted in Fig.~\ref{fig:n-p-focus}. 
  
  We wish to emphasize that this behavior is not generic, and is a special feature of the system under consideration. In fact, arbitarily small changes to the spatial setup, such as a non-straight barrier interface, or arbitarily small changes to the dispersion will ruin the perfect focus. In real graphene samples, we expect at least two corrections to the Hamiltonian~(\ref{eq:H-Dirac}) to contribute to the breaking down of the perfect focus. The first of these is the presence of next-nearest neighbor hopping,~\cite{Katsnelson13} which will slightly change the classical trajectories of the system. Furthermore, it destroys Klein tunneling, although its influence on the transmission through an \emph{n-p} junction was shown to be small.~\cite{Kretinin13} The second important correction to the Hamiltonian is trigonal warping,~\cite{Katsnelson13} the influence of which will become stronger as the energy increases. As with next-nearest neighbor hopping, trigonal warping will change the classical trajectories of the system. Furthermore, it also destroys Klein tunneling for almost all orientations.~\cite{Ando98,Logemann15}

  \begin{figure*}[htb]
  \includegraphics{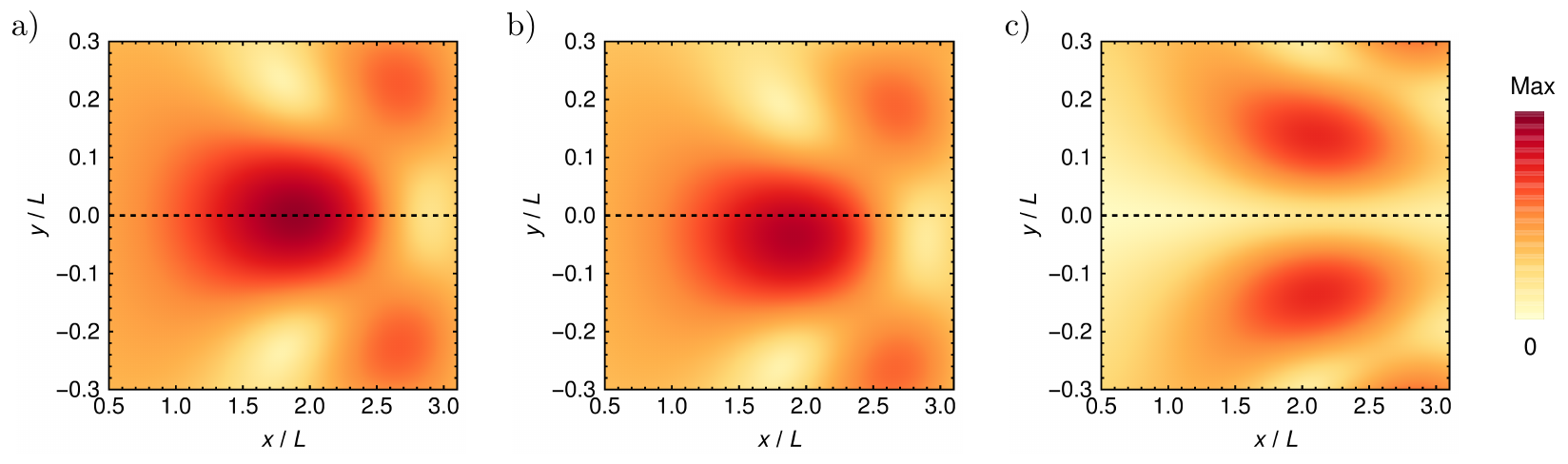}
  \caption{The density $\lVert\Psi\rVert$ computed by numerically evaluating the exact wavefunction~(\ref{eq:wf-Veselago}) for the dimensionless parameters $U_0=2.5$ and $h=0.0639$. For graphene, these numbers correspond to $E=100$ meV, $U_0=250$ meV and $L=100$ nm. We consider three different polarizations. (a) For $(\alpha_1,\alpha_2) = (1, 1)/\sqrt{2}$, the density is symmetric about the $x$-axis. (b) When $(\alpha_1,\alpha_2) = (1, 0)$, we see that the symmetry is broken and that the maximum lies at $y<0$. (c) For $(\alpha_1,\alpha_2) = (1, -1)/\sqrt{2}$, the density is symmetric again, but the central resonance has disappeared completely. The maximum of the color scale equals (a) 70, (b) 55 and (c) 22.}
  \label{fig:green-pols-large-h}
  \end{figure*}
  
  \begin{figure}[htb]
    \includegraphics[width=4.8cm]{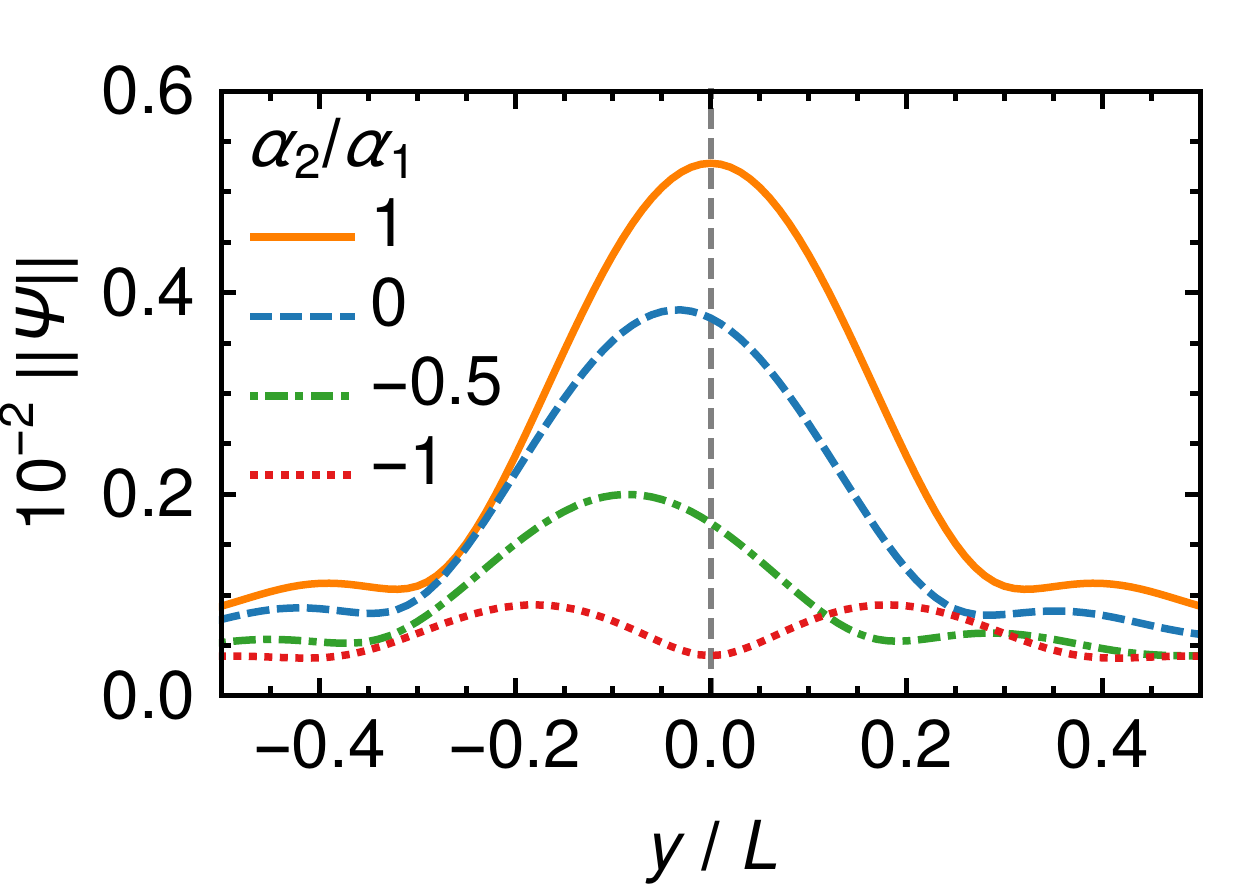}
    \caption{Sections of the norm $\lVert \Psi \rVert$ of the exact wavefunction~(\ref{eq:wf-Veselago}) on a line through the cusp point and parallel to the $y$-axis. The dimensionless parameters are equal to $U_0=2.5$ and $h=0.0639$. We clearly see that when the polarization $\alpha_2/\alpha_1$ decreases, the maximum is shifted to the left, while its size decreases. For $\alpha_2/\alpha_1=-1$, the wavefunction attains its minimum at the cusp point and the main focus has disappeared completely.}
    \label{fig:green-y-sections}
  \end{figure}

  \section{Quantum interference and symmetry breaking}      \label{sec:symm-breaking}
  
  In the previous section, we saw that the classical trajectories and the caustic are symmetric with respect to the $x$-axis. Let us now consider the symmetry of the Green's function. First, we note that the classical action~(\ref{eq:action-np}) satisfies $S_{np}(x,-y,-p_y)=S_{np}(x,y,p_y)$. Then, making the change of variables $p_y \to -p_y$ in the integral, it is easy to show that
  \begin{equation}
    G(x,-y,x_0,-y_0) = \sigma_x G(x,y,x_0,y_0) \sigma_x .
    \label{eq:green-reflection}
  \end{equation}
  Now let us consider the wavefunction~(\ref{eq:source-wf-dimless}) induced by a (pseudo)spin polarized point source. For its norm, \mbox{$\lVert \Psi \rVert=\sqrt{\Psi^\dagger \Psi}$}, we obtain the equality
  \begin{align}
    \lVert \Psi(x,-y) \rVert^2 &= \left\lVert \sigma_x G(x,y,x_s,0) \sigma_x \begin{pmatrix} \alpha_1 \\ \alpha_2 \end{pmatrix} \right\rVert^2 \nonumber \\
                               &= \left\lVert G(x,y,x_s,0) \begin{pmatrix} \alpha_2 \\ \alpha_1 \end{pmatrix} \right\rVert^2 \label{eq:wf-refl}
  \end{align}
  This equals $\lVert \Psi(x,y) \rVert^2$ only whenever $\alpha_1 = \pm \alpha_2$. Therefore, the wavefunction $\Psi(x,y)$ will in general not be symmetric, even though the classical trajectories are. 
  
  In Fig.~\ref{fig:green-pols-large-h}, we have plotted the density $\lVert\Psi\rVert$, given by Eq.~(\ref{eq:wf-Veselago}), near the cusp point for three different polarizations. For the polarizations $(1,1)/\sqrt{2}$ and $(1,-1)/\sqrt{2}$, the intensity is symmetric about the $x$-axis, in accordance with what we just showed. For $(1,0)/\sqrt{2}$, the symmetry is broken and we see that the maximum of the wavefunction is displaced. This shift is due to quantum interference and is an effect of the (pseudo)spin polarization of the source. In Fig.~{\ref{fig:green-y-sections}}, we show sections of the wavefunction along a line parallel to the $y$-axis and through $x_{\text{cusp}}$ for various polarizations. We see that as the ratio $\alpha_2/\alpha_1$ decreases, the position $y_{\text{max}}$ of the maximum shifts more and more towards negative $y$, while the intensity at the maximum decreases. When $\alpha_2=-\alpha_1$, the situation is once again symmetric, but the main focus has disappeared completely. Therefore, we conclude that we can markedly change the position of and the intensity at the central focus by changing the polarization.
  
  Briefly returning to the case of graphene, we see from Eqs.~(\ref{eq:graphene-Green-Kpr}), (\ref{eq:green-reflection}) and~(\ref{eq:source-wf-dimless}) that
  \begin{equation}
    \lVert \Psi(x,-y) \rVert^2 = \lVert \Psi_{K'}(x,y) \rVert^2 ,
  \end{equation}
  which means that the densities for the two valleys in graphene are related to each other by a reflection in the $x$-axis. In particular, $y_{\text{max}}$ changes sign, which means that the maxima for the two valleys are on opposite sides of the $x$-axis. In the following two sections, we investigate how large this asymmetry is and whether this may provide another way of realizing a valley filter in graphene.

  \section{Semiclassical evaluation of the wavefunction}      \label{sec:semiclassical-evaluation}
  
  To gain a better understanding of this asymmetry and the factors that influence it, we investigate the wavefunction~(\ref{eq:wf-Veselago}) with the semiclassical approximation. This will also give us more insight in the intensity at the central focus and in the way the size of the focus scales.
  
  Central to the semiclassical approximation is the dimensionless small parameter $h=\hbar v_F/(E_0 l)$ that we introduced in section~\ref{subsec:dimless-pars}. In section~\ref{sec:caustics}, we already saw that in the limit $h \to 0$, the main contribution to the integral~(\ref{eq:wf-Veselago}) is given by the stationary points of the action, which give rise to the classical trajectories (hence the name semiclassical approximation, as we are in a situation that is `almost' classical). In this limit, we can expand the wavefunction~(\ref{eq:wf-Veselago}) as an asymptotic series in powers of $h$.
  
  In the simplest case, we consider points $\mathbf{x}$ that are not on the caustic, which means that $\partial^2 S_{np}/\partial p_y^2$ does not vanish at any of the stationary points $p_{y,i}$. Such stationary points are called nondegenerate. Looking at Fig.~\ref{fig:n-p-focus}, we see that we can distinguish two regions. In the first region, each point $\mathbf{x}$ lies on a single trajectory, and hence the action only has one stationary point. In the second region, each point $\mathbf{x}$ lies on three trajectories, and the action has three stationary points. In appendix~\ref{appsub:WKB}, we discuss how the leading order contribution of a nondegenerate stationary point to the integral~(\ref{eq:wf-Veselago}) can be obtained by the conventional stationary phase approximation,~\cite{Fedoryuk77,Guillemin77,Maslov81} with the result given by Eq.~(\ref{eq:statphase-nondeg}). In the first region, this directly gives us the leading order term of the wavefunction. In the second region, we need to compute the contribution of each of the three stationary points, and then add these contributions to find the correct approximation to the wavefunction~(\ref{eq:wf-Veselago}).
  We will henceforth refer to these results as the Wentzel-Kramers-Brillouin (WKB) approximation.
  
  In section~\ref{sec:caustics}, we discussed that at the caustic the second derivative~$\partial^2 S_{np}/\partial p_y^2$ vanishes. Therefore, the result~(\ref{eq:statphase-nondeg}) diverges and we need to obtain the main contribution to the integral~(\ref{eq:wf-Veselago}) in a different way. In appendix~\ref{app:oscillatory-integrals}, we show that the simplest approximation for the wavefunction near a caustic can be obtained by making a Taylor expansion of the action~$S_{np}$ in $p_y$ up to the first nonvanishing term. 
  
  For the fold caustic, discussed in appendix~\ref{appsub:Airy}, this means that we have to expand up to third order, from which one obtains an expression in terms of the Airy function,~\cite{Airy38} see also e.g. Ref.~\onlinecite{Connor81b}. The final result, presented in Eq.~(\ref{eq:Airy-caustic}), is valid in an $\mathcal{O}(h^{5/6})$ neighborhood of the fold. We remark that expression~(\ref{eq:Airy-caustic}) was derived under the assumption that the limits of integration are infinite, whereas in Eq.~(\ref{eq:wf-Veselago}) they are finite. This is, however, not a problem, since the main contribution to the integral comes from a narrow vicinity of the stationary points.~\cite{Fedoryuk77,Guillemin77,Maslov81} Since all of the latter lie between the finite limits of integration in the integral, we can extend the limits of integration to infinity without changing the leading order term. Furthermore, we note that when we expand the action to even higher orders near the fold caustic, we will only get corrections beyond the leading order, i.e. terms in higher powers of $h$. Finally, a more accurate result can be obtained by using the uniform Airy approximation,~\cite{Chester57} see also e.g. Ref.~\onlinecite{Connor81b}, but we will not consider this approximation in this paper.
  
  The leading order approximation to Eq.~(\ref{eq:wf-Veselago}) for a point $\mathbf{x}$ near the fold caustic then consists of two terms. The first term is the one with the Airy function that we just discussed. The second term is a WKB term that comes from the third stationary point. In terms of the trajectories plotted in Fig.~\ref{fig:n-p-focus}, this term originates from the trajectory that is not tangent to the caustic near the point $\mathbf{x}$, but rather ``crosses'' the caustic. We henceforth refer to the sum of these two terms as the Airy approximation.
  
  Since our main interest in this paper is the asymmetry that is induced near the main focus, we now concentrate on the wavefunction near the cusp. Since the third derivative $\partial^3 S_{np}/\partial p_y^3$ vanishes at the cusp, one has to expand the action up to fourth order. In appendix~\ref{appsub:pearcey-leading-approx}, we review how this leads to an expression for the wavefunction near the cusp caustic that involves the Pearcey function,~\cite{Pearcey46,Connor81a,Connor82} see also e.g. Ref.~\onlinecite{Connor81b}, which is defined in Eq.~(\ref{eq:def-Pearcey}). The result, presented in Eq.~(\ref{eq:Pearcey-caustic}), contains the coefficients $a_i$ and $\mathbf{b}_i$, defined in Eqs.~(\ref{eq:q-exp-Pearcey}) and~(\ref{eq:q-exp-Airy}), which can be obtained by taking derivatives of the action $S_{np}$. As we already saw in section~\ref{sec:caustics}, the cusp corresponds to $p_y=0$, which considerably simplifies the calculations.
  After some calculus, we find that the nonzero coefficients $a_i$, which are the $i$-th derivatives of the action at the cusp point, are given by
  \begin{equation}
    a_0= -x_s \frac{U_0(2 E - U_0)}{E} , \quad a_4=-x_s\frac{3 U_0 (2 E - U_0)}{E^3 (E - U_0)^2} .  \label{eq:ai-Pearcey-Veselago}
  \end{equation}
  As we already discussed in section~\ref{sec:caustics}, we see that $a_4$ is positive for $U_0<2E$, and negative for $U_0>2E$. Furthermore, we obtain the coefficients $\mathbf{b}_i$ as
  \begin{equation}
    \begin{aligned}
      \langle \mathbf{b}_0 , \mathbf{z} \rangle &= -(U_0-E)(x-x_{\text{cusp}}),  \;\;\; &
      \langle \mathbf{b}_1 , \mathbf{z} \rangle &= y , \\
      \langle \mathbf{b}_2 , \mathbf{z} \rangle &= \frac{1}{U_0-E}(x-x_{\text{cusp}}) .
    \end{aligned}  \label{eq:bi-Pearcey-Veselago}
  \end{equation}
  Comparing Eqs.~(\ref{eq:wf-Veselago}) and~(\ref{eq:int-generic}), we see that the amplitude $f(\mathbf{x},\eta)$ does not depend on $\mathbf{x}$, and that
  \begin{equation}
    f(p_y) = \frac{i}{4\pi h^2} \frac{\alpha_1 e^{i\phi/2} + \alpha_2 e^{-i\phi/2}}{\cos[(\phi+\theta)/2]} \left( \begin{array}{c} e^{-i \theta/2} \\ e^{i \theta/2} \end{array} \right) .
    \label{eq:wf-Veselago-amp}
  \end{equation}
  Finally, combining the above results with the general result~(\ref{eq:Pearcey-caustic}), we find that the leading order approximation of the wavefunction in an $\mathcal{O}(h^{7/8})$ neighborhood of the cusp is given by
  \begin{multline}
    \Psi_{c0}(\mathbf{x}) = \frac{i(\alpha_1 + \alpha_2)}{4 \pi h^2} \sqrt[4]{\frac{24 h}{|a_4|}} \exp\left[\frac{i}{h}\left( a_0 + \langle \mathbf{b}_0 , \mathbf{z} \rangle \right)\right] \\ 
    \times \text{P}^{\pm} \left[ \sqrt{\frac{6}{h |a_4|}} \frac{x-x_{\text{cusp}}}{U_0-E} , \sqrt[4]{\frac{24}{h^3 |a_4|}} y \right] \left( \begin{array}{c} 1 \\ 1 \end{array} \right) .
    \label{eq:wf-cusp-Pearcey-Veselago}
  \end{multline}
  The first thing that should be noted about the result~(\ref{eq:wf-cusp-Pearcey-Veselago}) is that, regardless of the polarization, it is symmetric with respect to the $x$-axis, because of the fact that the Pearcey function is even in its second argument. Therefore, this approximation is insufficient if we want to understand the asymmetry. Second, we note that this approximation is not valid when the the potential $U_0$ is equal or close to $2E$, i.e. when we are close to the ideal focus, since in that case the coefficient $a_4$ vanishes or becomes very small, and the result~(\ref{eq:wf-cusp-Pearcey-Veselago}) diverges.
  
  Before we take a closer look at the asymmetry, we first want to see how well the approximation~(\ref{eq:wf-cusp-Pearcey-Veselago}) works for the symmetric polarization $(1,1)/\sqrt{2}$. To this end, we compare it with the exact wavefunction, Eq.~(\ref{eq:wf-Veselago}), which is evaluated by numerical integration. We also compare it with the result of the uniform approximation,~\cite{Ursell72,Connor81b} which is discussed in appendix~\ref{appsub:uniform-cusp}. In this approximation, we do not perform a Taylor expansion of the action, but instead bring the action to its normal form near the cusp by an exact change of variables. The final result, shown in Eq.~(\ref{eq:Pearcey-uni-result}), is given as a sum of the Pearcey function and its derivatives. In order to make the comparison complete, we also include the WKB approximation and the Airy approximation that we discussed before. These are not expected to work well in the vicinity of the cusp.
  
  \begin{figure*}[t]
  \includegraphics{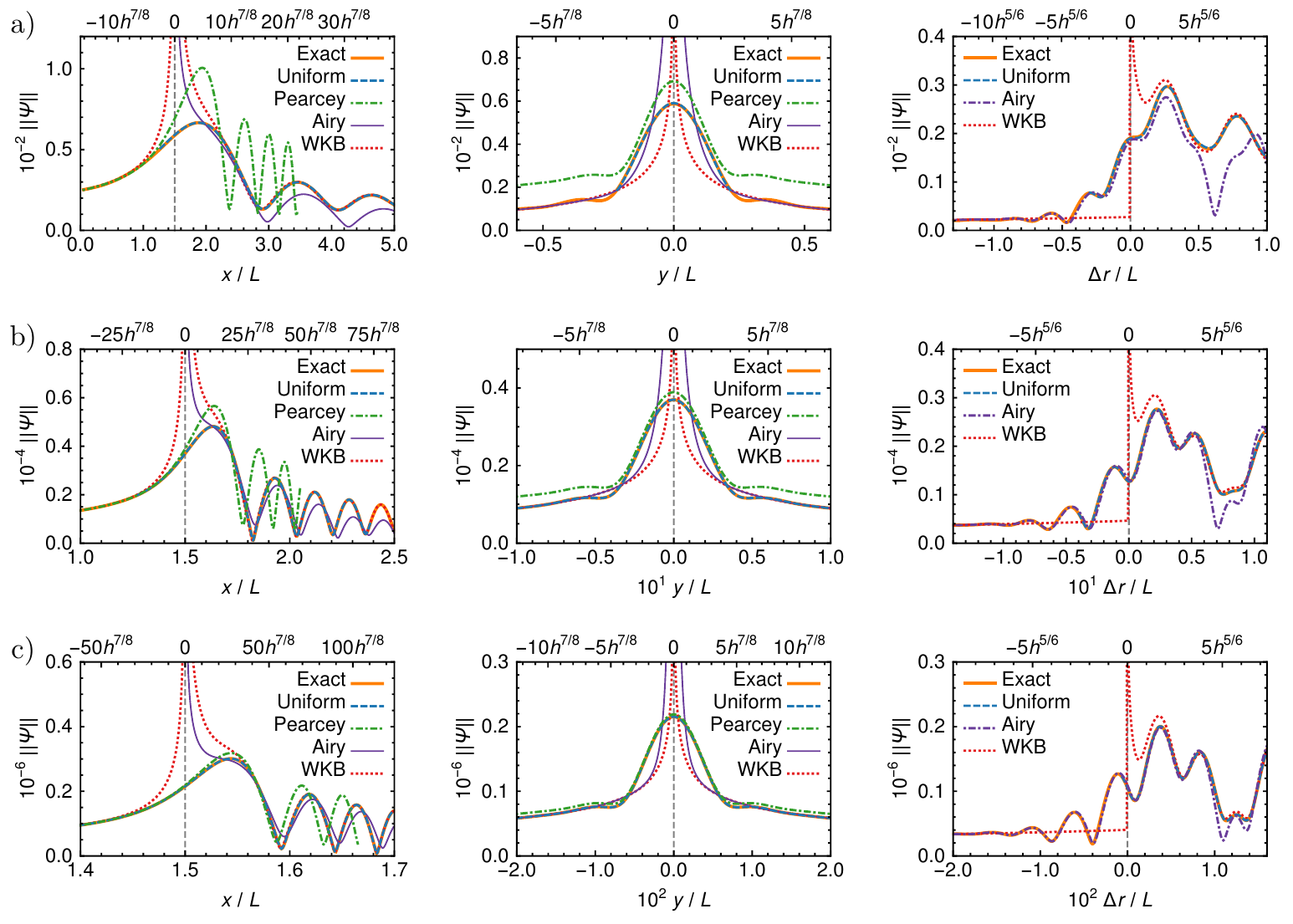}
  \caption{Comparison of different approximation schemes for the wavefunction near a caustic. For all figures, the dimensionless potential $U_0=2.5$, and the polarization $(\alpha_1,\alpha_2)=(1,1)/\sqrt{2}$. The dimensionless parameter is different for each of the three rows, namely (a) $h=0.0639$, (b) $h=0.00639$, (c) $h=0.000639$. For graphene, these numbers correspond to $E=100$ meV, $U_0=250$ meV and length scales of (a) $L=10^2$ nm, (b) $L=10^3$ nm and (c) $L=10^4$ nm. In the left column, we show a comparison along the $x$-axis, in which the position of the cusp is indicated by a vertical dashed grey line. Although the Pearcey approximation~(\ref{eq:wf-cusp-Pearcey-Veselago}) typically gives too large values for the wavefunction, especially for large $h$, it correctly predicts the position of the maximum for all three values of $h$. In the middle column, we show a comparison along the line that is parallel to the $y$-axis and passes through the cusp point. The cusp point is again indicated by a vertical dashed grey line. In the right column, we show a comparison along a line perpendicular to one of the points on the fold line. The fold point is indicated by a vertical dashed grey line. In all three cases, the Airy approximation works rather well for a large range of values, whereas the WKB approximation only works far away from the fold point. In all of the plots, we have not only indicated the dimensionless coordinates, but also the distance from the caustic in the relevant power of $h$, which is $h^{7/8}$ for the cusp caustic and $h^{5/6}$ for the fold caustic (see also appendix~\ref{app:oscillatory-integrals}).}
  \label{fig:comp-along-lines-diff-approx-no-corr}
  \end{figure*}  
  
  In Fig.~\ref{fig:comp-along-lines-diff-approx-no-corr}, we compare these five approximations for three different values of the small parameter $h$. We see that the uniform approximation perfectly coincides with the exact wavefunction~(\ref{eq:wf-Veselago}). Furthermore, we observe that the Pearcey approximation is not very accurate for large values of $h$, but becomes much better when we decrease $h$. Note that although it typically overestimates the magnitude of the wavefunction, it correctly predicts the position of the maximum for all three values of $h$. This implies that we may be able to find the position of the asymmetry by including higher order corrections, even for rather large $h$.
  As predicted, the WKB approximation works well far away from the cusp, but diverges as we come close to it, as does the Airy approximation.
  Near the fold caustic, the Airy approximation performs well for a large range of distances and for all three values of $h$, whereas the WKB approximation only works far away from the fold.
  
  As we just saw, the leading order Pearcey approximation~(\ref{eq:wf-cusp-Pearcey-Veselago}) is not enough to reproduce the asymmetry that we found in section~\ref{sec:symm-breaking}. Therefore, let us look at higher order corrections to the Pearcey approximation, which are discussed in appendix~\ref{appsub:pearcey-higher-order-approx}. The corrections can come from two different sources, namely from higher order terms in the Taylor expansion of the action $S_{np}$ and from higher order terms in the expansion of the amplitude, i.e. the part of the integrand in Eq.~(\ref{eq:wf-Veselago}) that precedes the exponent with the action. As we discussed in section~\ref{sec:caustics}, the cusp point lies on the line $y=0$, which implies that the action~(\ref{eq:action-np}) is symmetric with respect to $p_y$. Therefore, all terms in its Taylor expansion that are odd with respect to $p_y$ vanish at the cusp, and in particular the fifth order term vanishes. This means that the second term of $\mathcal{O}(h^{1/2})$ in Eq.~(\ref{eq:Pearcey-corrs}) is irrelevant, as $q_5(\mathbf{z}) = \mathcal{O}(\mathbf{z}) = \mathcal{O}(h^{7/8})$. Hence the only correction of $\mathcal{O}(h^{1/2})$ is given by Eq.~(\ref{eq:Pearcey-corr-1}). Using the results~(\ref{eq:ai-Pearcey-Veselago}),~(\ref{eq:bi-Pearcey-Veselago}) and~(\ref{eq:wf-Veselago-amp}), we obtain the first-order correction to the leading order term~(\ref{eq:wf-cusp-Pearcey-Veselago}) as
  \begin{multline}
    \Psi_{c1}(\mathbf{x}) = \frac{i}{8 \pi h^2} \left[ \frac{\alpha_1-\alpha_2}{E} \left( \begin{array}{c} 1 \\ 1 \end{array} \right) + \frac{\alpha_1+\alpha_2}{U_0-E} \left( \begin{array}{c} 1 \\ -1 \end{array} \right) \right] \\ 
    \times \left(\frac{24 h}{|a_4|}\right)^{1/2} \exp\left[\frac{i}{h}\left( a_0 + \langle \mathbf{b}_0 , \mathbf{z} \rangle \right)\right]  \\
    \times \text{P}_v^{\pm} \left[ \sqrt{\frac{6}{h |a_4|}} \frac{x-x_{\text{cusp}}}{U_0-E} , \sqrt[4]{\frac{24}{h^3 |a_4|}} y \right] ,
    \label{eq:wf-cusp-Pearcey-Veselago-corr1}
  \end{multline}
  where $\text{P}_v^{\pm}$ represents the derivative of the Pearcey function with respect to its second argument, as defined in Eq.~(\ref{eq:def-Pearcey-diff-v}). Since $\text{P}_v^{\pm}$ is odd in its second argument, the sum $\Psi_{c0}(\mathbf{x})+\Psi_{c1}(\mathbf{x})$ of the leading order term~(\ref{eq:wf-cusp-Pearcey-Veselago}) and the first-order correction~(\ref{eq:wf-cusp-Pearcey-Veselago-corr1}) does not necessarily have its maximum at $y=0$.
  
  \begin{figure*}[t]
  \includegraphics{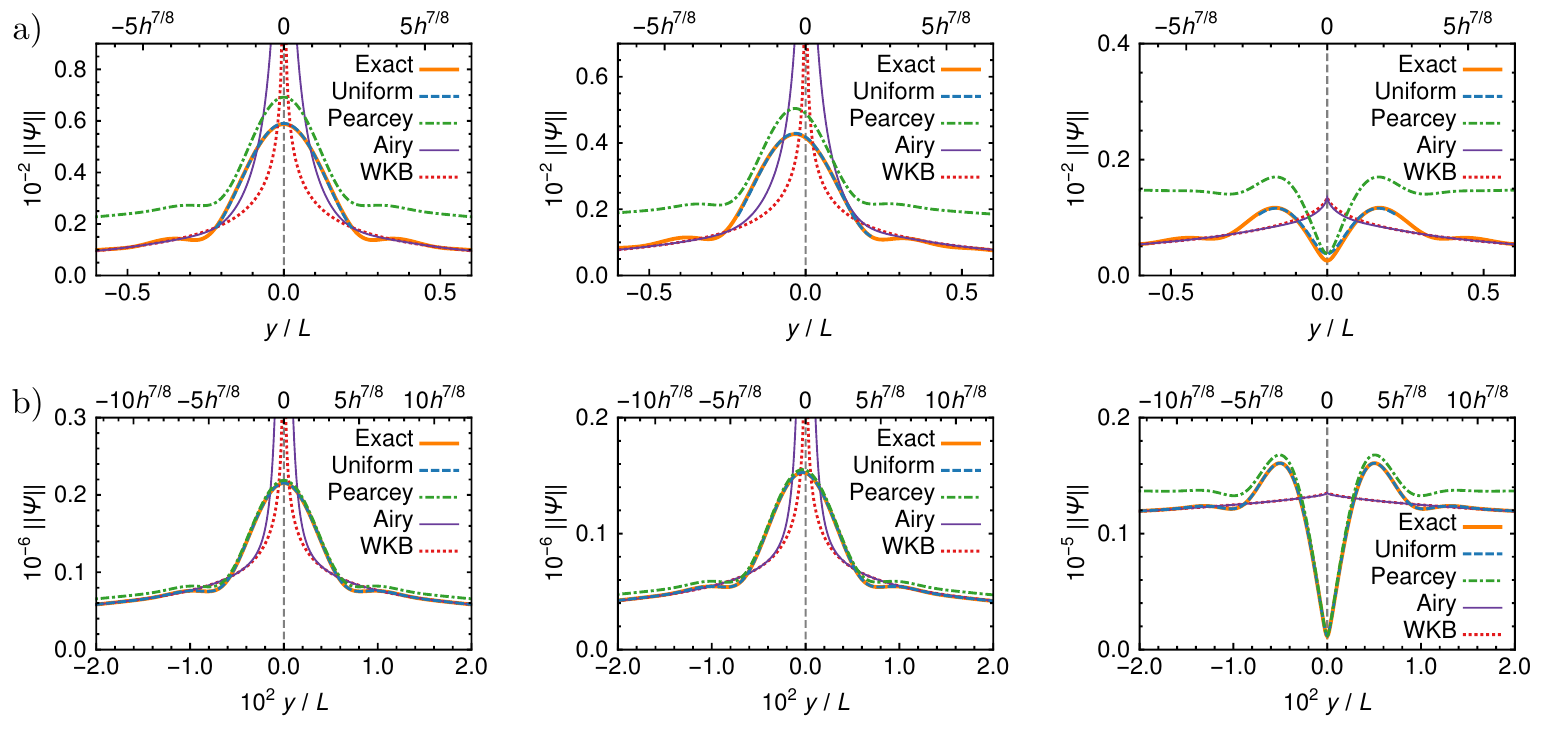}
  \caption{Comparison of different approximation schemes for the wavefunction on a line through the cusp point and parallel to the $y$-axis. The dimensionless semiclassical parameter is equal to (a) $h=0.0639$ and (b) $h=0.000639$, with $U_0=2.5$ for all figures. Left, middle and right panels correspond to the polarizations $(\alpha_1,\alpha_2) = (1, 1)/\sqrt{2}$, $(\alpha_1,\alpha_2) = (1, 0)$ and $(\alpha_1,\alpha_2) = (1, -1)/\sqrt{2}$, respectively. In the left and middle panels, we clearly see that the Pearcey approximation $\Psi_{c0}(\mathbf{x})+\Psi_{c1}(\mathbf{x})$ correctly predicts the positions of the maxima. In the right panels, the Pearcey approximation $\Psi_{c0}(\mathbf{x})+\Psi_{c1}(\mathbf{x})+\Psi_{c2}(\mathbf{x})$ correctly predicts the position of the maximum and adequately reproduces the value of $\lVert \Psi \rVert$ at $y=0$.}
  \label{fig:sections-y-asymmetry}
  \end{figure*}
  
  In Fig.~\ref{fig:sections-y-asymmetry}, we compare the Pearcey approximation including the first-order correction with the exact solution, the uniform approximation and the Airy and WKB approximations on the line that goes through the cusp point and is parallel to the $y$-axis. Comparing Fig.~\ref{fig:sections-y-asymmetry} with the middle panels of Fig.~\ref{fig:comp-along-lines-diff-approx-no-corr}, we see that for polarization $(1,1)/\sqrt{2}$ the result does not qualitatively differ from the leading order approximation, although the numerical values are slightly different. For polarization $(1,0)$, we see that with the first-order correction~(\ref{eq:wf-cusp-Pearcey-Veselago-corr1}) we correctly reproduce the position of the maximum, even though it is no longer at $y=0$. This holds for both the large and the small value of $h$.
  
  When the polarization equals $(1,-1)/\sqrt{2}$, we see from Eq.~(\ref{eq:wf-cusp-Pearcey-Veselago}) that the term proportional to $h^{1/4}$ vanishes. This makes sense, since we already saw in section~\ref{sec:symm-breaking} that the central resonance vanishes in this case. Hence, the leading order term for this case is given by Eq.~(\ref{eq:wf-cusp-Pearcey-Veselago-corr1}), which correctly reproduces the position of the two maxima that lie symmetrically on both sides of $y=0$. However, since $\text{P}_v^{\pm}$ vanishes at $y=0$, this approximation predicts that the wavefunction also vanishes on the $x$-axis, which is incorrect, as can be seen from Fig.~\ref{fig:sections-y-asymmetry}(c). Therefore, we have also included the second correction, which is of $\mathcal{O}(h^{3/4})$, in the Pearcey approximation plotted in Fig.~\ref{fig:sections-y-asymmetry}(c). Looking at Eq.~(\ref{eq:Pearcey-corrs}), and remembering that both $f(\mathbf{x},\eta_0)$ and $q_5(\mathbf{z})$ vanish in our case, we easily see that this correction is given by Eq.~(\ref{eq:Pearcey-corr-2}). Taking the results~(\ref{eq:ai-Pearcey-Veselago}),~(\ref{eq:bi-Pearcey-Veselago}) and~(\ref{eq:wf-Veselago-amp}) into account, we obtain the second order correction as
  \begin{multline}
    \Psi_{c2}(\mathbf{x}) = \frac{1}{8 \pi h^2} \frac{1}{E(E-U_0)} \left( \begin{array}{c} \alpha_1 \\ \alpha_2 \end{array} \right) \\
      \times \left(\frac{24 h}{|a_4|}\right)^{3/4} \exp\left[\frac{i}{h}\left( a_0 + \langle \mathbf{b}_0 , \mathbf{z} \rangle \right)\right]  \\
      \times \text{P}_u^{\pm} \left[ \sqrt{\frac{6}{h |a_4|}} \frac{x-x_{\text{cusp}}}{U_0-E} , \sqrt[4]{\frac{24}{h^3 |a_4|}} y \right]  ,
    \label{eq:wf-cusp-Pearcey-Veselago-corr2}
  \end{multline}
  where $\text{P}_u^{\pm}$ is the derivative of the Pearcey function with respect to its first argument, as defined in Eq.~(\ref{eq:def-Pearcey-diff-u}).
  With this second-order correction, we see from Fig.~\ref{fig:sections-y-asymmetry}(c) that we have a reasonable approximation for the value of the wavefunction at $y=0$. We remark that this correction does not substantially influence our prediction for the position of the maximum for this polarization.
  
  \begin{figure*}[t]
  \includegraphics{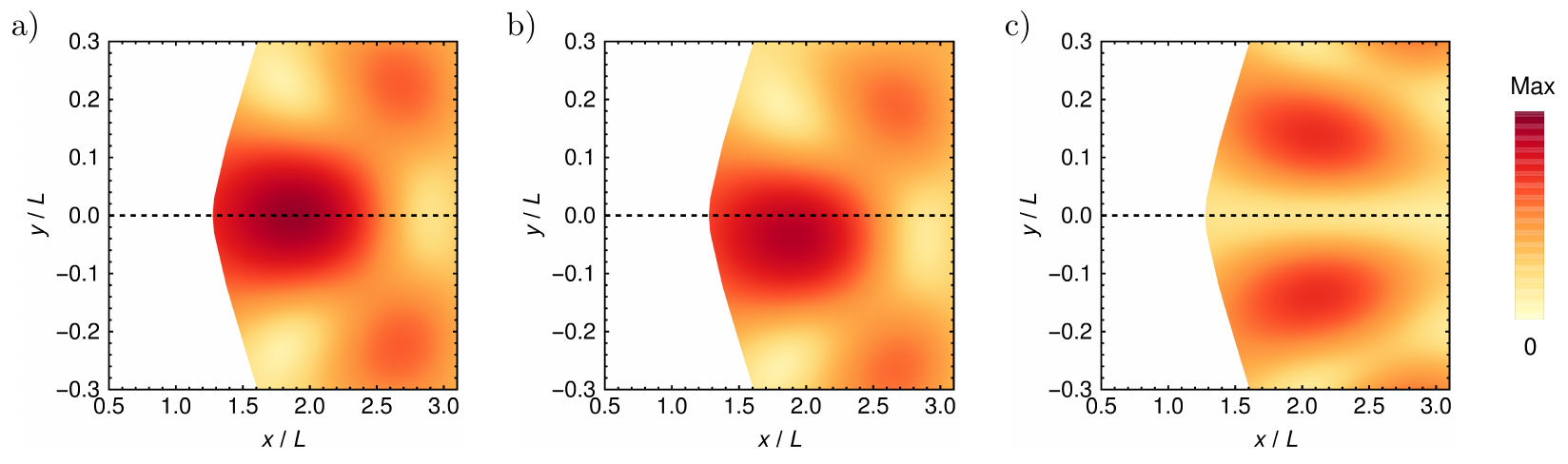}
  \caption{The density $\lVert\Psi\rVert$ obtained from the uniform approximation~(\ref{eq:Pearcey-uni-result}) for the dimensionless parameters $U_0=2.5$ and $h=0.0639$. We consider three different polarizations $(\alpha_1,\alpha_2)$, namely (a) $(1, 1)/\sqrt{2}$, (b) $(1, 0)$ and (c) $(1, -1)/\sqrt{2}$. In all cases, the exact result shown in Fig.~\ref{fig:green-pols-large-h} is accurately reproduced. As in Fig.~\ref{fig:green-pols-large-h}, the maximum of the color scale equals (a) 70, (b) 55 and (c) 22.}
  \label{fig:overview-uniform-large-h}
  \end{figure*}  
  
  Figure~\ref{fig:sections-y-asymmetry} clearly shows that we can greatly decrease the intensity at the central focus by changing the polarization from $(1,1)/\sqrt{2}$ to $(1,-1)/\sqrt{2}$, as discussed in the previous section. Let us now use the Pearcey approximation that we have developed to derive an equation for the ratio between the intensities $\lVert \Psi \rVert^2$ for these two polarizations. In order to arrive at a simple expression, we will use the value of the various Pearcey approximations at the cusp point $(x_{\text{cusp}},0)$. Although this is not the position of the main focus, the wavefunction at this point gives us a good indication of its value at the maximum. In Fig.~\ref{fig:comp-along-lines-diff-approx-no-corr}, we see that for the largest $h$, the exact value of the maximal intensity is approximately equal to the value of the Pearcey approximation at the cusp, due the fact that the latter gives too large values. For the smallest $h$, we see that the maximal intensity is about a factor of two larger than the Pearcey approximation at the cusp. For the polarization $(1,1)/\sqrt{2}$, we use the leading order Pearcey approximation, given by Eq.~(\ref{eq:wf-cusp-Pearcey-Veselago}). At the origin, the Pearcey function takes a particularly simple form, as $\text{P}^+(0,0)=2\exp(i\pi/8)\Gamma(5/4)$, where $\Gamma(x)$ is the gamma function.~\cite{Korn61} This identity can easily be proven directly using the definition~(\ref{eq:def-Pearcey}) of the Pearcey function and the definition of the gamma function. Since we also have $|\text{P}^+(0,0)|^2 = |\text{P}^-(0,0)|^2$, see also Eq.~(\ref{eq:symm-Pearcey}), we obtain
  \begin{equation}
    \lVert \Psi_{(\frac{1}{\sqrt{2}},\frac{1}{\sqrt{2}})}(x_{\text{cusp}},0) \rVert^2 = \frac{16}{(4 \pi h^2)^2} \left(\frac{24 h}{|a_4|}\right)^{1/2} \Gamma\left(\frac{5}{4}\right)^2
  \end{equation}
  For the polarization $(1,-1)/\sqrt{2}$ we use the second order correction~(\ref{eq:wf-cusp-Pearcey-Veselago-corr2}), since both the leading order term~(\ref{eq:wf-cusp-Pearcey-Veselago}) and the first order correction~(\ref{eq:wf-cusp-Pearcey-Veselago-corr1}) vanish on the $x$-axis. One can show that $\text{P}_u^+(0,0)=\exp(7i\pi/8)\Gamma(3/4)/2$ and $|\text{P}_u^+(0,0)|^2 = |\text{P}_u^-(0,0)|^2$. Therefore, we find that
  \begin{multline}
    \lVert \Psi_{(\frac{1}{\sqrt{2}},-\frac{1}{\sqrt{2}})}(x_{\text{cusp}},0) \rVert^2 = \frac{E^{-2}(E-U_0)^{-2}}{16(4 \pi h^2)^2} \left(\frac{24 h}{|a_4|}\right)^{3/2} \\ \times \Gamma(3/4)^2
  \end{multline}
  For the ratio between the two intensities at the cusp point, we then obtain
  \begin{equation}
    \frac{\lVert \Psi_{(\frac{1}{\sqrt{2}},-\frac{1}{\sqrt{2}})}(x_{\text{cusp}},0) \rVert^2}{\lVert \Psi_{(\frac{1}{\sqrt{2}},\frac{1}{\sqrt{2}})}(x_{\text{cusp}},0) \rVert^2} = \frac{3}{32} \frac{h}{|a_4|} \frac{1}{E^2(E-U_0)^2} \frac{\Gamma(3/4)^2}{\Gamma(5/4)^2} ,
    \label{eq:ratio-intensities}
  \end{equation}
  which shows that the relative decrease of the intensity at the main focus is proportional to the small semiclassical parameter $h$.
  
  We finish this section by showing the densities that the various approximation schemes give near the cusp point for two different values of $h$. Looking at the comparisons in Fig.~\ref{fig:comp-along-lines-diff-approx-no-corr}(a), we see that at $h=0.0639$ it is difficult to construct a global approximation for $\lVert \Psi(\mathbf{x}) \rVert$ by combining the different local approximations, since there is no region where the Pearcey approximation smoothly joins the stationary phase approximation. Therefore, we conclude that for a global approximation only the uniform approximation is adequate. In Fig.~\ref{fig:overview-uniform-large-h}, we show the results of this approximation $h=0.0639$. Comparing Figs.~\ref{fig:green-pols-large-h} and~\ref{fig:overview-uniform-large-h}, we see that the agreement is excellent, as we already inferred from the comparisons along the various sections.
  
  A few words about the implementation of the uniform approximation are in place here. In the region where each point $\mathbf{x}$ lies on three trajectories, the equation $\partial S_{np}/\partial p_y = 0$ has three real roots $p_{y,i}$. The values of these roots are restricted, since for $U_0-E<E$ we have $|p_y|<U_0-E$ and for $U_0-E>E$ we have $|p_y|<E$. We can obtain these roots numerically, and subsequently determine the action $S_{np}$, its second derivative and the amplitude at these points, from which we can obtain the parameters for the uniform approximation, as explained in detail in appendix~\ref{appsub:uniform-cusp}. When $\mathbf{x}$ only lies on a single trajectory, the equation $\partial S_{np}/\partial p_y$ still has three roots, but this time only one of them is real and two of them are complex. However, the absolute value of these complex roots is not necessarily restricted.
  We have found that when the complex roots become too large in absolute value, the performance of the uniform approximation becomes rather poor. When we impose on $p_y$ the same demands that hold for the case when $p_y$ is real, i.e. $|p_y|<U_0-E$ when $U_0-E<E$ and $|p_y|<E$ when $U_0-E>E$, we obtain good agreement. However, this means that we cannot use the uniform approximation far away from the caustic, which gives rise to the large white area in Fig.~\ref{fig:overview-uniform-large-h}.
  Note in particular the strange situation that occurs for $U_0>2E$, $y=0$ and $x<-x_s$, where we have three real roots, two of which have an absolute value larger than $U_0-E$.

  \begin{figure*}[t]
  \includegraphics{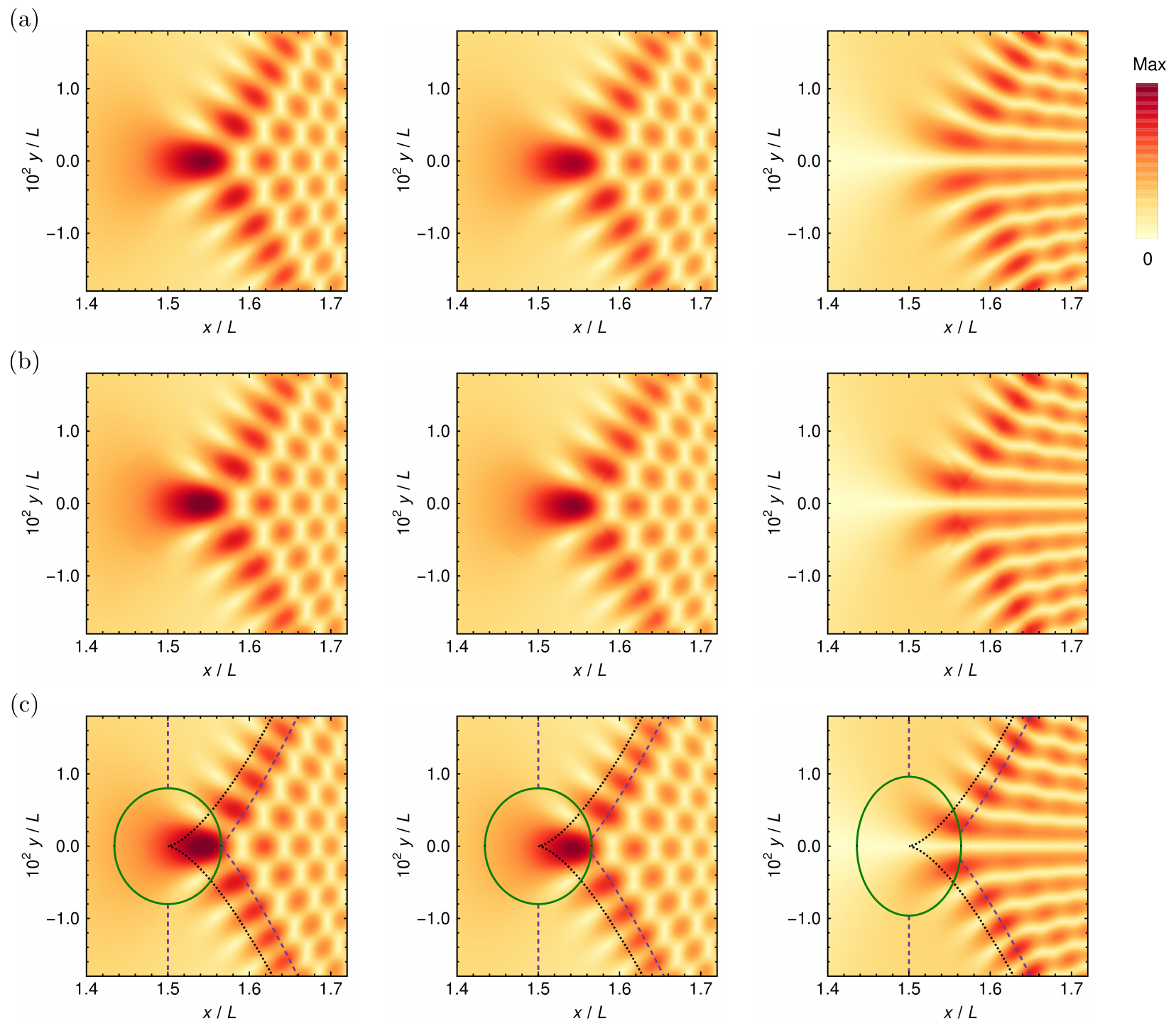}
  \caption{The density $\lVert\Psi\rVert$ for the dimensionless parameters $U_0=2.5$ and $h=0.000639$. For graphene, these numbers correspond to $E=100$ meV, $U_0=250$ meV and $L=10^4$ nm. (a) The exact result obtained by numerically evaluating the exact wavefunction~(\ref{eq:wf-Veselago}). (b) The result of combining the Pearcey approximation, the Airy approximation and the WKB approximation. (c) The region in which each of the approximations is used. The Pearcey approximation is used inside the green ellipse. Between the two (dashed) purple lines the Airy approximation is used, and outside both these regions the WKB approximation is used. The black (dotted) line represents the caustic. The left, middle and right panels correspond to three different polarizations $(\alpha_1,\alpha_2)$, to wit $(1, 1)/\sqrt{2}\,$; $(1, 0)$ and $(1, -1)/\sqrt{2}$. The maximum of the color scale equals $30\cdot 10^4$ for the left column, $23\cdot 10^4$ for the middle column and $5\cdot 10^4$ for the right column.}
  \label{fig:overview-approximations-large-h}
  \end{figure*}
  
  Coming back to the densities that the various approximations predict near the cusp point, we see that for $h=0.000639$, it is possible to construct a global approximation for $\lVert \Psi(\mathbf{x}) \rVert$ by combining various local approximations. From the comparisons in Fig.~\ref{fig:comp-along-lines-diff-approx-no-corr}, we see that for polarization $(1,1)/\sqrt{2}$, we can use the Pearcey approximation in an area around the cusp that is described by an ellipse with semi-major axis $41 h^{7/8}$ (along the $x$-direction) and semi-minor axis $5 h^{7/8}$ (along the $y$-direction). Furthermore, we can use the Airy approximation along a distance $10 h^{5/6}$ from the fold caustic in the outward direction (where there is only a single trajectory) and along a distance $3 h^{5/6}$ in the inward direction (where there are three trajectories). Outside of both these regions, we can use the WKB approximation. We remark that we do not need to patch the different approximations together by determining certain constants, since all of the different approximations are simplifications of the same wavefunction~(\ref{eq:wf-Veselago}) that are appropriate for a certain region.
  
  In Fig.~\ref{fig:overview-approximations-large-h}, we show the combination of the various approximations, as well as the region for each of the approximations. We see that the final result nicely coincides with the exact wavefunction~(\ref{eq:wf-Veselago}), which was evaluated numerically. Since the result of the uniform approximation perfectly coincides with the exact wavefunction~(\ref{eq:wf-Veselago}), it is not shown separately.

  \section{Derivation of the displacement from semiclassical considerations}  \label{sec:asymm-semiclassical}
  
  In the previous section, we saw that, for a certain set of parameters, we could reproduce the vertical position of the maximum, even though it was displaced from the $x$-axis, by using the first order correction~(\ref{eq:wf-cusp-Pearcey-Veselago-corr1}) to the leading order term~(\ref{eq:wf-cusp-Pearcey-Veselago}).
  In this section, we perform a more systematic study of the vertical displacement of the maximum, which is caused by the (pseudo)spin polarization, and obtain a simple formula for the shift.
  
  Let us therefore try to derive a formula for the $y$-coordinate of the maximum when $x$ equals $x_{\text{cusp}}$. To this end, we consider the sum of the leading order Pearcey approximation~(\ref{eq:wf-cusp-Pearcey-Veselago}) and its first correction~(\ref{eq:wf-cusp-Pearcey-Veselago-corr1}), i.e. $\Psi(\mathbf{x})=\Psi_{c0}(\mathbf{x})+\Psi_{c1}(\mathbf{x})$, at $x_{\text{cusp}}$. In order to find the maximum, we need to find the points where the first derivative $\partial \lVert \Psi \rVert^2/\partial y$ vanishes. Unfortunately, this equation cannot easily be solved, as it involves the Pearcey function and one of its partial derivatives. Therefore, let us approximate the Pearcey function by its Taylor expansion. First of all, we note that it is even in its second argument, see Eq.~(\ref{eq:symm-Pearcey}), which means that when we perform a Taylor expansion in the second argument, all terms of odd order vanish.
  In the previous section, we already determined the zeroth order coefficient $c_0$ of the Taylor expansion, which is equal to $\text{P}^+(0,0)$. Furthermore, from the definitions~(\ref{eq:def-Pearcey}) and~(\ref{eq:def-Pearcey-diff-u}), it is easy to see that $\partial^2 \text{P}^+(u,v)/\partial v^2 = i \partial \text{P}^+(u,v)/\partial u$. Using the result for $\partial \text{P}^+(u,v)/\partial u$ from the previous section, we can then obtain the second order coefficient $c_2$. Combining our results, we find that
  \begin{equation}
  \begin{aligned}
    c_0 = \text{P}^+(0,0) &= 2\exp(i\pi/8)\Gamma(5/4),  \\ 
    c_2 = \frac{1}{2}\text{P}^+_{vv}(0,0) &= -\frac{1}{4}\exp(3i\pi/8)\Gamma(3/4), \\
    c_4 = \frac{1}{24}\text{P}^+_{vvvv}(0,0) &= \frac{i}{96} c_0, 
  \end{aligned}
  \label{eq:Pearcey-Taylor-coeffs}
  \end{equation}
  where the last equality can be obtained by direct computation or by using the differential equation that is satisfied by the Pearcey function, see Ref.~\onlinecite{Connor81a}. From these equalities, we see that the fourth order coefficient $c_4$ is much smaller than both $c_0$ and $c_2$. Therefore, we obtain a rather accurate approximation by replacing the Pearcey function by its second order Taylor expansion, that is,
  \begin{equation}
    P^+(0,v) \approx c_0 + c_2 v^2, \quad P^-(0,v) \approx c_0^* + c_2^* v^2 ,
    \label{eq:Pearcey-Taylor}
  \end{equation}
  where the second relation is a consequence of the symmetries of the Pearcey function, see Eq.~(\ref{eq:symm-Pearcey}). We can use the same approximation for the derivative, which naturally gives
  \begin{equation}
    P_v^+(0,v) \approx 2 c_2 v, \quad P_v^-(0,v) \approx 2 c_2^* v ,
    \label{eq:Pearcey-der-v-Taylor}
  \end{equation}
  Using the approximations~(\ref{eq:Pearcey-Taylor}) and~(\ref{eq:Pearcey-der-v-Taylor}) for the Pearcey function and its derivative, we find, after some algebra, that
  \begin{multline}
    \frac{\partial \lVert \Psi \rVert^2}{\partial y} = \frac{96}{|a_4|} \bigg( i f_{p_y}^\dagger f \text{Re}(c_0 c_2^*) + \lVert f \rVert^2 \text{Re}(c_0 c_2^*) \frac{y}{h} + \\
      \qquad\qquad\qquad\quad \lVert f \rVert^2 |c_2|^2 q^2 \frac{y^3}{h^3} + 3 i f_{p_y}^\dagger f |c_2|^2 q^2 \frac{y^2}{h^2} + \\
       2 \lVert f_{p_y} \rVert^2 |c_2|^2 q^2 \frac{y}{h}  \bigg) =0 ,
    \label{eq:maximum-y-cubic-equation}
  \end{multline}
  where $q=\sqrt[4]{24 h/|a_4|}$, see also the equivalent definition in appendix~\ref{appsub:pearcey-leading-approx}, and the amplitude $f$ and its derivative $f_{p_y}$, see Eq.~(\ref{eq:wf-Veselago-amp}), are to be evaluated at $p_y=0$. We remark that this equation is valid for both $a_4>0$ and $a_4<0$, since $\text{Re}(c_0 c_2^*)=\text{Re}(c_0^* c_2)$.
  
  \begin{figure*}[t]
  \includegraphics{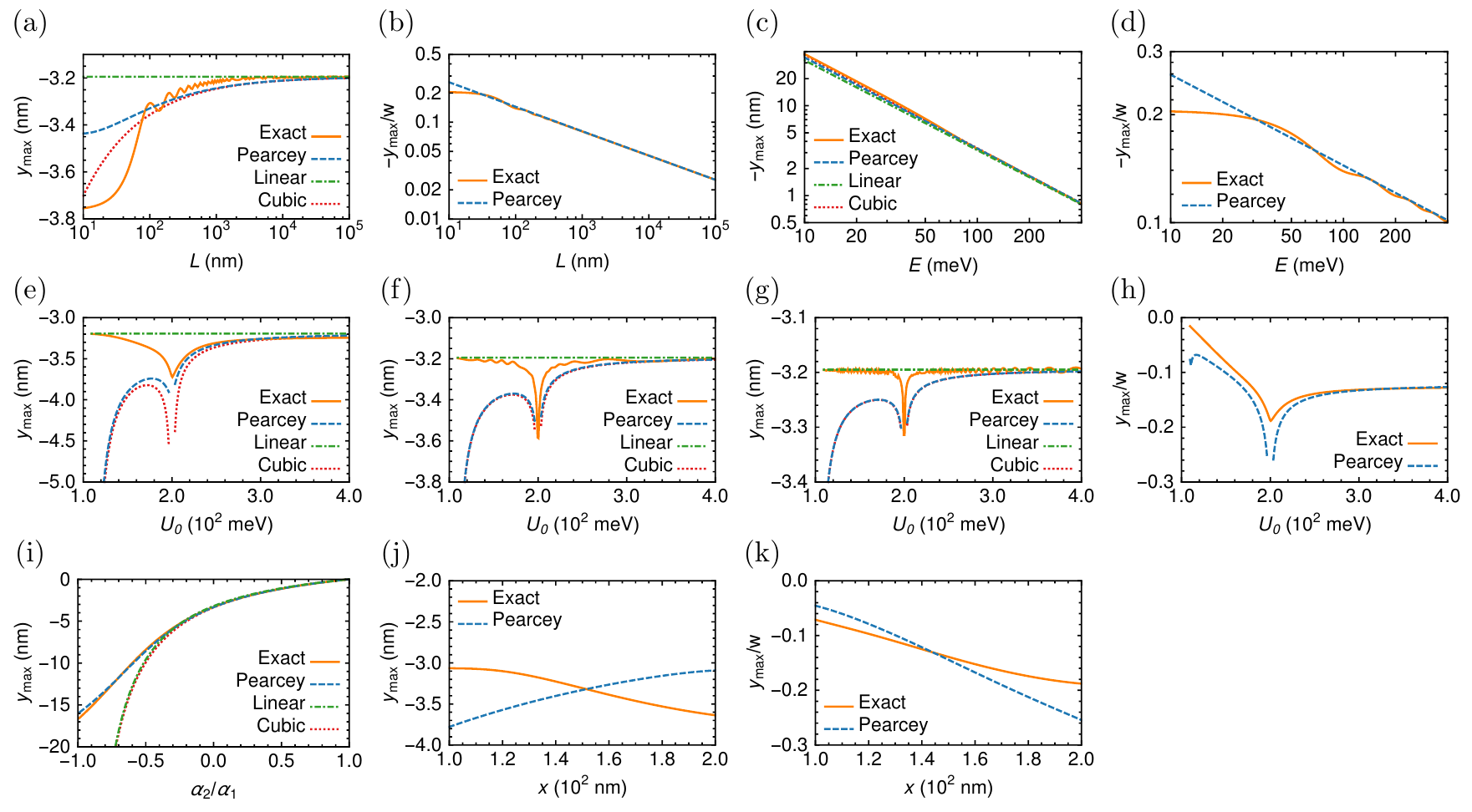}
  \caption{The dependence of the position $y_{\text{max}}$ of the maximum on various parameters. We compare the numerically obtained maxima for the exact wavefunction~(\ref{eq:wf-Veselago}) and the Pearcey approximation $\Psi_{c0}(\mathbf{x})+\Psi_{c1}(\mathbf{x})$ with the result~(\ref{eq:y0-max-simple-dim}) (labeled linear) and the solution of the third order equation~(\ref{eq:maximum-y-cubic-equation}) (labeled cubic). We consider graphene, where $\hbar v_F= 3 t a_{CC}/2$. (a) $E=100$ meV, $U_0=2.5 E$ and $\alpha_2/\alpha_1=1$. The maximum only weakly depends on the length $L$. (b) The relative position $y_{\text{max}}/w$, where $w$ is the FWHM of $\lVert \Psi \rVert^2$, approximately scales as $L^{-1/4}$. (c) $L=100$ nm, $U_0=2.5E$ and $\alpha_2/\alpha_1=1$. The position of the maximum $y_{\text{max}}$ approximately scales as $1/E$. (d) The relative position $y_{\text{max}}/w$ scales as $E^{-1/4}$ to a good approximation. (e) The dependence of $y_{\text{max}}$ on $U_0$ for $\alpha_2/\alpha_1=1$, $E=100$ meV and $L=100$ nm. (f) The same as (e) for $L=10^3$ nm. (g) The same as (e) for $L=10^4$ nm. In all cases there is a local maximum at $U_0=2E$. The accuracy of the Pearcey approximation and the solution to the cubic equation improve as $L$ increases. Our result~(\ref{eq:y0-max-simple-dim}) generally performs well. (h) The dependence of the relative position $y_{\text{max}}/w$ on $U_0$ for $L=100$~nm. (i) The dependence of $y_{\text{max}}$ on the polarization for $E=100$ meV, $U_0=2.5E$ and $L=100$ nm. All results for $y_{\text{max}}$ coincide when $\alpha_2/\alpha_1 \gtrapprox -0.4$. For smaller polarizations, only the Pearcey approximation stays close to the exact solution. (j) The dependence of $y_{\text{max}}$ on the position $x$ for $E=100$ meV, $U_0=2.5E$, $L=100$ nm and $\alpha_2/\alpha_1=1$. The result is roughly constant, with the two lines intersecting close to $x_{\text{cusp}}$. (k) The relative position $y_{\text{max}}/w$ as a function of $x$.}
  \label{fig:y-comparisons}
  \end{figure*}
  
  At this point we recall that the vertical displacement of the maximum is not a classical effect, since the classical focus lies at $y=0$. Therefore, the $y$-coordinate of the maximum $y_{\text{max}}$ cannot be of order unity. Instead, we expect $y_{\text{max}}$ to be of order $h$, since the effect is caused by (quantum) interference. Indeed, we see that Eq.~(\ref{eq:maximum-y-cubic-equation}) is a cubic equation in $y/h$, which can be solved analytically to find the maximum $y_{\text{max}}$. However, we can also do one additional approximation, using the main assumption of the semiclassical approximation, which is that the semiclassical parameter $h$ is small. Compared to the first two terms, the last three terms in Eq.~(\ref{eq:maximum-y-cubic-equation}) have an additional factor of $h^{1/2}$. Therefore, in a crude approximation, we can neglect them. The resulting linear equation in $y/h$ can easily be solved and we obtain
  \begin{equation}
    y_{\text{max}} = -h \frac{i f_{p_y}^\dagger f}{\lVert f \rVert^2} = -\frac{h}{2E} \frac{\alpha_1-\alpha_2}{\alpha_1+\alpha_2} .
    \label{eq:y0-max-simple-dimless}
  \end{equation}
  Interestingly, this result does not depend on the coefficients $c_0$ and $c_2$ of the Taylor expansion. Returning from dimensionless units to regular units, we find that
  \begin{equation}
    y_{\text{max}} = -\frac{\hbar v_F}{2E} \frac{\alpha_1-\alpha_2}{\alpha_1+\alpha_2} .
    \label{eq:y0-max-simple-dim}
  \end{equation}
  Note that the factors of $L$ that are present in both $h$ and $\tilde{y}_{\text{max}}$ have cancelled, yielding a result that does not depend on the length scale of the system.
  
  In the remainder of this section, we compare several results for the position of the maximum. The first of these is the exact value, obtained by numerically determining the maximum of the wavefunction~(\ref{eq:wf-Veselago}) at $x_{\text{cusp}}$. The second is the value obtained by numerically determining the maximum of the Pearcey approximation, composed of the leading order term~(\ref{eq:wf-cusp-Pearcey-Veselago}) and the first correction~(\ref{eq:wf-cusp-Pearcey-Veselago-corr1}). The third and fourth results are the result~(\ref{eq:y0-max-simple-dim}) and the solution of the cubic equation~(\ref{eq:maximum-y-cubic-equation}), respectively. To gain understanding of the numbers involved, we now specialize to the case of graphene, for which $\hbar v_F = 3 t a_{CC}/2$, see section~\ref{subsec:dimless-pars}.
  
  Figure~\ref{fig:y-comparisons}(a) shows the dependence of $y_{\text{max}}$ on the length scale $L$ of the system. We see that $y_{\text{max}}$ is indeed largely independent of length, as predicted by Eq.~(\ref{eq:y0-max-simple-dim}), with the exact solution showing only a slight variation. In Fig.~\ref{fig:y-comparisons}(b), we consider the dependence of the ratio $y_{\text{max}}/w$ on length. For this purpose, we define the width $w$ of the peak as 
  the full width at half maximum (FWHM) of $\lVert \Psi \rVert^2$.
  Looking at the leading-order Pearcey approximation~(\ref{eq:wf-cusp-Pearcey-Veselago}), we expect the peak width to scale as $L^{1/4}$, since both $h$ and $\tilde y$ contain a factor $1/L$, and the coefficient $a_4$ does not depend on $L$. In Fig.~\ref{fig:y-comparisons}(b), we indeed see a clear power law scaling of $-y_{\text{max}}/w$, and from a fit we find a value close to $-1/4$ for the power, as predicted. So although the position of the maximum roughly stays the same with increasing length, the displacement from the $x$-axis will be harder to see because the width of the maximum increases. This effect can be clearly seen when comparing the middle panels of Fig.~\ref{fig:sections-y-asymmetry}.
  
  In Fig.~\ref{fig:y-comparisons}(c) and (d), we consider the dependence of $y_{\text{max}}$ and $y_{\text{max}}/w$ on the electron energy. It is clear that our result~(\ref{eq:y0-max-simple-dim}) performs very well. In fact, all results are quite close to each other and show a clear power law behavior \mbox{$y_{\text{max}} \propto E^p$}, with power $p \approx -1$. Looking at Eq.~(\ref{eq:wf-cusp-Pearcey-Veselago}), we expect the width $w$ to scale as $E^{-3/4}$. Indeed, we see from Fig.~\ref{fig:y-comparisons}(d) that $-y_{\text{max}}/w$ shows a clear power law behavior, and from a fit we find that the power is close to $-1/4$. This implies that although we can increase $y_{\text{max}}$ by lowering the energy, this will also increase the width of the peak, yielding only a small increase in the ratio $y_{\text{max}}/w$.
  
  The dependence of $y_{\text{max}}$ on the potential $U_0$ is shown in Figs.~\ref{fig:y-comparisons}(e)--(h). From the exact value, we see that $y_{\text{max}}$ is largely independent of the potential, being only slightly larger at the point $U_0=2E$. Interestingly, our result~(\ref{eq:y0-max-simple-dim}) outperforms the other two approximations for $U_0<2E$, with the difference becoming smaller as $L$ increases.
  
  In Fig.~\ref{fig:y-comparisons}(i), the dependence of $y_{\text{max}}$ on the polarization is shown. As already noted in section~\ref{sec:symm-breaking}, the displacement from the $x$-axis becomes larger as the ratio $\alpha_2/\alpha_1$ decreases. It is clearly seen that our approximation~(\ref{eq:y0-max-simple-dim}) and the solution of the cubic equation~(\ref{eq:maximum-y-cubic-equation}) give good results when $\alpha_2/\alpha_1$ is larger than approximately $-0.4$, but fail for smaller ratios. The reason for this is that for large values of $v$, we can no longer approximate the Pearcey function $P^\pm(u,v)$ by its second order Taylor expansion around $v=0$. This is indicated by the fact that the Pearcey approximation, consisting of the leading order term~(\ref{eq:wf-cusp-Pearcey-Veselago}) and the first correction~(\ref{eq:wf-cusp-Pearcey-Veselago-corr1}), gives good results for all polarizations. Adding the second correction~(\ref{eq:wf-cusp-Pearcey-Veselago-corr2}) does not substantially change the result.
  Note that when we invert the polarization, that is, when we consider $\alpha_1/\alpha_2$ in the range minus one to one, the position $y_{\text{max}}$ of the maximum changes sign with respect to Fig.~\ref{fig:y-comparisons}. This can be seen directly from Eq.~(\ref{eq:wf-refl}) and is particularly clear from our result~(\ref{eq:y0-max-simple-dim}).
  
  Finally, we show the dependence of $y_{\text{max}}$ and $y_{\text{max}}/w$ on the coordinate $x$ in Fig.~\ref{fig:y-comparisons}(j) and (k). We see that $y_{\text{max}}$ is roughly constant, and that the width varies somewhat. Interestingly, the exact solution and the Pearcey approximation follow slightly different trends and intersect around $x_{\text{cusp}}$. The fact that the position of the maximum does not show a large variation with $x$ means that we can safely use our results to obtain an estimate for the asymmetry at the main focus, which is generally not located at the cusp point.
  
  Looking at all the different dependencies in Fig.~\ref{fig:y-comparisons}, we conclude that our approximation~(\ref{eq:y0-max-simple-dim}) gives quite accurate predictions for the position $y_{\text{max}}$ of the maximum, even though it was derived using several approximations. It only fails when the ratio $\alpha_2/\alpha_1$ comes close to minus one, which is the point where the central resonance disappears completely.

  In section~\ref{sec:symm-breaking}, we showed that, for the case of graphene, the displacement of the maximum in the $K'$-valley is opposite to the displacement in the $K$-valley. Since the effect is rather large, on the order of a few nanometers for energies around 100~meV, we believe that it would be possible to observe it experimentally by measuring the spatial profile of the wavefunction with the help of an STM. Another possibility would be to try to place a tiny contact near the predicted maximum and to measure the valley composition of the current using the valley Hall effect.~\cite{Xiao07,Gorbachev14} Since a typical laser beam is larger than a few nanometers in size, we believe that it would not be possible to measure the effect using second harmonic generation.~\cite{Wehling15}
  
  An important remark is that, for typical energy and length scales, $y_{\text{max}}/w$ does not exceed 0.5, as can be seen in Fig.~\ref{fig:y-comparisons}(b) and (d). So although the peak displacement is rather large, the peaks are also rather broad, making it much harder to identify them. This ratio improves as the energy and length of the device become smaller, although it should be noted that both the peak displacement and its width increase as the energy decreases. Because of the rather small value of $y_{\text{max}}/w$, we do not think that the effect is large enough to create an effective valley filter in graphene.

  \section{Current entering from a lead}  \label{sec:curr-edge-samp}
  
  In this section, we no longer consider the Green's function, but discuss the related problem where current flows into the sample through a lead on one of its sides. We first construct the wavefunction for the general case and subsequently specialize to the case of a graphene sample with graphene leads. In the second subsection, we consider the symmetries of the wavefunction. In the final subsection, we consider the semiclassical evaluation of the wavefunction.

  \subsection{Derivation of the wavefunction}
  
  For definiteness, we henceforth assume that current enters the sample from the left, through a lead of width $w$ which is located between the points $(x_s, -w/2)$ and $(x_s, w/2)$. As before, we consider a sample with a potential that consists of a single step, see Eq.~(\ref{eq:pot-step}). This gives rise to a setup that is qualitatively similar to the Green's function, but instead of considering a single point source, we now consider a lead with a finite width, which, following Huygens' principle,~\cite{Arnold89} may be regarded as a collection of point sources.
  
  Defining the characteristic length scale of the system by $L=|x_s|$, we can define the dimensionless quantities $h$, $\tilde{\mathbf{x}}$, $\hat{\tilde{\mathbf{p}}}$, $\tilde{E}$ and $\tilde{U}$ in the same way as in section~\ref{subsec:dimless-pars}. Omitting tildes, this leads to the Hamiltonian~(\ref{eq:H-Dirac}). A new dimensionless parameter in the problem is the lead width, which is naturally defined as $\tilde{w} = w/L$. Because of the translational symmetry of the lead, the wavefunction of each mode in the lead can naturally be decomposed as the product of a phase factor $e^{i p_x x/h}$ and a transversal wavefunction $\Psi_0(y)$. We henceforth assume that the each of the modes is normalized in such a way that it carries unit current, which means that in dimensionless units
  \begin{equation}
    \int^\infty_{-\infty} \tilde{\Psi}_0(y)^\dagger \sigma_x \tilde{\Psi}_0(y) \diff \tilde{y} = 1 .
    \label{eq:curr-conservation-initial-value}
  \end{equation}
  In order for this equality to hold in units with dimensions as well, we set $\Psi(\mathbf{x})=\tilde{\Psi}(\mathbf{x})/\sqrt{L}$. As before, we omit the tildes from here on and deal exclusively with these newly defined quantities, unless otherwise indicated.
  
  In mathematical terms, we can now formulate the problem at hand as an initial value problem, namely
  \begin{equation}
    \left[ {\bm \sigma} \cdot \hat{\mathbf{p}} + U(\mathbf{x}) \right] \Psi(x,y) = E \Psi(x,y) , \quad \Psi(x_s,y) = \Psi_0(y) ,
    \label{eq:Dirac-initial-value}
  \end{equation}
  where the dimensionless $x_s$ equals minus one.
  In appendix~\ref{app:initial-value-problem}, we solve this problem for an arbitrary initial wavefunction $\Psi_0(y)$. The general solution~(\ref{eq:wavefunction-initial-value}) is a linear combination of the independent solutions $\overline{\Psi}_>(x)$ and $\overline{\Psi}_<(x)$, defined in Eqs.~(\ref{eq:Psi-inc-right}) and~(\ref{eq:Psi-inc-left}), which correspond to waves coming in from minus infinity and infinity, respectively. When we consider the case where no current flows into the sample from the right, the coefficient in front of $\overline{\Psi}_<(x)$ should be zero, and it is sufficient to consider only the term proportional to $\overline{\Psi}_>(x)$. Note that in a realistic sample, the former coefficient is not necessarily zero, as the finite length and width of the sample will introduce scattering between various modes. Nevertheless, we expect the approximation to hold in reasonably sized samples, as the induced scattering will be small. In the appendix, we show that in that case the wavefunction is approximately given by
  \begin{multline}
    \Psi(x,y) = \frac{e^{i \pi/4}}{\sqrt{2\pi h}} \int^{E}_{-E} \diff p_y \; e^{i p_y y/h} e^{-i \sqrt{E^2-p_y^2} x_s/h} \, \overline{\Psi}_>(x) \\
      \times \frac{1}{\sqrt{2 \cos\phi}} \Big( e^{-i \phi/2} \; \; \; e^{i \phi/2} \Big) \overline{\Psi}_0(p_y) ,
    \label{eq:sol-fwd-initial-value}
  \end{multline}
  where $( e^{-i \phi/2} \; \; e^{i \phi/2} )$ is a row vector and $\overline{\Psi}_0(p_y)$ is the Fourier transform of $\Psi_0(y)$, defined in Eq.~(\ref{eq:def-FT}).

  In the remainder of this section, we consider the specific example of a graphene sample, with current entering through a graphene lead. In particular, we consider a graphene lead with zigzag edges, which do not mix the two valleys $K$ and $K'$. Within the continuum approximation, which is valid for sufficiently broad leads, we can then obtain the wavefunction in the lead by setting the boundary conditions~\cite{Brey06}
  \begin{equation}
    \Psi_A(y=-w/2)=0, \quad \Psi_B(y=w/2)=0 ,
    \label{eq:bc-ZZ}
  \end{equation}
  which are valid for both valleys.
  Within the graphene lead, we allow for the presence of a constant mass, which can for instance arise in the context of chemical functionalization,~\cite{Katsnelson13} or for graphene on a substrate, such as h--BN.~\cite{Sachs11,Bokdam14,VanWijk14} In the Hamiltonian~(\ref{eq:H-Dirac}) for the $K$-valley, it manifests itself as an additional term $m \sigma_z$. Solving the eigenvalue equation for a constant potential and a constant mass, and imposing the boundary conditions~(\ref{eq:bc-ZZ}), we find that the wavefunction within the lead equals~\cite{Brey06}
  \begin{equation}
    \Psi_n^{\text{lead}}(x,y) = \frac{e^{i p_x x/h}}{\sqrt{J_n}}
      \begin{pmatrix}
        \sin\left( p_n (y+w/2)/h \right) \\
        \alpha_n \sin\left( p_n (y-w/2)/h \right)
      \end{pmatrix} B\left(\frac{y}{w}\right) ,
    \label{eq:wf-lead}
  \end{equation}
  where $B(x)$ is the so-called boxcar function:
  \begin{equation}
    B(x) = \left\{ 
      \begin{array}{ll}
        1,  \quad\; |x| \leq 1/2, \\
        0,  \quad\; |x| > 1/2 .
      \end{array} 
      \right.
  \end{equation}
  The momenta $p_x$ and $p_n$ are defined by the relations $p_x^2 = E^2-m^2-p_n^2$ and
  \begin{equation}
    \tan(p_n w/h) = - \frac{p_n}{p_x} .
    \label{eq:pn-lead-roots}
  \end{equation}
  Furthermore, the factor $\alpha_n$ is defined by
  \begin{equation}
    \alpha_n = -\frac{p_n}{(E+m) \sin(p_n w/h)} .
    \label{eq:def-alpha-lead}
  \end{equation}
  When $m=0$, it is easy to show that $\alpha_n=\pm 1$, and that its value alternates between successive bands. Finally, the normalization factor $J_n$ ensures that the mode carries unit current, i.e. that Eq.~(\ref{eq:curr-conservation-initial-value}) is satisfied. We remark that in the above computation we have disregarded the surface states,~\cite{Brey06} and hence do not consider very low energies. The computations for the $K'$-valley are entirely analogous.
  
  The last issue that we need to consider is the relation between $\Psi_0(y)$ and $\Psi_n^{\text{lead}}$. Let us first look at the case where $m=0$ and consider a single incoming mode. Since the lead and the left side of the sample have the same potential and the same mass, we can expect that there will be very little backreflection into the lead. Though small, this reflection will in reality however not be zero, because of the finite width of the lead. Furthermore, note that when we completely neglect the backreflection, the coefficient $c_2$ in front of $\overline{\Psi}_<(x)$ no longer vanishes, as a computation using Eq.~(\ref{eq:wavefunction-initial-value}) shows. Nevertheless, this coefficient is still small, and we consider the approximation of $\Psi_0(y)$ by $\Psi_n^{\text{lead}}$ feasible.
  
  When the mass inside the lead does not vanish, i.e. $m\neq 0$, the situation is rather different. In this case, the dispersion in the lead, $p_x^2=E^2-m^2-p_n^2$ differs from the dispersion in the sample, $p_x^2=E^2-p_y^2$. We therefore expect significant backreflection into the lead, which increases as the mass increases. This means that we can no longer approximate $\Psi_0(y)$ by $\Psi_n^{\text{lead}}(0,y)$, but that we should include multiple left-moving modes with appropriate reflection coefficients.

  \subsection{Symmetries of the wavefunction}
  
  Let us first consider the symmetry of the wavefunction in the absence of a mass term, i.e. for $m=0$. In this case $\alpha_n=\pm 1$ and it is easy to show that
  \begin{equation}
    \Psi_n^{\text{lead}}(x,-y) = - \alpha_n \sigma_x \Psi_n^{\text{lead}}(x,y) .
    \label{eq:symm-lead}
  \end{equation}
  As we discussed in the previous subsection, we can approximate $\Psi_0(y)$ by $\Psi_n^{\text{lead}}(0,y)$ in this case. When we consider its Fourier transform, we see that it has the same symmetry, i.e
  \begin{equation}
    \overline{\Psi}_0(-p_y) = - \alpha_n \sigma_x \overline{\Psi}_0(p_y) .
  \end{equation}
  Using this identity, we can show that the wavefunction~(\ref{eq:sol-fwd-initial-value}) also possesses this symmetry:
  \begin{equation}
    \Psi(x,-y) = - \alpha_n \sigma_x \Psi(x,y) .
    \label{eq:symm-fwd-initial}
  \end{equation}
  The main ingredient of the calculation is the change of variables $p_y \to -p_y$ in the integral. Under this transformation, $\overline{\Psi}_>(x)$ becomes $\sigma_x \overline{\Psi}_>(x)$, as can be seen from Eq.~(\ref{eq:Psi-inc-right}). We conclude from Eq.~(\ref{eq:symm-fwd-initial}) that $\Psi(x,y)$ is symmetric in the $x$-axis when $m=0$. 
  
  Before considering the case of a nonzero mass, let us first consider a second symmetry of the lead wavefunction. From Eq.~(\ref{eq:wf-lead}), we see that $\Psi_n^{\text{lead}}$ is real, irrespective of the mass $m$ as long as $m<E$. Because of the properties of the Fourier transform~(\ref{eq:def-FT}), this means that
  \begin{equation}
    \left[ \overline{\Psi}_n^{\text{lead}}(x,-p_y) \right]^* =  e^{i \pi/2} \overline{\Psi}_n^{\text{lead}}(x,p_y) .
    \label{eq:symm-FT}
  \end{equation}
  This symmetry implies that $\lVert \overline{\Psi}_n^{\text{lead}}(x,-p_y) \rVert = \lVert \overline{\Psi}_n^{\text{lead}}(x,p_y) \rVert $, which, roughly speaking, means that the amount of current that has positive $p_y$ is the same as the amount of current that has negative $p_y$. For $m=0$, the latter equality is also implied by Eq.~(\ref{eq:symm-fwd-initial}). However, the more general statement~(\ref{eq:symm-FT}) is also true for nonzero masses.
  
  When $m\neq 0$, the symmetry~(\ref{eq:symm-lead}) is clearly broken, since one sees from Eqs.~(\ref{eq:wf-lead}) and~(\ref{eq:def-alpha-lead}) that a mass term creates a difference in the amplitudes on the two sublattices. However, something more fundamental is going on when $m\neq 0$, since at the point where the lead and the sample join, the dispersion relation changes. As discussed in the previous subsection, we therefore expect significant backreflection into the lead and we cannot simply approximate $\Psi_0(y)$ by the incoming mode $\Psi_n^{\text{lead}}(0,y)$. Instead, we need to take a linear combination of the incoming mode and several reflected modes, with appropriate coefficients. These reflection coefficients are in general complex, meaning that $\Psi_0(y)$ is no longer a real function. Hence, the symmetry~(\ref{eq:symm-FT}) is broken, and we can expect the amount of current that is emitted with positive transversal momentum to be different from the amount of current that is emitted with negative transversal momentum. Therefore, we expect the effect of sublattice polarization for this case to be quite different from the effect for the case of the Green's function, which we discussed elaborately in the previous sections. Since the determination of the reflection coefficients is in general not an easy task, we do not pursue this problem further in this paper. However, from our previous considerations it is clear that a sublattice polarization, originating from a mass term within the lead, should lead to an asymmetry.

  \subsection{Semiclassical evaluation}
  
  In this final subsection, we consider the semiclassical evaluation of the wavefunction~(\ref{eq:sol-fwd-initial-value}) for a sample where the current enters from a lead with zero mass, i.e. $\Psi_0(y)$ is given by $\Psi_n^{\text{lead}}(0,y)$, Eq.~(\ref{eq:wf-lead}), with $m=0$. We make the dependence on the lead mode explicit by including the mode number $n$ in the notation, i.e. we write $\Psi_{n,0}$ and $\Psi_n$.
  
  Let us first consider the ``deep'' semiclassical limit, where both $h \ll 1$ and the dimensionless parameter $\hbar v_F/(E_0 w) \ll 1$. When we want to apply the stationary phase approximation to the solution~(\ref{eq:sol-fwd-initial-value}), we should be aware of the dependence of $\overline{\Psi}_{n,0}(p_y)$ on $h$. Explicitly writing down the Fourier transform, we obtain
  \begin{multline}
    \overline{\Psi}_{n,0}(p_y) = \frac{e^{-i \pi/4}}{\sqrt{2\pi h}} \int_{-\infty}^{\infty} \frac{\diff y_0}{2 i} \left[ 
      e^{i p_n y_0/h} \begin{pmatrix} e^{i p_n w/2 h} \\ \alpha_n e^{-i p_n w/2 h} \end{pmatrix} - 
      \right. \\ \left. 
      e^{-i p_n y_0/h} \begin{pmatrix} e^{-i p_n w/2 h} \\ \alpha_n e^{i p_n w/2 h} \end{pmatrix} 
    \right] \frac{e^{-i p_y y_0/h}}{\sqrt{J_n}} B\left(\frac{y_0}{w}\right) .
    \label{eq:Psi-lead-FT}
  \end{multline}
  We insert this Fourier transform into the wavefunction~(\ref{eq:sol-fwd-initial-value}) and specialize to the case $x>0$, whence $\overline{\Psi}_>(x)$ is given by Eq.~(\ref{eq:Psi-inc-right}). We then see that we are dealing with a sum of two two-dimensional integrals, that should be considered separately. The actions for these two integrals are given by
  \begin{multline}
    S_{np}^\pm(x,y,p_y,y_0) = p_y (y - y_0) \pm p_n y_0 - x_s \sqrt{E^2-p_y^2} - \\ x \sqrt{(U_0-E)^2-p_y^2} .
    \label{eq:action-lead-pm}
  \end{multline}
  The stationary points correspond to those points where the partial derivatives with respect to $p_y$ and $y_0$ vanish. The condition $\partial S_{np}^\pm/\partial p_y = 0$ yields the condition
  \begin{equation}
    y-y_0 +x_s \frac{p_y}{\sqrt{E^2-p_y^2}} + x \frac{p_y}{\sqrt{(E-U_0)^2-p_y^2}} = 0 ,
    \label{eq:statpoint-py}
  \end{equation}
  which is very similar to the condition~(\ref{eq:trajectories}) that we had for the Green's function. Furthermore, $\partial S_{np}^\pm/\partial y_0 = 0$ yields
  \begin{equation}
    -p_y \pm p_n = 0.
    \label{eq:statpoint-y0}
  \end{equation}
  Together, these two conditions determine the classical trajectories of the system. We see that, as in the case of the Green's function, the trajectories are straight lines and are focussed by the \emph{n-p} junction. However, this time they are not emitted from a single point, but from a line, parametrized by the variable $y_0$. This can be seen as an illustration of Huygens' principle: each point of the lead acts as a point source. However, in this situation the transversal momenta $p_y$ of the trajectories are strongly constrained and can take only two values, namely $\pm p_n$, with $p_n$ the transversal momentum of the mode in the lead. We also note that, because of this constraint, the set of trajectories covers only a limited region of space.
  
  Caustics in the system arise when the Hessian matrix $A$, the matrix of second derivatives of the action, is degenerate, i.e. $\det A=0$, see appendix~\ref{appsub:WKB}. Computing the second derivatives of the action~(\ref{eq:action-lead-pm}), we see that $\partial^2 S/\partial y_0^2 = 0$ and that $\partial^2 S/\partial y_0 \partial p_y = -1$. Hence $\det A=-1$, and we always have one positive and one negative eigenvalue. We therefore conclude that, as long as we are in the deep semiclassical limit, there are no caustics in the system.
  
  In this limit, we can therefore construct an approximation for the wavefunction, given by Eqs.~(\ref{eq:sol-fwd-initial-value}) and~(\ref{eq:Psi-lead-FT}), by employing the WKB approximation, as explained in appendix~\ref{appsub:WKB}. The calculation is rather involved, and in particular requires a careful analysis of the transversal momenta $p_n$, defined by Eq.~(\ref{eq:pn-lead-roots}). Numbering the modes from the lowest value of $p_n$ up and starting at one, one can show that $\sqrt{\alpha_n} = e^{i \pi n/2}$. The final result is then given by  
  \begin{multline}
    \Psi_n(x,y) = \frac{t}{\sqrt{J_n}} \frac{\sqrt{\cos\phi}}{\sqrt{\cos\theta}}
      e^{-i (\sqrt{E^2-p_n^2} x_s +\sqrt{(U_0-E)^2-p_n^2} x)/h} \\
      \times \frac{\sqrt{\alpha_n}}{2 i} \left[
      e^{i p_n y/h} \begin{pmatrix} e^{-i\theta/2} \\ e^{i\theta/2} \end{pmatrix} B\left(\frac{y_0(x,y,p_n)}{w}\right) - \right. \\
      \left. \alpha_n e^{-i p_n y/h} \begin{pmatrix} e^{i\theta/2} \\ e^{-i\theta/2} \end{pmatrix} B\left(\frac{y_0(x,y,-p_n)}{w}\right) 
      \right] ,
    \label{eq:wf-lead-statphase}
  \end{multline}
  where $t$ is the transmission coefficient~(\ref{eq:t-potstep}). We note that $y_0$ is a function of $x$, $y$ and $p_y$ (which equals $\pm p_n$) through the condition~(\ref{eq:statpoint-py}) for a stationary point.
  
  From the equations for the stationary point, we see that the two trajectories that emerge from the point $(x_s,0)$ meet each other in the point $x_{i,n} = x_s \tan\phi_{p_n}/\tan\theta_{p_n}$. Around this point there is a rhombus shaped region where interference occurs. Furthermore, we note that the point $x_{i,n}$ is different for each mode $n$, unless $U_0=2E$, in which case the point $x_{i,n}$ is the same for all modes, as discussed in section~\ref{sec:caustics}. We also remark that the WKB approximation~(\ref{eq:wf-lead-statphase}) has the symmetry~(\ref{eq:symm-fwd-initial}). This can be easily seen when one uses the identity $B(y_0(x,-y,p_n)/w)=B(y_0(x,y,-p_n)/w)$. Note that this symmetry implies that within the interference region, both components of the wavefunction are given by a cosine for modes with $\alpha_n=-1$, whereas they are given by a sine for modes with $\alpha_n=1$.
  
  Secondly, let us consider the case where $h=\hbar v_F/(E_0 L)$ is small, and $\hbar v_F/(E_0 w)$ is rather large, which means that $w/L$ is small, i.e. we are dealing with a relatively narrow lead. In that case, the physical situation is slightly different from the one sketched in the previous paragraphs. In order to understandy why, one needs to consider the width of the Fourier transform $\overline{\Psi}^{\text{lead}}_n(p_y)$ of the modes in the lead. To determine it, we first note that each of the components of the wavefunction $\Psi_n^{\text{lead}}$ is the product of a trigonometric function and a boxcar function. Therefore, the Fourier transform of each of these components is the convolution of the Fourier transforms of the two functions that make up the product. Since the Fourier transform of a trigonometric function is the sum of two delta functions, it is the Fourier transform of the boxcar function that determines the width of the Fourier transform of the lead wavefunction. This Fourier transform is easy to compute, and we obtain
  \begin{align}
    \overline{B}(p_y) &= \frac{e^{-i \pi/4}}{\sqrt{2\pi h}} \int_{-\infty}^\infty B(y/w) e^{-i p_y y/h} \diff y \nonumber \\
                      &= e^{-i \pi/4} \frac{2 h}{\sqrt{2\pi h}} \frac{\sin(p_y w/2 h)}{p_y}
    \label{eq:boxcar-FT}
  \end{align}
  To obtain an estimate of the full width at half maximum (FWHM) of this Fourier transform, we expand the sine up to third order in its argument. Solving for the point where the absolute value of the function is half of its maximal value and subtracting the two solutions, we find that the FWHM is approximately given by
  \begin{equation}
    \Delta p_y = p_{y,+}-p_{y,-} = 4\sqrt{3} \frac{h}{w},
    \label{eq:leadwf-width-FT}
  \end{equation}
  where all units are dimensionless. Going back to units with dimensions, we see that the width is determined by the dimensionless parameter $\hbar v_F/(E_0 w)$.

  When this dimensionless semiclassical parameter in the lead is rather large, the Fourier transform is broad. In that case, it is less appropriate to consider the solution~(\ref{eq:sol-fwd-initial-value}) as a double integral to which one should apply the WKB approximation. Instead, one should rather think of it as a single integral and consider $\overline{\Psi}_{n,0}(p_y)$ as a function that does not depend on $h$. From a physical perspective, one might say that, from each point in the lead, trajectories come out at all angles, instead of at just two.
  This means that we enter a regime for which the classical picture is qualitatively more similar to the one discussed in section~\ref{sec:caustics}, albeit with a lead as the source of electrons instead of a single point. In particular, we expect the formation of caustics when $U_0\neq 2E$.
  
  \begin{figure}[t]
  \includegraphics{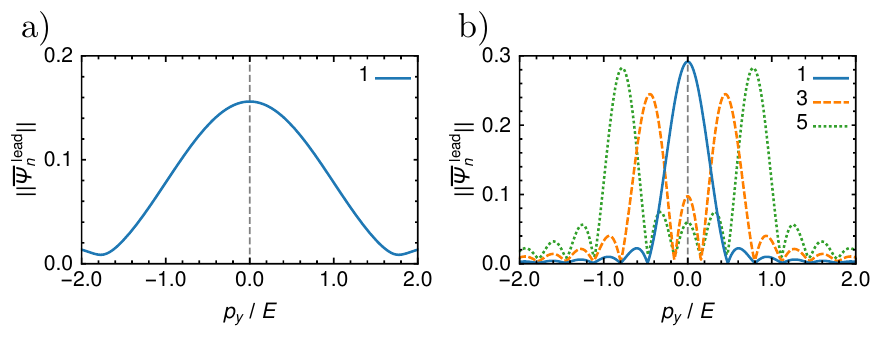}
  \caption{The norm of the Fourier transform $\overline{\Psi}^{\text{lead}}_n$ for various modes in a graphene lead of width $30$~nm. (a) The energy $E=100$~meV and $\hbar v_F/(E_0 w)=0.213$. There is only one mode in the lead, with a broad Fourier transform. (b) $E=400$~meV and $\hbar v_F/(E_0 w)=0.0533$. There are six modes in the lead, which have a rather narrow Fourier transform. One can speak about characteristic momenta.}
  \label{fig:ft-modes-lead}
  \end{figure}
  
  To illustrate this, let us consider a particular situation. In Fig.~\ref{fig:ft-modes-lead}, we show the norm of the Fourier transform of the wavefunction $\Psi_n^{\text{lead}}(x,y)$ for various modes in a graphene lead of width $30$ nm. In Fig.~\ref{fig:ft-modes-lead}(a), the energy of the electrons is $100$ meV, which means that $\hbar v_F/(E_0 w)=0.213$. One indeed sees that the Fourier transform of the only mode in the lead is very broad and that it does not make much sense to speak about characteristic momenta. When we raise the energy of the electrons to $400$ meV, the dimensionless parameter $\hbar v_F/(E_0 w)=0.0533$ and there are six modes in the lead. Now the Fourier transform of the various modes is much narrower, as can be seen in Fig.~\ref{fig:ft-modes-lead}(b), and it does make sense to speak about characteristic momenta.
  
  \begin{figure*}[t]
  \includegraphics{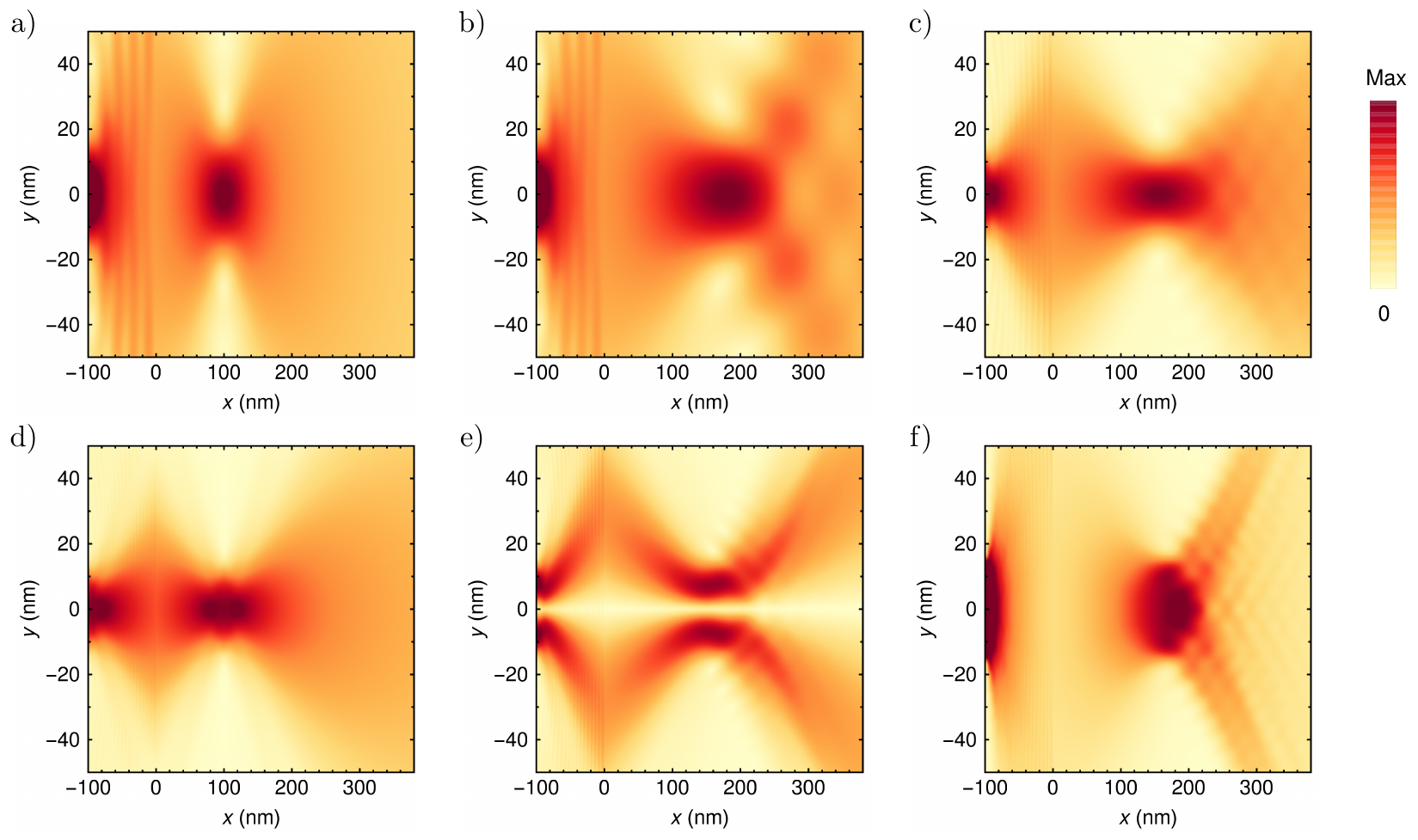}
  \caption{The norm $\lVert \Psi_n \rVert$ of the wavefunction that results from current entering through a lead of width $w=30$~nm on the left side of a sample with $L=100$~nm. The \emph{n-p} junction is located at $x=0$. (a) For $E=100$~meV, there is only one mode in the lead and the trajectories exit the lead at all angles. Since $U_0=200$~meV, they are focussed in a single point, which makes the situation qualitatively similar to the one for the Green's function. (b) For $E=100$~meV and $U_0=250$~meV, we see a caustic. (c) When $E=250$~meV and $U_0=625$~meV, the interference pattern that results from the first mode in the lead is much less pronounced, and we are moving closer towards the deep semiclassical limit. (d) For $E=400$~meV, the lead contains six modes, each of which has a rather sharp Fourier transform and gives rise to a well-defined transversal momentum. Depicted here is the wavefunction that results from the first mode for $U_0=800$~meV. (e) We clearly see that for $E=400$~meV, the second mode carries a well-defined transveral momentum. For $U=1000$~meV, we see almost no interference pattern, implying that we are close to the deep semiclassical limit. (f) The total density $\lVert \Psi \rVert^2_{\text{tot}} = \sum_n \lVert \Psi_n \rVert^2$,
  for $E=400$~meV and $U_0=1000$~meV. We see that we have a sharp focussing spot. The maximum of the color scale equals (a), (b) 0.18, (c), (d), (e) 0.25, (f) 0.135.}
  \label{fig:overview-current-lead}
  \end{figure*}
  
  In Fig.~\ref{fig:overview-current-lead}, we show the norm $\lVert \Psi_n \rVert$ of the wavefunction~(\ref{eq:sol-fwd-initial-value}) for various electron energies and potential heights for fixed length scales $L=100$~nm and $w=30$~nm. Comparing Figs.~\ref{fig:overview-current-lead}(a) and~(b), we see that for $E=100$ meV, we are indeed in a situation that is qualitatively similar to the Green's function, since trajectories come out of the lead at all angles. When $U_0=2E$, they are focussed in a single point, and when $U_0 \neq 2E$, a caustic occurs and we see the characteristic interference pattern. When $E=250$~meV, this pattern is already much less pronounced, see Fig.~\ref{fig:overview-current-lead}(c). For an electron energy $E=400$~meV, the various modes in the lead carry designated momenta, as is particularly clear from Fig.~\ref{fig:overview-current-lead}(e), where the wavefunction that results from the second mode in the lead is shown. For $U_0=2E$, Fig.~\ref{fig:overview-current-lead}(d), all trajectories are once again focussed in a single point, whereas for $U_0 \neq 2E$ this is not the case. However, in Fig.~\ref{fig:overview-current-lead}(f), where we show the intensity $\lVert \Psi \rVert^2_{\text{tot}}=\sum_n \lVert \Psi_n \rVert^2$, i.e. the sum of the intensities $\lVert \Psi_n \rVert^2$ that result from the separate modes in the lead, we do not see a clear interference pattern characteristic of a caustic. Instead, we have a rather sharp focussing spot, indicating that we are in the regime where we can approximate the wavefunction~(\ref{eq:sol-fwd-initial-value}) by its WKB approximation~(\ref{eq:wf-lead-statphase}).
  
  \begin{figure*}[t]
  \includegraphics{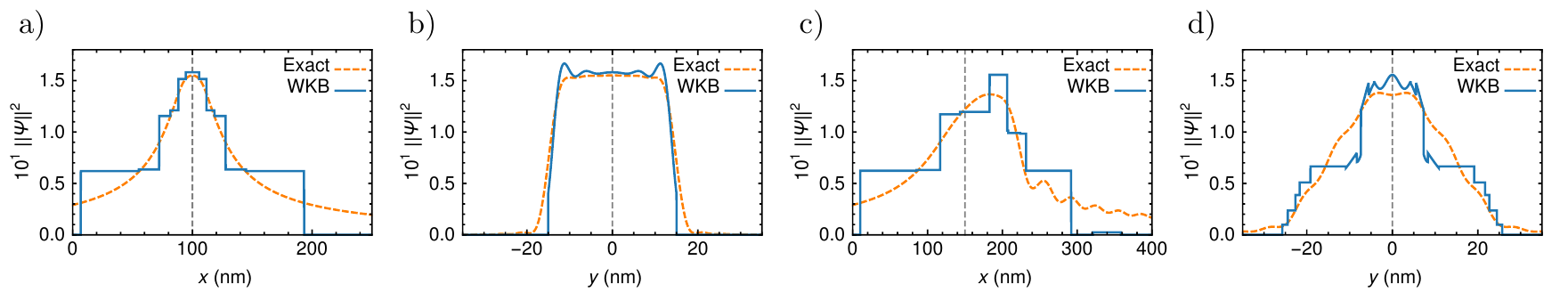}
  \caption{Comparison of the solution~(\ref{eq:sol-fwd-initial-value}) with the stationary phase approximation~(\ref{eq:wf-lead-statphase}). We consider $\lVert \Psi \rVert^2_{\text{tot}}$, meaning that we sum over all modes in the lead. The lead width is $w=30$~nm and $L=100$~nm; the electron energy $E=400$~meV. (a) Comparison along the $x$-axis for $U_0=800$ meV. (b) Comparison along the line $x=L$ for the same potential. Both sections show quite good agreement. (c) Comparison along the $x$-axis for $U_0=1000$~meV. (d) Comparison along the line $x=190$~nm for the same potential. The agreement is not as good as for $U_0=2E$, but the stationary phase approximation still captures the essential features of the wavefunction.}
  \label{fig:statphase-lead-comparisons}
  \end{figure*}
  
  Comparing Fig.~\ref{fig:overview-current-lead}(f) with the numerical results from Ref.~\onlinecite{Milovanovic15}, we see that there is qualitative agreement between the two approaches. Unfortunately, their numerical simulations use rather high energies, typically 0.4~eV, in combination with rather wide leads, typically 50~nm, so their pictures only show one of the two regimes that we have identified, i.e. the WKB regime. However, going to lower energies or narrower leads, we believe that it should be straightforward to observe the other regime, in which caustics occur, as well. In their paper, they also observe a lowered transmission for leads with zigzag edges as compared to leads with armchair edges, which they attribute to a poorer lensing ability of an armchair \emph{p-n} interface as compared to a zigzag \emph{p-n} interface. As expected, our study, which stays within the framework of the continuum approximation, does not offer any alternative explanations for this effect and we believe that further research would be necessary to elucidate its nature.

  In Fig.~\ref{fig:statphase-lead-comparisons}, we compare the total density $\lVert \Psi \rVert^2_{\text{tot}}$ for the wavefunction~(\ref{eq:sol-fwd-initial-value}) and its WKB approximation~(\ref{eq:wf-lead-statphase}) for $E=400$~meV, $w=30$~nm and $L=100$~nm. We see that the WKB-approximation captures the essential behavior of the wavefunction, but that the discrepancy is rather large, especially away from the maximum. We ascribe this discrepancy, which is notably smaller for $U_0=2E$ than for $U_0\neq 2E$, to the fact that the width of the Fourier transform of the lead wavefunction is still rather large. For $E=400$~meV and $w=30$~nm, our estimate~(\ref{eq:leadwf-width-FT}) gives $\Delta p_y = 0.37$ in dimensionless units, which means that for each mode, apart from $\pm p_n$, there are still a lot of other values of $p_y$ that contribute to the scattering.

  \section{Conclusion} \label{sec:conclusion}
  
  In this paper, we have studied two realizations of electronic Veselago lenses for massless Dirac fermions. We have found that in both cases the presence of (pseudo)spin polarization leads to symmetry breaking. By comparing the exact solutions with various semiclassical approximations, we have established that the semiclassical approximation is an effective tool to study focussing in graphene.
  
  For the case of the Green's function, we have demonstrated that, depending on the (pseudo)spin polarization, the main focus can either be vertically displaced, or can vanish completely. When the polarization equals $(1,0)$, the main focus is displaced from the $x$-axis, on which it lies in the absence of (pseudo)spin polarization. Specializing to the case of graphene, the size of this effect is typically on the order of several nanometers and the effect is opposite for electrons in the $K$-valley and in the $K'$-valley. However, since the ratio of the displacement and the peak width is typically smaller than 0.25, we believe that this effect is not strong enough to create an effective valley filter in graphene. Nonetheless, we think that the effect could be measured experimentally. An initial sublattice polarization of $(1,0)$ could be realized by injecting electrons onto a single site with an STM, and a smaller amount of symmetry breaking could perhaps be attained by considering tunneling through hexagonal boron nitride.~\cite{Sachs11,Bokdam14,VanWijk14} Subsequently, one could for instance try to measure the displacement using an STM, or with the valley Hall effect.~\cite{Xiao07,Gorbachev14} 
  When the polarization equals $(1,-1)$, the main focus vanishes completely. Although we believe that such a polarization would be hard to realize in graphene, it is likely to be attainable in topological insulators, where we are dealing with real spin instead of pseudospin.
  
  To gain more insight into the effect of different polarizations, we have studied the Green's function using various semiclassical approximations. We have demonstrated that the vertical position of the main focus can be well predicted using the Pearcey approximation, provided that we include its first correction. This approximation also gives good results for the horizontal position of the maximum, even for large values of the semiclassical parameter. Furthermore, we have shown that the uniform approximation shows very good agreement with the exact solution, making it the preferred approximation when one is not only interested in the position of the maximum, but also in its value. Using the Pearcey approximation with various corrections, we have derived Eq.~(\ref{eq:ratio-intensities}), which shows that the ratio between the peak intensities for polarizations $(1,-1)/\sqrt{2}$ and $(1,1)/\sqrt{2}$ is proportional to the dimensionless semiclassical parameter $h$ of the system. Finally, we have derived Eq.~(\ref{eq:y0-max-simple-dim}), which reveals how the displacement of the main focus depends on the different system parameters. We have demonstrated that it shows excellent agreement with the exact solution.
  
  For the case of current entering a graphene sample through a narrow graphene lead, we have used the semiclassical approximation to identify two different regimes. 
  When the dimensionless semiclassical parameter in the lead is rather large, while the semiclassical parameter of the system is small, which happens for instance for low energies or narrow leads, we expect caustics to be formed in the system. On the other hand, when both semiclassical parameters are small, we have shown that a rather sharp focussing spot will occur. We believe that the transition between these two regimes should be visible both in experiment and in numerical simulations. The effects of symmetry breaking in this system are less clear, since a mass term in the lead not only breaks the reflection symmetry in the $x$-axis, but also the symmetry of the Fourier transform of the total wavefunction in the lead, due to the presence of reflected waves. Because of this, we can generally expect the amount of current with positive transversal momentum to differ from the amount of current with negative transversal momentum.

  \section*{Acknowledgements}
  
  We are grateful to Timur Tudorovskiy for stimulating discussions that initiated this work, and to Sergey Dobrokhotov for valuable discussions on the semiclassical evaluation of the wavefunction. Furthermore, we are grateful to Kostya Novoselov, Alex Khajetoorians, Misha Titov and Erik van Loon for helpful discussions about the project.
  
  The authors acknowledge support from the ERC Advanced Grant 338957 FEMTO/NANO and from the NWO via the Spinoza Prize.
  
  \appendix

  \section{Green's function} \label{app:Green}

  In this appendix, we derive the Green's function defined by Eq.~(\ref{eq:Dirac-Green-dimless}) with the potential~(\ref{eq:pot-step}). In the first subsection, we define the scattering states for the problem without a source term, which we will need in the construction later on. In appendix~\ref{appsub:green-rigorous}, we present a rigorous derivation of the Green's function, based on the method of variation of constants.

  \subsection{Scattering states}  \label{appsub:scat-states}
  
  Let us consider the eigenstates of the matrix Hamiltonian~(\ref{eq:H-Dirac}). Since the potential step~(\ref{eq:pot-step}) is independent of $y$, the transversal momentum $p_y$ is constant. Therefore, we can define $\Psi(\mathbf{x})=\overline{\Psi}(x)\exp(i p_y y/h)$. In the electron region ($U(x)=0$), we define the right-moving and left-moving states by
  \begin{align}
    \overline{\Psi}_{e,r}(x) &= \frac{1}{\sqrt{2 \cos \phi}} \left( \begin{array}{c} e^{-i\phi/2} \\ e^{i\phi/2} \end{array} \right) e^{i \sqrt{E^2-p_y^2} x/h} , \label{eq:Psi-el-r} \\
    \overline{\Psi}_{e,l}(x) &= \frac{1}{\sqrt{2 \cos \phi}} \left( \begin{array}{c} -e^{i\phi/2} \\ e^{-i\phi/2} \end{array} \right) e^{-i \sqrt{E^2-p_y^2} x/h} , \label{eq:Psi-el-l}
  \end{align}
  where $\phi$ is defined by $\sqrt{E^2-p_y^2}=E\cos\phi$, $p_y = E\sin\phi$.
  In the hole region ($U(x)=U_0>E$), we define the right-moving and left-moving states by
  \begin{align}
    \overline{\Psi}_{h,r}(x) &= \frac{1}{\sqrt{2 \cos \theta}} \left( \begin{array}{c} e^{-i\theta/2} \\ e^{i\theta/2} \end{array} \right) e^{-i \sqrt{(U_0-E)^2-p_y^2} x/h} , \label{eq:Psi-h-r} \\
    \overline{\Psi}_{h,l}(x) &= \frac{1}{\sqrt{2 \cos \theta}} \left( \begin{array}{c} -e^{i\theta/2} \\ e^{-i\theta/2} \end{array} \right) e^{i \sqrt{(U_0-E)^2-p_y^2} x/h} , \label{eq:Psi-h-l}
  \end{align}
  where $\theta$ is defined by $\sqrt{(U_0-E)^2-p_y^2}=(U_0-E)\cos\theta$, $p_y = -(U_0-E)\sin\theta$.
  After some calculus, we find that for both right-moving states, the probability current $j_x = \overline{\Psi}^\dagger \sigma_x \overline{\Psi}$ equals 1, whereas for the left-moving states it equals -1. Furthermore, we find that $\overline{\Psi}_{e,r}^\dagger \sigma_x \overline{\Psi}_{e,l}$ vanishes.
  
  With these definitions, we can define two independent solutions for scattering by a potential step. One solution incoming from the left,
  \begin{equation}
  \begin{aligned}
    \overline{\Psi}_{>}(x) &= \overline{\Psi}_{e,r}(x) + r \overline{\Psi}_{e,l}(x), && \quad x \leq 0 , \\
    \overline{\Psi}_{>}(x) &= t \overline{\Psi}_{h,r}(x), && \quad x \geq 0 ,
  \end{aligned} \label{eq:Psi-inc-right}
  \end{equation}
  and one solution incoming from the right,
  \begin{equation}
  \begin{aligned}
    \overline{\Psi}_{<}(x) &= t' \overline{\Psi}_{e,l}(x), && \quad x \leq 0 , \\
    \overline{\Psi}_{<}(x) &= \overline{\Psi}_{h,l}(x) + r' \overline{\Psi}_{h,r}(x) , && \quad x \geq 0 .
  \end{aligned} \label{eq:Psi-inc-left}
  \end{equation}
  Matching each of these solutions at the barrier interface gives
  \begin{align}
    t = t' &= \frac{\sqrt{\cos\phi \cos\theta}}{\cos[(\phi+\theta)/2]} , \label{eq:t-potstep} \\
    r = r' &= i \frac{\sin[(\theta-\phi)/2]}{\cos[(\phi+\theta)/2]} . \label{eq:r-potstep}
  \end{align}
  Note that the conservation of probability current automatically ensures that $|r|^2+|t|^2=1$.

  \subsection{Derivation of the Green's function} \label{appsub:green-rigorous}
  
  In order to obtain the solution to Eq.~(\ref{eq:Dirac-Green-dimless}), we need the Fourier transform and its inverse, which we define as
  \begin{equation} \label{eq:def-FT}
  \begin{aligned}
    f(x) &= \frac{e^{i\pi/4}}{(2\pi h)^{1/2}} \int_{-\infty}^{\infty} \overline{f}(p) e^{i p x/h} \, \diff p , \\
    \overline{f}(p) &= \frac{e^{-i\pi/4}}{(2\pi h)^{1/2}} \int_{-\infty}^{\infty} f(x) e^{-i p x/h} \, \diff x .
  \end{aligned}
  \end{equation}
  Performing the Fourier transform with respect to $p_y$, we obtain
  \begin{multline} \label{eq:green-1FT}
    \left[ \sigma_x \hat{p}_x + \sigma_y p_y + U(x) - E \right] \overline{G}(x,p_y) = \\ \frac{e^{-i \pi/4}}{\sqrt{2\pi h}} e^{-i p_y y_0/h} \delta(x-x_0) 1_2 .
  \end{multline}
  
  We now construct the solution $\overline{\Phi}(x)$ to
  \begin{equation} \label{eq:green-aux}
    \left[ \sigma_x \hat{p}_x + \sigma_y p_y + U(x) - E \right] \overline{\Phi}(x) = f(x)
  \end{equation}
  for arbitrary $f(x)$, coming back to Eq.~(\ref{eq:green-1FT}) only at the very end. Following the method of variation of constants,~\cite{Korn61} we can seek the solution as
  \begin{equation} \label{eq:green-trialsol}
    \overline{\Phi}(x) = c_1(x) \overline{\Psi}_>(x) + c_2(x) \overline{\Psi}_<(x) ,
  \end{equation}
  where $\overline{\Psi}_>$ and $\overline{\Psi}_<$ are the solutions to the homogeneous equation with $f(x) = 0$, see Eqs.~(\ref{eq:Psi-inc-right}) and~(\ref{eq:Psi-inc-left}).
  Since we are looking for the Green's function, we demand that there are no waves incoming from $\pm\infty$. Hence, we pose the boundary conditions
  \begin{equation} \label{eq:green-bc}
    c_1(-\infty) = 0, \qquad c_2(\infty) = 0 .
  \end{equation}
  Inserting the trial solution~(\ref{eq:green-trialsol}) into~(\ref{eq:green-aux}), we find
  \begin{equation} \label{eq:green-trialsol-result}
    c_1'(x) \sigma_x \overline{\Psi}_>(x) + c_2'(x) \sigma_x \overline{\Psi}_<(x) = \frac{i}{h} f(x) .
  \end{equation}
  
  We now confine ourselves to the situation where the source term $f(x)$ vanishes at $x\geq 0$. Then, because of the linear independence of the solutions, $c_1'(x) = c_2'(x) = 0$ for positive $x$, and we only need to consider the region $x<0$, i.e. we only have a source in the electron region. Multiplying Eq.~(\ref{eq:green-trialsol-result}) by $\overline{\Psi}_{e,r}^\dagger(x)$ and using our previous results for the probability current, we find that
  \begin{equation} \label{eq:green-c1}
    c_1'(x) = \frac{i}{h} \overline{\Psi}_{e,r}^\dagger(x) f(x) \equiv h_1(x)f(x),
  \end{equation}
  where the last equality defines $h_1(x)$ for $x<0$.
  Multiplying Eq.~(\ref{eq:green-trialsol-result}) by $\overline{\Psi}_{e,l}^\dagger(x)$, we find that
  \begin{equation} \label{eq:green-c2-in-c1}
    -r c_1'(x) - t' c_2'(x) = \frac{i}{h} \overline{\Psi}_{e,l}^\dagger(x) f(x) .
  \end{equation}
  Using the result~(\ref{eq:green-c1}), we obtain
  \begin{equation} \label{eq:green-c2}
    c_2'(x) = -\frac{i}{h} \frac{1}{t'} \Big( \overline{\Psi}_{e,l}^\dagger(x) + r \overline{\Psi}_{e,r}^\dagger(x) \Big) f(x) \equiv h_2(x)f(x) ,
  \end{equation}
  where we have defined $h_2(x)$ for $x<0$ in the last equality.
  
  Having obtained the derivatives $c_1'(x)$ and $c_2'(x)$, we can find the coefficients themselves by integrating. Taking the boundary conditions~(\ref{eq:green-bc}) into account, we have
  \begin{align}
    \int_{-\infty}^x c_1'(\xi) \diff \xi &= c_1(x) - c_1(-\infty) = c_1(x) , \\
    \int_{x}^{\infty} c_2'(\xi) \diff \xi &= c_2(\infty) - c_2(x) = -c_2(x) .
  \end{align}
  Inserting this into Eq.~(\ref{eq:green-trialsol}), using the results~(\ref{eq:green-c1}) and~(\ref{eq:green-c2}), and using that $c_1'(\xi)=c_2'(\xi)=0$ for $\xi>0$, we obtain
  \begin{align}
    \overline{\Phi}(x) &= \int_{-\infty}^x \overline{\Psi}_>(x) c_1'(\xi) \diff \xi - \int_{x}^\infty \overline{\Psi}_<(x) c_2'(\xi) \diff \xi \\
    &= \int_{-\infty}^0 g(x,\xi) f(\xi) \diff \xi ,
  \end{align}
  where
  \begin{equation}
    g(x,\xi) = \left\{ \begin{array}{ll} \phantom{-}\overline{\Psi}_>(x)h_1(\xi), \;\; & -\infty < \xi < x, \; \xi<0, \\ -\overline{\Psi}_<(x)h_2(\xi),  & x < \xi < 0 ,  \end{array} \right.
  \end{equation}
  is the Green's function.
  Note that the zero upper boundary for $\xi$ is not a fundamental limitation. Rather, it is a result of the fact that we have confined our attention to the situation where $f(x)$ vanishes for $x \geq 0$. If desired, one can expand the description and determine the derivatives~$c_1'(x)$ and~$c_2'(x)$ for positive $x$ by multiplying Eq.~(\ref{eq:green-trialsol-result}) by $\overline{\Psi}_{h,r}^\dagger(x)$ and $\overline{\Psi}_{h,l}^\dagger(x)$. This gives a natural way of expanding the definitions of $h_1(x)$ and $h_2(x)$ to the region $x \geq 0$.
  
  Now let us come back to the original problem. Comparing Eqs.~(\ref{eq:green-1FT}) and~(\ref{eq:green-aux}), we see that
  \begin{equation}
    \overline{G}(x,p_y) = \frac{e^{-i \pi/4}}{\sqrt{2\pi h}} e^{-i p_y y_0/h} g(x,x_0)
  \end{equation}
  We are mainly interested in the Green's function in the hole region, i.e. the region $x>0$. Using the definition of~$h_1(\xi)$ and of~$\Psi_>(x)$ for $x>0$, we obtain
  \begin{widetext}
  \begin{equation}
    \overline{G}(x,p_y) = \frac{e^{-i \pi/4}}{\sqrt{2 \pi h}} \frac{i}{h} \frac{1}{2 \cos[(\phi+\theta)/2]} \begin{pmatrix} e^{i\phi/2} e^{-i \theta/2} & e^{-i\phi/2} e^{-i \theta/2} \\ e^{i\phi/2} e^{i \theta/2} & e^{-i\phi/2} e^{i \theta/2} \end{pmatrix}
    e^{-i \sqrt{E^2-p_y^2} x_0/h} e^{-i \sqrt{(E-U_0)^2-p_y^2} x/h} e^{-i p_y y_0/h} .
    \label{eq:green-p-final-pspace}
  \end{equation}
  \end{widetext}
  Finally, applying the inverse Fourier transform to the Green's function~(\ref{eq:green-p-final-pspace}), we find the solution to Eq.~(\ref{eq:Dirac-Green-dimless}) as
  \begin{multline}
    G(\mathbf{x},\mathbf{x}_0) = \frac{i}{4 \pi h^2} \int_{-p_{y,\text{max}}}^{p_{y,\text{max}}} \begin{pmatrix} e^{i\phi/2} e^{-i \theta/2} & e^{-i\phi/2} e^{-i \theta/2} \\ e^{i\phi/2} e^{i \theta/2} & e^{-i\phi/2} e^{i \theta/2} \end{pmatrix} \\ 
    \times \frac{1}{\cos[(\phi+\theta)/2]} e^{i S_{np}(p_y,x,y)/h} \diff p_y ,
    \label{eq:Green-np-app}
  \end{multline}
  where the classical action $S_{np}(p_y,x,y)$ is given by
  \begin{multline}
    S_{np}(p_y,x,y) = -x_0\sqrt{E^2-p_y^2} - x\sqrt{(E-U_0)^2-p_y^2} \\ + (y-y_0)p_y . \label{eq:action-np-app}
  \end{multline}
  Note that we have set the integration limits in Eq.~(\ref{eq:Green-np-app}) to $\pm p_{y,\text{max}}$, which is defined in the last paragraph of section~\ref{subsec:class-focussing}, since the action becomes imaginary for larger values of $p_y$. Classically, $\pm p_{y,\text{max}}$ corresponds to the maximal angle $\phi_\text{max}$ for which an electron emitted by the source can propagate to the hole region. For $U_0-E>E$, we have $p_{y,\text{max}} = E$ and larger momenta do not give rise to propagating waves, whence we can ignore their contribution far from the source. For $U_0-E<E$, $\phi_{\text{max}}$ is defined by Eq.~(\ref{eq:boundary-angle}). Modes with momentum larger than $\pm p_{y,\text{max}}$ will (classically) be reflected by the barrier, since they cannot propagate in the hole region, and can therefore be ignored sufficiently far away from the barrier.

  \section{Evaluation of oscillatory integrals}   \label{app:oscillatory-integrals}
  
  In this appendix, we consider the prototype integral
  \begin{equation}
    I(\mathbf{x},h) = \int_{-\infty}^\infty \text{d} \eta \, f(\mathbf{x},\eta) e^{i S(\mathbf{x},\eta)/h} , \label{eq:int-generic}
  \end{equation}
  in the limit $h\to 0$. The vector $\mathbf{x}$ is two-dimensional and we assume that $f(\mathbf{x},\eta)$ vanishes for sufficiently large $|\eta|$. In the first subsection, we allow $\eta$ to be a vector; in the second and third subsections, we only consider scalar $\eta$. The scalar function $S(\mathbf{x},\eta)$ is henceforth referred to as the action.
  When $h$ becomes small, we can apply the so-called stationary phase approximation~\cite{Fedoryuk77,Guillemin77,Maslov81} to the integral~(\ref{eq:int-generic}). In this appendix, we briefly discuss this method for regular points and near caustics. In particular, we will discuss in detail how we can obtain good results near a cusp caustic. Although we will also give some derivations, the emphasis will be on the results.

  \subsection{WKB approximation}      \label{appsub:WKB}
  In this subsection, we allow the variable of integration $\bm{\eta}$ to be an $n$-dimensional vector. In the limit $h \to 0$, the main contribution to the integral~(\ref{eq:int-generic}) is given by the critical points,~\cite{Fedoryuk77,Guillemin77,Maslov81} where the gradient of the phase function $S(\mathbf{x}, \bm{\eta})$ with respect to $\bm{\eta}$ vanishes, i.e.
  \begin{equation}
    \left. \frac{\partial S}{\partial \eta_i} \right|_{(\mathbf{x}_0, \bm{\eta}_0)} = 0, \quad i=1 \ldots n .
  \end{equation}  
  Let us start by considering the simplest case, where the action $S(\mathbf{x},\bm{\eta})$ has a nondegenerate critical point $(\mathbf{x}_0, \bm{\eta}_0)$:
  \begin{equation}
    \det A(\mathbf{x}_0,\bm{\eta}_0) \equiv \det \left. \frac{\partial^2 S}{\partial \eta_i \partial \eta_j} \right|_{(\mathbf{x}_0, \bm{\eta}_0)} \neq 0 , \label{eq:crit-nondeg}
  \end{equation}
  which means that the Hessian matrix $A$ is invertible at the critical point. In this case, the implicit function theorem states that there exists a neighborhood of $(\bm{\eta}_0,\mathbf{x}_0)$ and a function $\bm{\eta}=\bm{\eta}(\mathbf{x})$ such that Eq.~(\ref{eq:crit-nondeg}) holds for all points in this neighborhood. One can then show that, for a nondegenerate critical point $(\mathbf{x}_0, \eta_0)$, one has~\cite{Fedoryuk77,Guillemin77,Maslov81}
  \begin{multline}
    I(\mathbf{x},h) = (2 \pi h)^{n/2} \frac{f(\mathbf{x}_0, \bm{\eta}_0)}{\sqrt{|\det A(\mathbf{x}_0,\bm{\eta}_0)|}} e^{i \pi \, \text{sgn}(A(\mathbf{x}_0,\bm{\eta}_0))/4} \\
    \times e^{i S(\mathbf{x}_0, \bm{\eta}_0)/h} \big( 1 + \mathcal{O}(h^{n/2+1}) \big) . \label{eq:statphase-nondeg}
  \end{multline}
  In the physical literature, these kind of approximations, notably for the one-dimensional case, are usually referred to as the Wentzel-Kramers-Brillouin (WKB) approximation. Although we do not discuss it here, we remark that the need to make a consistent choice for the sign of $\sqrt{\det A}$ naturally leads to the notion of the Maslov index, see Refs.~\onlinecite{Maslov73, Maslov81}. Furthermore, because of the aforementioned implicit function theorem, the result~(\ref{eq:statphase-nondeg}) can be extended to a neighborhood of the point $\mathbf{x}_0$.
  
  When there are multiple critical points $\bm{\eta}_{0,j}$ for a given value of $\mathbf{x}_0$, one has to compute the right-hand side of Eq.~(\ref{eq:statphase-nondeg}) for each of them. The integral~(\ref{eq:int-generic}) then equals the sum of these results. From a physical perspective, this means that we have interference between multiple trajectories.

  \subsection{Fold caustic: Airy approximation}   \label{appsub:Airy}
  
  From here on, we consider only scalar $\eta$. When the critical point $(\mathbf{x}_0, \eta_0)$ is degenerate, that is,
  \begin{equation}
    \left. \frac{\partial S}{\partial \eta} \right|_{(\mathbf{x}_0, \eta_0)} = 0 , \; \text{and} \;
    \left. \frac{\partial^2 S}{\partial \eta^2} \right|_{(\mathbf{x}_0, \eta_0)} = 0 , \label{eq:crit-deg}
  \end{equation}
  the approximation~(\ref{eq:statphase-nondeg}) diverges and is no longer valid. In this appendix, and in appendix~\ref{appsub:pearcey-leading-approx}, we show how the leading order term of the asymptotic expansion of $I(\mathbf{x},h)$ for $\mathbf{x}$ near $\mathbf{x}_0$ can be obtained near a fold and a cusp caustic. We remark that there is an extensive body of literature on approximating the integral $I(\mathbf{x},h)$ near a caustic, and that most approximations were gradually developed. A good overview of the various approximations in the context of semiclassical collission theory, a subject in which they have been used extensively, is given in Ref.~\onlinecite{Connor81b}. The derivations that we present in this appendix and in appendix~\ref{appsub:pearcey-leading-approx} closely follow appendix 2 of Ref.~\onlinecite{Dobrokhotov14}, only extending some of their arguments. In turn, the derivation presented there makes extensive use of the ideas of catastrophe theory and the stationary phase approximation, as presented in Refs.~\onlinecite{Fedoryuk77, Arnold82}. We have nevertheless chosen to include these derivations in this appendix, in order to make the paper self-contained and to make appendix~\ref{appsub:pearcey-higher-order-approx} more accessible to the reader.
  
  When the degenerate stationary point lies on a fold caustic, we know~\cite{Arnold82} that the third derivative of the action $S(\mathbf{x},\eta)$ does not vanish. Let us therefore consider the Taylor expansion of the action up to third order in $\eta$ around $\eta_0$, i.e.
  \begin{equation}
    \begin{aligned}
      S(\mathbf{x},\eta) &= S^{(3)}(\mathbf{x},\eta) + \mathcal{O}(\beta^4) \\
      & = q_0(\mathbf{z}) + q_1(\mathbf{z}) \beta + \frac{q_2(\mathbf{z})}{2} \beta^2 + \frac{q_3(\mathbf{z})}{6} \beta^3 + \mathcal{O}(\beta^4) ,
    \end{aligned} \label{eq:S-exp-Airy}
  \end{equation}
  where $\beta=\eta-\eta_0$ and $\mathbf{z}=\mathbf{x}-\mathbf{x}_0$. We note that $\eta_0$ and $\mathbf{x}_0$ are related through Eq.~(\ref{eq:crit-deg}).
  Subsequently, we expand the coefficients $q_i(\mathbf{z})$ up to first order in $\mathbf{z}$, that is,
  \begin{equation}
    \begin{aligned}
      q_0(\mathbf{z}) &= a_0 + \langle \mathbf{b}_0 , \mathbf{z} \rangle + \mathcal{O}(z^2), &
      q_1(\mathbf{z}) &= \langle \mathbf{b}_1 , \mathbf{z} \rangle + \mathcal{O}(z^2), \\
      q_2(\mathbf{z}) &= \langle \mathbf{b}_2 , \mathbf{z} \rangle + \mathcal{O}(z^2), &
      q_3(\mathbf{z}) &= a_3 + \mathcal{O}(z).
      \label{eq:q-exp-Airy}
    \end{aligned}
  \end{equation}
  Note that the constant parts of $q_1$ and $q_2$ vanish due to Eq.~(\ref{eq:crit-deg}).
  
  We now show how to express the leading order term of the asymptotic expansion of the integral~(\ref{eq:int-generic}) near the fold caustic in terms of the Airy-function~\cite{Airy38} $\text{Ai}(x)$, which has the integral representation
  \begin{equation}
    \text{Ai}(u) = \frac{1}{2\pi} \int_{-\infty}^\infty \exp\left( \frac{i}{3}t^3 + i u t \right) \, \diff t .
  \end{equation}
  In the integral $I(\mathbf{x},h)$, we first make the substitution $\eta=\beta+\eta_0$, and subsequently $\beta = q \gamma - q_2/q_3$,  where $q=\sqrt[3]{2 h/q_3}$. We also make a Taylor expansion of $f(\mathbf{x},\eta)$ in $\eta$ around $\eta_0$. We then see that the leading order term in the asymptotic expansion is $\mathcal{O}(h^{1/3})$, whereas the terms of $\mathcal{O}(\beta^4)$ in the action give a contribution of $\mathcal{O}(h^{2/3})$. Similarly, the first order term in the expansion of $f(\mathbf{x},\eta)$ gives a contribution of $\mathcal{O}(h^{2/3})$. Using Eq.~(\ref{eq:S-exp-Airy}) and making the above substitutions, we therefore arrive at
  \begin{align}
    I(\mathbf{x},h) &= \int_{-\infty}^\infty \text{d} \eta \, f(\mathbf{x},\eta_0) e^{i S^{(3)}(\mathbf{x},\eta)/h} + \mathcal{O}(h^{2/3}) , \label{eq:asymp-Airy-prep-0} \\
             &= 2 \pi f(\mathbf{x},\eta_0) \sqrt[3]{\frac{2 h}{|q_3|}} \exp\left[\frac{i}{h} \left( q_0+\frac{q_2^3}{3 q_3^2}-\frac{q_1 q_2}{q_3} \right) \right] \nonumber \\
             & \qquad    \times \text{Ai}\left[\frac{2^{1/3}}{h^{2/3} q_3^{1/3}} \left( q_1 - \frac{q_2^2}{2 q_3} \right) \right] + \mathcal{O}(h^{2/3}) . \label{eq:asymp-Airy-prep}
  \end{align}
  
  We now need to determine in which neighborhood of the fold caustic we can use this formula. To this end, we note that
  \begin{align}
    &q_0+\frac{q_2^3}{3 q_3^2}-\frac{q_1 q_2}{q_3} = a_0 + \langle \mathbf{b}_0 , \mathbf{z} \rangle + \mathcal{O}(z^2), \label{eq:S0-Airy} \\
    &\frac{2^{1/3}}{h^{2/3} q_3^{1/3}} \left( q_1 - \frac{q_2^2}{2 q_3} \right) = \frac{2^{1/3} \langle \mathbf{b}_1 , \mathbf{z} \rangle}{h^{2/3} a_3^{1/3}} + \frac{\mathcal{O}(z^2)}{h^{2/3}} \label{eq:arg-Airy}.
  \end{align}
  The integral $I(\mathbf{x},h)$ is heavily oscillating for small $h$ and has different asymptotic expansions for different values of $\mathbf{x}$. In particular, we cannot use Eq.~(\ref{eq:asymp-Airy-prep}) when we are far away from the point $\mathbf{x}_0$ on the fold caustic, since the coefficient $q_2(\mathbf{z})$ will be too large to justify the equality~(\ref{eq:asymp-Airy-prep-0}). More specifically, we could say that the argument of the Airy function should not be large. If it were large, we could expand the Airy function for large arguments and we would be in the regime of the WKB approximation. Therefore, a safe estimate seems to be to demand that the argument of the Airy function is $\mathcal{O}(h^\delta)$, with $\delta>0$. Setting for instance $\delta=1/6$, we find from Eq.~(\ref{eq:arg-Airy}) that we can use Eq.~(\ref{eq:asymp-Airy-prep}) in an $\mathcal{O}(h^{5/6})$ neighborhood of the fold caustic. Of course, this is an estimate and it may be possible to use the approximation in a larger neighborhood. This is however dependent on the details of the problem, for instance on the values of the coefficients $a_3$ and $\mathbf{b}_1$.
  
  Since for $z = \mathcal{O}(h^{5/6})$, we have $\mathcal{O}(z^2)h^{-2/3}=\mathcal{O}(h)$ and $\mathcal{O}(z^2)h^{-1}=\mathcal{O}(h^{2/3})$, we see that the errors that are introduced by neglecting the second order terms in the Taylor expansions of the coefficients $q_i(\mathbf{z})$ are smaller than those introduced in Eq.~(\ref{eq:asymp-Airy-prep-0}). Using Eqs.~(\ref{eq:S0-Airy}) and~(\ref{eq:arg-Airy}) and keeping only the zeroth order term of the Taylor expansion of $f(\mathbf{x},\eta_0)$ around $\mathbf{x}_0$, we can then simplify Eq.~(\ref{eq:asymp-Airy-prep}) to
  \begin{multline}
    I(\mathbf{x},h) = 2 \pi f(\mathbf{x}_0,\eta_0) \sqrt[3]{\frac{2 h}{|a_3|}} \exp\left[ \frac{i}{h} \left(a_0 + \langle \mathbf{b}_0 , \mathbf{z} \rangle \right)\right] \\
    \times \text{Ai}\left( \frac{2 \langle \mathbf{b}_1 , \mathbf{z} \rangle}{2^{2/3}h^{2/3}a_3^{1/3}} \right) + \mathcal{O}(h^{2/3}) .
    \label{eq:Airy-caustic}
  \end{multline}

  \subsection{Cusp caustic: Pearcey approximation} \label{appsub:Pearcey}
  
  \subsubsection{Leading order approximation} \label{appsub:pearcey-leading-approx}
   
  For a point on the cusp caustic, the third derivative of the action vanishes as well, but the fourth derivative does not. We therefore expand the action up to fourth order in $\eta$, similar to Eq.~(\ref{eq:S-exp-Airy}):
  \begin{multline}
    S(\mathbf{x},\eta) = S^{(4)}(\mathbf{x},\eta) + \mathcal{O}(\beta^5) = q_0(\mathbf{z}) + q_1(\mathbf{z}) \beta + \\ \frac{q_2(\mathbf{z})}{2} \beta^2 + \frac{q_3(\mathbf{z})}{6} \beta^3 + \frac{q_4(\mathbf{z})}{24} \beta^4 + \mathcal{O}(\beta^5)
     , \label{eq:S-exp-Pearcey}
  \end{multline}
  where $\beta=\eta-\eta_0$. As in the previous section, we expand the coefficients $q_i(\mathbf{z})$ up to first order in $z$. The expansions of $q_0(\mathbf{z})$, $q_1(\mathbf{z})$ and $q_2(\mathbf{z})$ are equal to those in Eq.~(\ref{eq:q-exp-Airy}). For the other coefficients, we have
  \begin{equation}
    q_3(\mathbf{z}) = \mathcal{O}(z), \qquad q_4(\mathbf{z}) = a_4 + \mathcal{O}(z) . \label{eq:q-exp-Pearcey}
  \end{equation}

  For $\mathbf{x}$ near $\mathbf{x}_0$, we can then express the leading order term of the asymptotic expansion of the integral~(\ref{eq:int-generic}) in terms of the Pearcey function~\cite{Pearcey46} $\text{P}^\pm(x)$, which is defined by the integral
  \begin{equation}
    \text{P}^{\pm}(u,v) = \int_{-\infty}^\infty \exp\left( \pm i t^4 + i u t^2 + i v t \right) \, \diff t , \label{eq:def-Pearcey}
  \end{equation}
  where the superscript plus or minus corresponds to the sign in front of the $t^4$ term. This function has two important symmetries, namely~\cite{Connor81a,Connor82}
  \begin{equation}
    \text{P}^\pm(u,-v)=\text{P}^\pm(u,v), \quad \text{P}^-(u,v) = [\text{P}^+(-u,-v)]^* . \label{eq:symm-Pearcey}
  \end{equation}
  These can be easily verified using the definition~(\ref{eq:def-Pearcey}). Furthermore, its two partial derivatives satisfy
  \begin{equation}
    \text{P}_v^\pm(u,-v) = -\text{P}_v^\pm(u,v), \quad \text{P}_u^\pm(u,-v) = \text{P}_u^\pm(u,v) .
  \end{equation}
  Although the Pearcey function is not implemented in most computer algebra systems, it can be efficiently computed using the methods in Refs.~\onlinecite{Connor81a,Connor82}.
  
  As in the previous section, we first make the substitution $\eta=\beta+\eta_0$, and subsequently $\beta = q \gamma - q_3/q_4$, where $q = \sqrt[4]{24 h/ |q_4|}$, in the integral $I(\mathbf{x},h)$. Making a Taylor expansion of $f(\mathbf{x},\eta)$ in $\eta$ around $\eta_0$, we see that the leading order term in the asymptotic expansion is $\mathcal{O}(h^{1/4})$. The terms of $\mathcal{O}(\beta^5)$ in the action give a contribution of $\mathcal{O}(h^{1/2})$, as does the first order term in the Taylor expansion of $f(\mathbf{x},\eta)$. Therefore,
  \begin{equation}
    I(\mathbf{x},h) = \int_{-\infty}^\infty \text{d} \eta \, f(\mathbf{x},\eta_0) e^{i S^{(4)}(\mathbf{x},\eta)/h} + \mathcal{O}(h^{1/2}) . \label{eq:asymp-Pearcey-prep-0}
  \end{equation}
  After performing the aforementioned substitution, we obtain the following expression for $I(\mathbf{x},h)$:
  \begin{widetext}
  \begin{equation}
    f(\mathbf{x},\eta_0) \sqrt[4]{\frac{24 h}{|q_4|}} \exp\left[\frac{i}{h}\left( q_0 -\frac{q_1 q_3}{q_4} + \frac{q_2 q_3^2}{2 q_4^2} - \frac{q_3^4}{8 q_4^3} \right)\right] 
    \text{P}^{\pm} \left[ \sqrt{\frac{6}{h |q_4|}} \left(q_2 - \frac{q_3^2}{2 q_4} \right) , \sqrt[4]{\frac{24}{h^3 |q_4|}} \left( q_1 + \frac{q_3^3}{2 q_4^2} - \frac{q_2 q_3}{q_4} \right) \right]+\mathcal{O}(h^{1/2}) \label{eq:asymp-Pearcey-prep}
  \end{equation}
  \end{widetext}
  The sign in $\text{P}^{\pm}$ is taken as the sign of $q_4$, and hence as the sign of $a_4$.
  
  Making use of the expansions of the $q_i(\mathbf{z})$, given in Eqs.~(\ref{eq:q-exp-Pearcey}) and~(\ref{eq:q-exp-Airy}), we find that
  \begin{align}
    & q_0 -\frac{q_1 q_3}{q_4} + \frac{q_2 q_3^2}{2 q_4^2} - \frac{q_3^4}{8 q_3^3} = a_0 + \langle \mathbf{b}_0 , \mathbf{z} \rangle + \mathcal{O}(z^2) , \label{eq:S0-Pearcey} \\
    & \frac{q_2 - \frac{q_3^2}{2 q_4}}{h^{1/2} |q_4|^{1/2}} = \frac{\langle \mathbf{b}_2 , \mathbf{z} \rangle}{h^{1/2} |a_4|^{1/2}} + \frac{\mathcal{O}(z^2)}{h^{1/2}} \label{eq:arg1-Pearcey} \\
    & \frac{q_1 + \frac{q_3^3}{2 q_4^2} - \frac{q_2 q_3}{q_4}}{h^{3/4} |q_4|^{1/4}} = \frac{\langle \mathbf{b}_1 , \mathbf{z} \rangle}{h^{3/4} |a_4|^{1/4}} + \frac{\mathcal{O}(z^2)}{h^{3/4}} \label{eq:arg2-Pearcey} .
  \end{align}
  Following the reasoning in the previous section, we then demand that both arguments of the Pearcey function are $\mathcal{O}(h^\delta_i)$, with $\delta_i>0$. Setting for example $\min(\delta_i)=1/8$, we see from Eqs.~(\ref{eq:arg1-Pearcey}) and~(\ref{eq:arg2-Pearcey}) that a safe estimate for the neighborhood in which we can use Eq.~(\ref{eq:asymp-Pearcey-prep}) is $\mathcal{O}(h^{7/8})$.
  
  Assuming that $\mathbf{z} = \mathcal{O}(h^{7/8})$, we find that $\mathcal{O}(z^2) h^{-3/4} = \mathcal{O}(h)$, $\mathcal{O}(z^2) h^{-1/4} = \mathcal{O}(h^{3/2})$ and $\mathcal{O}(z^2) h^{-1} = \mathcal{O}(h^{3/4})$. Therefore, for this neighborhood, the largest error that we introduce in making a first order Taylor expansion of the coefficients $q_i(\mathbf{z})$ is $\mathcal{O}(h)$, which is much smaller than the errors of $\mathcal{O}(h^{1/2})$ that are introduced in Eq.~(\ref{eq:asymp-Pearcey-prep-0}).
  
  Using the above results, and replacing $f(\mathbf{x},\eta_0)$ by its zeroth order Taylor approximation around $\mathbf{x}_0$, we can simplify Eq.~(\ref{eq:asymp-Pearcey-prep}) to
  \begin{multline}
    I(\mathbf{x},h)= f(\mathbf{x}_0,\eta_0) \sqrt[4]{\frac{24 h}{|a_4|}} \exp\left[\frac{i}{h}\left( a_0 + \langle \mathbf{b}_0 , \mathbf{z} \rangle \right)\right] \quad \\ 
    \quad \times \text{P}^{\pm} \left[ \sqrt{\frac{6}{h |a_4|}} \langle \mathbf{b}_2 , \mathbf{z} \rangle , \sqrt[4]{\frac{24}{h^3 |a_4|}} \langle \mathbf{b}_1 , \mathbf{z} \rangle \right] + \mathcal{O}(h^{1/2}). \label{eq:Pearcey-caustic}
  \end{multline}
  We can use this approximation in an $\mathcal{O}(h^{7/8})$-neighborhood of the point $\mathbf{x}_0$.

  \subsubsection{Higher order corrections}  \label{appsub:pearcey-higher-order-approx}
  
  Let us now look at higher order corrections to Eq.~(\ref{eq:Pearcey-caustic}). In the previous section, we identified two sources of corrections of $\mathcal{O}(h^{1/2})$, namely the terms of $\mathcal{O}(\beta^5)$ in the action, and the first order term in the Taylor expansion of $f(\mathbf{x},\eta)$ in $\eta$. We also saw that corrections that come from the Taylor approximations in $\mathbf{x}$ are of $\mathcal{O}(h)$. Therefore, let us look at higher order terms in the Taylor expansion of the action, i.e.~\cite{DobrokhotovPrivateCommunication}
  \begin{equation}
    S(\mathbf{x},\eta) = S^{(4)}(\mathbf{x},\eta) + \frac{q_5(\mathbf{z})}{5!}\beta^5 + \frac{q_6(\mathbf{z})}{6!}\beta^6 + \mathcal{O}(\beta^7) , \label{eq:S-exp-Pearcey-corr}
  \end{equation}
  where $S^{(4)}$ was defined in Eq.~(\ref{eq:S-exp-Pearcey}). Let us also consider higher order terms in the Taylor expansion of the amplitude $f(\mathbf{x},\eta)$ in $\eta$, that is,~\cite{DobrokhotovPrivateCommunication}
  \begin{equation}
    f(\mathbf{x},\eta) = f(\mathbf{x},\eta_0) + f_\eta(\mathbf{x},\eta_0) \beta + \frac{1}{2} f_{\eta\eta}(\mathbf{x},\eta_0) \beta^2 + \mathcal{O}(\beta^3) . \label{eq:amp-exp-Pearcey-corr}
  \end{equation}
  When we make the substitutions $\eta=\beta+\eta_0$ and $\beta = q \gamma - q_3/q_4$, where $q = \sqrt[4]{24 h/ |q_4|}$, in Eq.~(\ref{eq:int-generic}), we see that $q_5(\mathbf{z})\beta^5/h$ is $\mathcal{O}(h^{1/4})$ and that $q_6(\mathbf{z})\beta^6/h$ is $\mathcal{O}(h^{1/2})$. Therefore, these terms are small compared to $S^{(4)}(\mathbf{x},\eta)/h$, which is of order one, and we can make a Taylor expansion of the exponent. Gathering all terms of the same order, we obtain
  \begin{widetext}
    \begin{equation}
      \begin{aligned}
        I(\mathbf{x},h) = &
        \underbrace{\int_{-\infty}^\infty \diff\beta f(\mathbf{x}_0,\eta_0) e^{i S^{(4)}(\mathbf{x},\eta)/h}}_{\mathcal{O}(h^{1/4})} +
        \underbrace{\int_{-\infty}^\infty \diff\beta f_\eta(\mathbf{x}_0,\eta_0) \beta e^{i S^{(4)}(\mathbf{x},\eta)/h} + 
          \int_{-\infty}^\infty \diff\beta f(\mathbf{x}_0,\eta_0)\frac{i}{h}\frac{q_5(\mathbf{z})}{5!}\beta^5 e^{i S^{(4)}(\mathbf{x},\eta)/h}
          }_{\mathcal{O}(h^{1/2})} + \\
        &\underbrace{\int_{-\infty}^\infty \diff\beta f_{\eta\eta}(\mathbf{x}_0,\eta_0) \frac{\beta^2}{2} e^{i S^{(4)}(\mathbf{x},\eta)/h} +
        \int_{-\infty}^\infty \diff\beta f_\eta(\mathbf{x}_0,\eta_0) \beta \frac{i}{h}\frac{q_5(\mathbf{z})}{5!}\beta^5 e^{i S^{(4)}(\mathbf{x},\eta)/h}
          }_{\mathcal{O}(h^{3/4})} + \\
        &\underbrace{\int_{-\infty}^\infty \diff\beta f(\mathbf{x}_0,\eta_0) \frac{1}{2}\left(\frac{i}{h}\frac{q_5(\mathbf{z})}{5!}\beta^5\right)^2 e^{i S^{(4)}(\mathbf{x},\eta)/h} +
        \int_{-\infty}^\infty \diff\beta f(\mathbf{x}_0,\eta_0)\frac{i}{h}\frac{q_6(\mathbf{z})}{6!}\beta^6 e^{i S^{(4)}(\mathbf{x},\eta)/h}
          }_{\mathcal{O}(h^{3/4})} + \mathcal{O}(h) .
      \end{aligned}
      \label{eq:Pearcey-corrs}
    \end{equation}
  \end{widetext}
  We know from the discussion in the previous section that the first term becomes Eq.~(\ref{eq:Pearcey-caustic}). Let us now look at the first term of $\mathcal{O}(h^{1/2})$. After we have performed the substitutions and have discarded all terms of $\mathcal{O}(h)$ and higher, we find that it equals
  \begin{widetext}
    \begin{align}
      \int_{-\infty}^\infty \diff\beta f_\eta(\mathbf{x}_0,\eta_0) \beta e^{i S^{(4)}(\mathbf{x},\eta)/h} 
        &= q^2 f_\eta(\mathbf{x}_0,\eta_0) \int_{-\infty}^\infty \diff\gamma \, \gamma \, e^{i S^{(4)}(\mathbf{x},\eta)/h} \nonumber \\
        & = -i f_\eta(\mathbf{x}_0,\eta_0) \left(\frac{24 h}{|a_4|}\right)^{1/2} \exp\left[\frac{i}{h}\left( a_0 + \langle \mathbf{b}_0 , \mathbf{z} \rangle \right)\right] \text{P}_v^{\pm} \left[ \sqrt{\frac{6}{h |a_4|}} \langle \mathbf{b}_2 , \mathbf{z} \rangle , \sqrt[4]{\frac{24}{h^3 |a_4|}} \langle \mathbf{b}_1 , \mathbf{z} \rangle \right] .
      \label{eq:Pearcey-corr-1}
    \end{align}
  \end{widetext}
  One can prove the last equality by using the same arguments as in the previous subsection. By $\text{P}_v^{\pm}$ we mean the derivative of the Pearcey function with respect to its second argument, given by  
  \begin{equation}
    \text{P}_v^{\pm}(u,v) = i \int_{-\infty}^\infty t \exp\left( \pm i t^4 + i u t^2 + i v t \right) \, \diff t .  \label{eq:def-Pearcey-diff-v}
  \end{equation}
  In a similar way, one can express the first term of $\mathcal{O}(h^{3/4})$ in Eq.~(\ref{eq:Pearcey-corrs}) in terms of the derivative of the Pearcey function with respect to its first argument, given by
  \begin{equation}
    \text{P}_u^{\pm}(u,v) = i \int_{-\infty}^\infty t^2 \exp\left( \pm i t^4 + i u t^2 + i v t \right) \, \diff t . \label{eq:def-Pearcey-diff-u}
  \end{equation}
  After some algebra, we obtain
  \begin{widetext}
    \begin{align}
      \int_{-\infty}^\infty \diff\beta f_{\eta\eta}(\mathbf{x}_0,\eta_0) \frac{\beta^2}{2} e^{i S^{(4)}(\mathbf{x},\eta)/h}
       = -\frac{i}{2} f_{\eta\eta}(\mathbf{x}_0,\eta_0) \left(\frac{24 h}{|a_4|}\right)^{3/4} \exp\left[\frac{i}{h}\left( a_0 + \langle \mathbf{b}_0 , \mathbf{z} \rangle \right)\right] \text{P}_u^{\pm} \left[ \sqrt{\frac{6}{h |a_4|}} \langle \mathbf{b}_2 , \mathbf{z} \rangle , \sqrt[4]{\frac{24}{h^3 |a_4|}} \langle \mathbf{b}_1 , \mathbf{z} \rangle \right]
      \label{eq:Pearcey-corr-2}
    \end{align}
  \end{widetext}
  We will not go into the other terms in Eq.~(\ref{eq:Pearcey-corrs}) at this point, as they will prove to be irrelevant for the problem that we discuss in the main text.

  \subsection{Uniform approximation near the cusp}  \label{appsub:uniform-cusp}
  
  In the previous sections, we performed a Taylor expansion of the action until the first nonvanishing term, and constructed an approximation for the integral~(\ref{eq:int-generic}) based on this expansion. Using the theorems of catastrophe theory, a uniform approximation of $I(\mathbf{x},h)$ can be constructed. For points near the fold caustic this was first done in Ref.~\onlinecite{Chester57}, and for the cusp caustic in Refs.~\onlinecite{Ursell72,Connor81b}. In this appendix, we summarize the construction of the uniform approximation near a cusp caustic as presented in Refs.~\onlinecite{Ursell72,Connor81b}, using slightly different conventions.
  
  Consider a cusp point $(\mathbf{x}_0, \eta_0)$, at which the first three derivatives of the action $S$ with respect to $\eta$ vanish. Then, for points $\mathbf{x}$ in the vicinity of this cusp point $\mathbf{x}_0$, a transformation $\chi=\chi(\mathbf{x},\eta)$ exists, with inverse transformation $\eta=\eta(\mathbf{x},\chi)$, such that the action in the new variable $\chi$ has the form~\cite{Poston78,Arnold82,Ursell72}
  \begin{equation}
    S(\mathbf{x},\eta) = \pm \chi^4 + w_2(\mathbf{x}) \chi^2 + w_1(\mathbf{x}) \chi + w_0(\mathbf{x}) ,   \label{eq:action-normal-form}
  \end{equation}
  where the sign in front of $\chi^4$ equals the sign of $\partial^4 S/\partial \eta^4$ and both $w_1(\mathbf{x}_0)=0$ and $w_2(\mathbf{x}_0)=0$. Note that this is an exact transformation and that we are no longer using a truncated Taylor expansion here.
  
  Changing our integration variable in the integral~(\ref{eq:int-generic}) from $\eta$ to $\chi$, we obtain
  \begin{equation}
    I(\mathbf{x},h) = e^{\frac{i}{h} w_0(\mathbf{x})} \int_{-\infty}^\infty \diff \chi \, F(\mathbf{x},\chi) e^{\frac{i}{h} (\pm \chi^4 + w_2(\mathbf{x}) \chi^2 + w_1(\mathbf{x}) \chi)} ,
    \label{eq:I-changevar-normal-form}
  \end{equation}
  where we have introcuded the new amplitude function
  \begin{equation}
    F(\mathbf{x},\chi) = \left| \frac{\diff \eta}{\diff \chi} \right| f(\mathbf{x},\eta(\chi)) .
    \label{eq:def-amp-changevar}
  \end{equation}
  Subsequently, we expand $F(\mathbf{x},\chi)$ up to second order in $\chi$, i.e.
  \begin{equation}
    F(\mathbf{x},\chi) = A_0(\mathbf{x}) + B_0(\mathbf{x}) \chi + C_0(\mathbf{x}) \chi^2 + \mathcal{O}(\chi^3).
    \label{eq:uni-amp-expansion}
  \end{equation}
  One can then show that the following equality holds~\cite{Ursell72}
  \begin{widetext}
    \begin{equation}
      I(\mathbf{x},h) = e^{\frac{i}{h} w_0} \left[ h^{1/4} A_0 P^\pm\left(\frac{w_2}{h^{1/2}},\frac{w_1}{h^{3/4}}\right) 
          -i h^{1/2} B_0 P^\pm_v\left(\frac{w_2}{h^{1/2}},\frac{w_1}{h^{3/4}}\right)
          -i h^{3/4} C_0 P^\pm_v\left(\frac{w_2}{h^{1/2}},\frac{w_1}{h^{3/4}}\right) \right] + \mathcal{O}(h^{5/4}) ,
      \label{eq:Pearcey-uni-result}
    \end{equation}
  \end{widetext}
  where the derivatives of the Pearcey function were defined in Eqs.~(\ref{eq:def-Pearcey-diff-v}) and~(\ref{eq:def-Pearcey-diff-u}) and the sign in the definition of the Pearcey function corresponds to the sign in front of $\chi^4$ in Eq.~(\ref{eq:action-normal-form}). When higher order terms in the expansion~(\ref{eq:uni-amp-expansion}) are taken into account, the constants $A_0$, $B_0$ and $C_0$ in Eq.~(\ref{eq:Pearcey-uni-result}) are replaced by series in integer powers of $h$, as proven in Ref.~\onlinecite{Ursell72}.
  
  In order to compute the solution~(\ref{eq:Pearcey-uni-result}), we need to obtain the parameters $w_0$, $w_1$ and $w_2$, and $A_0$, $B_0$ and $C_0$ for a given point $\mathbf{x}$. In order for the mapping $\chi=\chi(\mathbf{x},\eta)$ to be one-to-one, the stationary points on the left-hand side of Eq.~(\ref{eq:action-normal-form}) should correspond to those on the right-hand side. Concerning the left-hand side of the equation, let us assume that we know the action $S_i=S(\mathbf{x},\eta_i)$ at the three stationary points $\eta_{1,2,3}$ of $S(\mathbf{x},\eta)$ for a given $\mathbf{x}$. On the right-hand side, the stationary points are defined by the equation
  \begin{equation}
    \pm 4\chi^3 + 2 w_2(\mathbf{x}) \chi + w_1(\mathbf{x}) .  \label{eq:roots-chi}
  \end{equation}
  When the discriminant
  \begin{equation}
    \Delta = \mp 2^7 w_2^3 - 2^4 3^3 w_1^2   \label{eq:discriminant}
  \end{equation}
  is positive, this cubic equation has three distinct real roots. When $\Delta$ is negative, the equation has one real root and two complex conjugate roots and when the discriminant vanishes, all roots are real, but there is a multiple root. We call these three roots of Eq.~(\ref{eq:roots-chi}) $\chi_{1,2,3}$. Note that in a practical implementation, it is important that both sets of stationary points are ordered in the same way, e.g. from small to large when all numbers are real. When this is not the case, one could for instance take the first stationary point to be the real one, followed by the two complex ones ordered by their imaginary part.
  
  Requiring that the stationary points on both sides of Eq.~(\ref{eq:action-normal-form}) coincide, we find that the following set of equalities has to hold.
  \begin{equation}
    \begin{aligned}
      S_1 &= \pm \chi_1^4 + w_2(\mathbf{x}) \chi_1^2 + w_1(\mathbf{x}) \chi_1 + w_0(\mathbf{x}), \\
      S_2 &= \pm \chi_2^4 + w_2(\mathbf{x}) \chi_2^2 + w_1(\mathbf{x}) \chi_2 + w_0(\mathbf{x}), \\
      S_3 &= \pm \chi_3^4 + w_2(\mathbf{x}) \chi_3^2 + w_1(\mathbf{x}) \chi_3 + w_0(\mathbf{x}).
      \label{eq:actions-match-stationary}
    \end{aligned}
  \end{equation}
  Let us, for simplicity, disregard the case of a degenerate critical point for a moment. Then we can subtract the second equation in~(\ref{eq:actions-match-stationary}) from the first and the third from the first to eliminate $w_0$, which gives
  \begin{equation}
    \begin{aligned}
      S_1-S_2 &= \pm (\chi_1^4-\chi_2^4) + 2 w_2 (\chi_1^2-\chi_2^2) + w_1 (\chi_1-\chi_2) , \\
      S_1-S_3 &= \pm (\chi_1^4-\chi_3^4) + 2 w_2 (\chi_1^2-\chi_3^2) + w_1 (\chi_1-\chi_3) .
      \label{eq:self-consistency-chi}
    \end{aligned}
  \end{equation}
  Given initial guesses for the parameters $w_1$ and $w_2$, one can find the three stationary points $\chi_{1,2,3}$ from Eq.~(\ref{eq:roots-chi}). These can be inserted into Eq.~(\ref{eq:self-consistency-chi}) to find new values for $w_1$ and $w_2$ and this process can be iterated until self-consistency is reached. As initial guesses for the parameters $w_1$ and $w_2$ one can use the result of the Taylor expansion of the action,
  \begin{equation}
    w_{2,0}=\sqrt{\frac{6}{|a_4|}} \langle \mathbf{b}_2 , \mathbf{z} \rangle, \quad w_{1,0} = \sqrt[4]{\frac{24}{|a_4|}} \langle \mathbf{b}_1 , \mathbf{z} \rangle ,
  \end{equation}
  cf. Eq.~(\ref{eq:Pearcey-caustic}). For a degenerate critical point, we can use the fact that the discriminant $\Delta$ vanishes to obtain a relation between $w_2$ and $w_1$. Inserting this into Eq.~(\ref{eq:roots-chi}), one obtains expressions~\cite{Connor81b} for $\chi_{1,2,3}$ in terms of $w_1/w_2$. We can then obtain $w_2$ from Eq.~(\ref{eq:self-consistency-chi}). For more details, we refer to Ref.~\onlinecite{Connor81b}. Alternatively, one can obtain $w_2$, $w_1$ and $w_0$ by using an algebraic method, that was described in Ref.~\onlinecite{Connor81b}. However, this method requires tracing certain solutions across the caustic, and we therefore find it less convenient.
  
  Finally, one needs to determine the parameters $A_0$, $B_0$ and $C_0$ in the expansion~(\ref{eq:uni-amp-expansion}). By combining Eqs.~(\ref{eq:def-amp-changevar}) and~(\ref{eq:uni-amp-expansion}) and evaluating the result at the three critical points $\chi_i$, we arrive at a system of three linear equations
  \begin{equation}
    \begin{aligned}
      \left| \frac{\diff \eta}{\diff \chi} \right|_{\chi=\chi_i} f(\mathbf{x},\eta_i) = A_0 + B_0 \chi_i + C_0 \chi_i^2 ,
    \end{aligned}
    \label{eq:uni-amp-system}
  \end{equation}
  from which $A_0$, $B_0$ and $C_0$ can easily be found. However, this requires the computation of the derivative $\diff \eta/\diff \chi$ at the stationary points. By taking the second derivative with respect to $\chi$ on both sides of Eq.~(\ref{eq:action-normal-form}), we find that 
  \begin{equation}
    \frac{\partial^2 S}{\partial \eta^2} \left(\frac{\diff \eta}{\diff \chi}\right)^2 + \frac{\partial S}{\partial \eta}\frac{\diff^2 \eta}{\diff \chi^2} = \pm 12 \chi^2 + 2 w_2 .
    \label{eq:action-normal-form-second-deriv}
  \end{equation}
  We are interested in the value of this derivative at the critical points, where the second term on the left-hand side vanishes by definition. When the stationary point is nondegenerate, the first term on the left-hand side is nonzero and we obtain
  \begin{equation}
    \left. \frac{\diff \eta}{\diff \chi} \right|_{\chi=\chi_i} = \left( \frac{\pm 12 \chi_i^2 + 2 w_2}{\partial^2 S/\partial \eta^2} \right)^{1/2} .
    \label{eq:deriv-eta-chi}
  \end{equation}
  When we are dealing with a fold point, where $\partial^2 S/\partial \eta^2$ vanishes as well, we can take the derivative of expression~(\ref{eq:action-normal-form-second-deriv}) with respect to $\chi$ once more to obtain an equation for $\diff\eta/\diff\chi$. For the cusp point, we can obtain $\diff\eta/\diff\chi$ by differentiating Eq.~(\ref{eq:action-normal-form-second-deriv}) twice.
  Combining Eqs.~(\ref{eq:uni-amp-system}) and~(\ref{eq:deriv-eta-chi}), we see that we can obtain $A_0$, $B_0$ and $C_0$ when we know the values of $f(\mathbf{x},\eta)$ and $\partial^2 S/\partial \eta^2$ at the critical points.

  \section{Initial value problem} \label{app:initial-value-problem}
  
  In this appendix, we construct the solution of the initial value problem~(\ref{eq:Dirac-initial-value}), considered in section~\ref{sec:curr-edge-samp}. First, we take the Fourier transform of both the equation and the initial condition with respect to $y$, which gives
  \begin{multline}
    \left[ \sigma_x \hat{p}_x + \sigma_y p_y + U(\mathbf{x}) \right] \overline{\Psi}(x,p_y) = E \overline{\Psi}(x,p_y) , \\ 
    \overline{\Psi}(x_s,p_y) = \overline{\Psi}_0(p_y) .
    \label{eq:Dirac-initial-value-FT}
  \end{multline}
  In appendix~\ref{appsub:scat-states}, we constructed two linearly independent solutions of the eigenvalue problem, $\overline{\Psi}_>(x)$ and $\overline{\Psi}_<(x)$, see Eqs.~(\ref{eq:Psi-inc-right}) and~(\ref{eq:Psi-inc-left}). The solution of Eq.~(\ref{eq:Dirac-initial-value-FT}) is a linear combination of these two solutions, i.e.
  \begin{equation}
    \overline{\Psi}(x,p_y) = c_1 \overline{\Psi}_>(x) + c_2 \overline{\Psi}_<(x) = 
      \Big(
        \overline{\Psi}_>(x) \;\;\; \overline{\Psi}_<(x)
      \Big)
      \begin{pmatrix}
        c_1 \\ c_2
      \end{pmatrix} , \label{eq:Phi-initial-value}
  \end{equation}
  that satisfies the initial condition. Here, $\big(\overline{\Psi}_>(x) \;\;\; \overline{\Psi}_<(x)\big)$ denotes the matrix with columns $\overline{\Psi}_>(x)$ and $\overline{\Psi}_<(x)$. Inserting the initial condition $\overline{\Psi}(x_s,p_y) = \overline{\Psi}_0(p_y)$ into Eq.~(\ref{eq:Phi-initial-value}) and multiplying by the inverse of the matrix on the right-hand side, we find that
  \begin{equation}
    \begin{pmatrix}
        c_1 \\ c_2
    \end{pmatrix} =
    \Big(
        \overline{\Psi}_>(x_s) \;\;\; \overline{\Psi}_<(x_s) 
    \Big)^{-1} \; \overline{\Psi}_0(p_y) .
  \end{equation}
  Combining the previous results and taking the inverse Fourier transform with respect to $p_y$, we find the solution of the initial value problem~(\ref{eq:Dirac-initial-value}) as
  \begin{multline}
    \Psi(x,y) = \frac{e^{i \pi/4}}{\sqrt{2\pi h}} \int_{-E}^E \diff p_y \;
      e^{i p_y y/h}
      \Big(
        \overline{\Psi}_>(x) \;\;\; \overline{\Psi}_<(x)
      \Big) \\
      \times
      \Big(
        \overline{\Psi}_>(x_s) \;\;\; \overline{\Psi}_<(x_s) 
      \Big)^{-1} \; \overline{\Psi}_0(p_y) ,
    \label{eq:wavefunction-initial-value}
  \end{multline}
  where the integration limits are $\pm E$, as states with higher transversal momentum do not propagate in the sample, see also the discussion at the end of appendix~\ref{app:Green}.
  
  When the electronic current only flows into the sample from the left-hand side, we can confine our attention to the term proportional to $\overline{\Psi}_>(x)$ (see the main text). This means that we are only interested in the first component of the vector $\big( \overline{\Psi}_>(x_s) \;\;\; \overline{\Psi}_<(x_s) \big)^{-1} \; \overline{\Psi}_0(p_y)$. After some calculus, we find that this first component equals
  \begin{multline}
    \frac{-i}{\det \big( \overline{\Psi}_>(x_s) \;\;\; \overline{\Psi}_<(x_s) \big)} \overline{\Psi}_<(x_s)^T \sigma_y  \overline{\Psi}_0(p_y) = \\
      \frac{1}{t'} \frac{1}{\sqrt{2\cos\phi}} t' \Big( e^{-i \phi/2} \;\;\; e^{i \phi/2} \Big) e^{-i \sqrt{E^2-p_y^2} x_s/h} \overline{\Psi}_0(p_y) . \nonumber
  \end{multline}
  Therefore, we can approximate the full solution~(\ref{eq:wavefunction-initial-value}) by
  \begin{multline}
    \Psi(x,y) = \frac{e^{i \pi/4}}{\sqrt{2\pi h}} \int_{-E}^E \diff p_y \; e^{i p_y y/h} e^{-i \sqrt{E^2-p_y^2} x_s/h} \, \overline{\Psi}_>(x) \\
      \times \frac{1}{\sqrt{2 \cos\phi}} \Big( e^{-i \phi/2} \;\;\; e^{i \phi/2} \Big) \overline{\Psi}_0(p_y) .
    \label{eq:sol-fwd-initial-value-app}
  \end{multline}
  When one considers a situation with $U_0-E<E$, and is interested only in the hole region, the integration limits should be reduced to $\pm p_{y,\text{max}}$, as discussed at the end of appendix~\ref{app:Green}.

\end{document}